\documentclass[11pt]{article}
\usepackage{jheppub}
\usepackage{graphicx} 
\usepackage{color}
\usepackage{amssymb}
\usepackage{amsmath}
\usepackage{mathtools}
\usepackage{hyperref}
\usepackage{tocloft}
\usepackage{float}
\usepackage{pgfplotstable}
\usepackage{booktabs}
\usepackage{array}
\usepackage{colortbl}
\usepackage[stable]{footmisc}   
\usepackage[normalem]{ulem}               
\usepackage{braket}
\usepackage{url}
\usepackage{upgreek}
\usepackage{caption}
\usepackage{cleveref}
\usepackage{bm}
\usepackage{ifthen}             
\newcommand\ord[1]{\mathcal{O}\left(#1\right)}

\newcommand\ignore[1]{}

\newcommand{\gap}{\mathrm{gap}}

\newcommand{\msout}[1]{\text{\sout{\ensuremath{#1}}}}  
\newcommand{\be}{\begin{equation}}
\newcommand{\ee}{\end{equation}}
\newcommand{\bea}{\begin{eqnarray}}
\newcommand{\eea}{\end{eqnarray}}
\newcommand{\bgd}{\begin{gather}}
\newcommand{\egd}{\end{gather}}
\newcommand{\bsp}{\begin{split}}
\newcommand{\esp}{\end{split}}

\newcommand{\smfrac}[2]{ {\scriptstyle \frac{#1}{#2} } }
\newcommand{\half}{ {\scriptstyle \frac{1}{2} } }

\newcommand{\g}{\gamma}
\newcommand{\f}{\frac}
\newcommand\lr[1]{{\left({#1}\right)}}

\newcommand\n{ \nonumber}

\renewcommand{\epsilon}{\varepsilon}
\renewcommand{\phi}{\varphi}

\newcommand{\dd}{\,\mathrm{d}}

\newcommand{\mr}{\mathrm}
\renewcommand{\vec}[1]{\ifthenelse{\equal{#1}{\ell}}{\boldsymbol{#1}}{\mathbf{#1}}}
\newcommand{\abs}[1]{\left\lvert #1\right\rvert}
\renewcommand{\l}{\left}
\renewcommand{\r}{\right}

\DeclareFontFamily{OT1}{pzc}{}
\DeclareFontShape{OT1}{pzc}{m}{it}{<-> s * [1.10] pzcmi7t}{}
\DeclareMathAlphabet{\mathpzc}{OT1}{pzc}{m}{it}

\numberwithin{equation}{section}
\numberwithin{figure}{section}

\makeatletter
\DeclareRobustCommand{\textsupsub}[2]{{%
  \m@th\ensuremath{%
    ^{\mbox{\fontsize\sf@size\z@#1}}%
    _{\mbox{\fontsize\sf@size\z@#2}}%
  }%
}}
\makeatother

\newlength\Myfigwd

\usepackage{cleveref}
\crefname{figure}{}{figures}
\creflabelformat{figure}{#2\txtkublue{\textsf{\textbf{\footnotesize Figure #1}}}#3}
\crefname{table}{}{tables}
\creflabelformat{table}{#2\txtkublue{\textsf{\textbf{\footnotesize Table #1}}}#3}
\crefname{section}{}{sections}
\creflabelformat{table}{#2\txtkublue{\textsf{\textbf{\footnotesize Section #1}}}#3}

\usepackage{pbox}
\usepackage{hhline}

\graphicspath{{graphics/}}

\title{First computation of Mueller Tang processes using the full NLL BFKL approach}

\author[1,2]{Dimitri Colferai}
\author[3]{Federico Deganutti}
\author[3,4]{Timothy G Raben}
\author[3]{Christophe Royon}

\affiliation[1]{Istituto Nazionale di Fisica Nucleare, Sezione di Firenze}
\affiliation[2]{Dipartimento di Fisica ed Astronomia, Universit\`a degli Studi di Firenze}
\affiliation[3]{Department of Physics and Astronomy, University of Kansas}
\affiliation[4]{Department of Physics and Astronomy, Michigan State University}

\emailAdd{colferai@fi.infn.it}
\emailAdd{fedeganutti@ku.edu}
\emailAdd{rabentim@msu.edu}
\emailAdd{christophe.royon@ku.edu}

\abstract{
We present the full next-to-leading order (NLO) prediction for the jet-gap-jet cross section at the LHC within the BFKL approach. We implement, for the first time, the NLO impact factors in the calculation of the cross section.
We provide results for differential cross sections as a function of the difference in rapidity and azimuthal angle betwen the two jets and the second leading jet transverse momentum.
The NLO corrections of the impact factors induce an overall reduction of the cross section with respect to the corresponding predictions with only LO impact factors.

We note that NLO impact factors feature a logarithmic dependence of the cross section on the total center of mass energy which formally violates BFKL factorization.
We show that such term is one order of magnitude smaller than the total contribution, and thus can be safely included in the current prediction without a need of further resummation of such logarithmic terms.

Fixing the renormalization scale $\mu_R$ according to the principle of minimal sensitivity, suggests $\mu_R$ about 4 times the sum of the transverse jet energies and provides smaller theroretical uncertainties with respect to the leading order case.}

\begin{document}

\maketitle

\tableofcontents

\section{Introduction}

In the high-energy regime, when the center-of-mass energy $\sqrt{s}$ greatly surpasses the typical transverse scale $\sqrt{-t}$ ($s\gg -t$), QCD is predicted to exhibit new dynamical behaviors. Provided that the scale driving the strong coupling $\alpha_s(-t)$ is well within the perturbative domain, i.e., $-t\gg
\Lambda^2_{\rm QCD}$, the high-energy limit can be understood under the
framework of the \emph{Balitsky-Fadin-Kuraev-Lipatov} (BFKL) expansion~\cite{Fadin1975, Balitsky78}: a
reformulation of the QCD perturbative series following a modified hierarchy where
the discrimination of the approximation order cannot be relegated to mere
coupling counting. Scattering amplitudes and cross-sections can involve large
logarithms of the ratio $s/(-t)\gg 1$ which can compensate the smallness
of the strong coupling $\alpha_s\ll 1$.  The perturbative series is thus
affected by large coefficients, so that all perturbative orders are
important. The BFKL framework consists precisely of the resummation of the
leading logarithmic (LL) terms of order $(\alpha_s\log s)^n$ and possibly of the
next-to-leading logarithmic (NLL) terms of order $\sim\alpha_s(\alpha_s\log s)^n$.

\begin{figure}[bh]
  \centering
\includegraphics[width=0.4\linewidth]{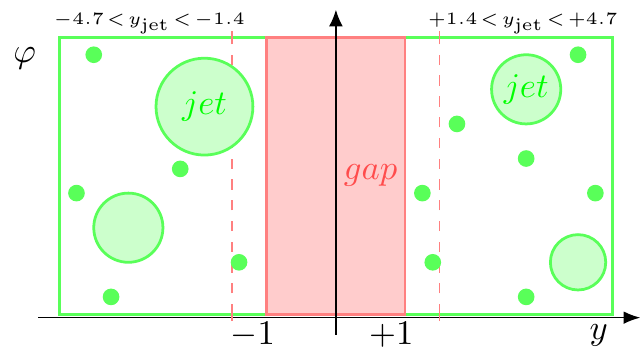}\hspace{10pt}
\vspace{-10pt}
\caption{Schematic of the gap and selection algorithm in rapidity-azimuthal space. }
  \label{fig:gap}
\end{figure}

Since the advent of modern colliders, great interest has been dedicated towards the search of a process able to clearly give away the influence of underlining BFKL dynamics. In the context of hadron-hadron colliders, Mueller and Tang (MT)~\cite{MuellerTang92} proposed to
investigate particular dijet processes where the interaction is conveyed by a
color-singlet exchange. These events are characterized by two hard jets which
are essentially back to back in the transverse (with respect to the beam axis) plane
(\(\vec{p}_{j_1}\simeq-\vec{p}_{j_2}\gtrsim 40\) GeV) and have a large
(pseudo-)rapidity separation $Y_{j_{12}}\equiv y_{j_1}-y_{j_2}$, with a sizable
part of this rapidity interval --- the so called gap --- devoid of
any detectable activity as shown in Fig.~\ref{fig:gap}.
More precisely, MT argued that the colourless BFKL hard pomeron becomes the
favored mean of interaction to produce jets at large rapidity separation
(\(Y_{j_{12}}\gtrsim 3\)) when no emission is allowed between them. On the contrary,
exchanges of coloured objects are very rarely associated with a rapidity gap,
tending to emit conspicuously in the central region. For such reason,
jet-gap-jet events are good candidates for the observation of BFKL dynamics.

The jet-gap-jet process is a very clean observable in the sense that there are very few background events. Its experimental signature
has been searched for in hadron collisions by the CDF and D0 Collaborations at the $\sqrt{s}=1.8$ TeV
Tevatron~\cite{abe1998dijet}, the D0 Collaboration at the $\sqrt{s}=630$ GeV Tevatron~\cite{abbott1998probing},  the joint
CMS and TOTEM Collaborations at the $\sqrt{s}=2.76$ TeV LHC~\cite{cms2021study} and the $\sqrt{s}=13$ TeV
LHC~\cite{collaborations2021hard}, and finally by the CMS Collaboration at the $\sqrt{s}=7$ TeV LHC~\cite{sirunyan2018study}.
However, since the first experimental studies at the
Tevatron~\cite{abe1998dijet, abbott1998probing}, the LL predictions failed to
accurately reproduce the observed cross-sections. Including the corrections to the gluon Green's function (GGF) at NLL shows an improved agreement with data from D0,
CDF at the Tevatron, and at the 7 TeV LHC~\cite{RoyonChevallier09:GapsBetweenJets,RoyonKepka10:GapsBetweenJets}---but substantial discrepancies still exist. However, the original Mueller-Tang description does not seem to describe the most recent measurements by the CMS Collaboration at 13 TeV~\cite{Baldenegro21:ColorSingletExchange13TeV}, especially the dependence on the difference in rapidity between the two jets.
It was recently shown~\cite{BaldenegroKlasenRoyon22:JetsSeparatedLargeRapidityGap} that Monte-Carlo implementations of the MT predictions can be very sensitive to the gap definition.

With this paper we aim at refining the theoretical predictions by including the next-to-leading order (NLO) impact factors (IF)
\footnote{We use the notation LL and NLL for the degree of accuracy of the GGF which resums the logarithms of the energy, while we use LO and NLO for impact factors which are computed as an ordinary expansion in $\alpha_s$.}
and, thus, completing the BFKL predictions of the Mueller Tang (MT) processes at NLL accuracy. The goal of our study is thus to complete the phenomenological analysis in the NLL approximation (NLLA). We thus present an improved NLL analysis of jet-gap-jet cross-sections by including contributions from the recently calculated NLO impact factors~\cite{HentschinskiSabioVera14:QuarkImpactFactor,HentschinskiSabioVera14:GluonImpactFactor}
on top of the usual approximated treatment of the NLL GGF\footnote{The exact description of the NLL BFKL GGF in the non-forward case has not been achieved yet.}.

In order to comply with the actual experimental procedure of jet-gap-jet event selection where low and non-detectable energy can flow into the gap, the original NLO impact factor~\cite{HentschinskiSabioVera14:QuarkImpactFactor,HentschinskiSabioVera14:GluonImpactFactor} has to be modified. As a result, additional logarithmic terms appear which are formally not expected as part of the BFKL factorization structure.
As we will discuss, this fact can be interpreted as a breakdown of BFKL factorization for MT jets in NLLA.
Although it represents a major theoretical issue, which deserves a separate study~\cite{lavoroDaFare}, we will show that such logarithmic terms are relatively small from the phenomenological point of view, so that the overall quantitative description of the jet-gap-jet observables remains reliable.

Another issue is that exact theoretical descriptions
of jet-gap-jet events are rather sensitive to the soft rescattering of proton remnants. This, and other non-perturbative phenomena (e.g., for instance the soft color interaction model ~\cite{EkstedtEnbergIngelman17:HardColorSingletGapsBetweenJetsLHC}), tends to contaminate the gap. These
soft interactions are often expected to suppress the overall jet-gap-jet rate without much effect on the shapes of the distributions.
Nonetheless, models that contradict that expectation have been proposed~\cite{BabiarzStaszewskiSzczurek17:MultiPartonInterationRapidityGap}.

This paper is organized as follows: in Sec.~\ref{sec:MTJ} we define the conventions and set up the general framework. In Sec.~\ref{sec:lead-log-appr} we describe the LL solution and identify the lowest order IFs and GGFs. In Sec.~\ref{sec:MTJ-NLO} we describe the NLL contributions. This section is mainly devoted to examining the NLO IFs and adapting them to the jet-gap-jet observable. We identify the ``anomalous'' $\log(s)$ enhanced contributions responsible for the violation of BFKL factorization. In Sec.~\ref{sec:background} we describe additional possible theoretical sources of jet-gap-jet-like events, which could affect our predictions. Finally, in Sec.~\ref{sec:results} we define the relevant experimental parameters and present the numerical results relevant at LHC energies.

\section{Mueller-Tang jets: definitions and conventions\label{sec:MTJ}}

Our main aim is to describe the semi-inclusive dijet hadro-production
\be
h(p_a)+h(p_b)\to J_1+X_1+\gap+X_2+J_2\, ,
\ee
where two hadrons ($h$) with momentum $p_{a/b}$ scatter to jets $J_{1/2}$.%
\footnote{All definitions and expressions are taken from ref.~\cite{HentschinskiSabioVera14:QuarkImpactFactor,HentschinskiSabioVera14:GluonImpactFactor}.
To avoid repetitions many details will be omitted but can be found in those
references.  The reader should assume that all arguments and relations developed
there continue to be valid throughout the present analysis if not explicitly
stated otherwise.}
We denote with $J_i=(y_{j_i},\vec{p}_{j_i})$ the rapidity and
transverse momentum of each jet.%
\footnote{We use bold fonts to indicate Euclidean 2-dimensional transverse vectors, and
also explicitly label the integration dimension to avoid any confusion.}
The two jets are required to be ``hard'', i.e., with
$|\vec{p}_{j_i}|\gg\Lambda_{QCD}$, and to have a large rapidity difference
$Y_{j_{12}}$ which encompasses the rapidity gap, a central region devoid of radiation with an extension of \emph{at least} $Y_\gap$.%
\footnote{No radiation is allowed in the gap region. Naturally, events where the void region extends beyond the prohibited gap region are perfectly valid and very common.}
The dijet emission is ``semi-inclusive'' because any radiation $X_{1/2}$ outside the central rapidity gap, in addition to the two jets, is permitted in both longitudinal hemispheres. By definition, $Y_{j_{12}}>Y_\gap$ since the gap stands between the edges of the jets.
More precisely, a dijet scattering event recorded at the LHC is considered to be a Mueller-Tang process if it passes the following selection algorithm~\cite{Baldenegro21:ColorSingletExchange13TeV} (see Fig.~\ref{fig:gap}):
\begin{enumerate}
\item Tag the hardest among all the jets in each longitudinal hemisphere fulfilling   \[
\begin{aligned}
 &p_{j_{\{1,2\}}}>40\ \mr{GeV}\;,
 &|y_{j_{\{1,2\}}}| > y_{j_{\rm min}} = 1.4\;,
 &&y_{j_1}y_{j_2}<0.\\
\end{aligned}
\]
The minimum allowed jet rapidity is \(y_{j_{\rm min}}>y_{\gap}\) to leave enough room between the ``jet cone'' edge (as defined by the jet radius) and the gap region.
\item Reject the event if any radiation, hadronic, electromagnetic, etc., is found in the prohibited gap region
\[
 \begin{aligned}
\abs{\vec{k}}>&E_{\rm th}\approx 1\ \mr{GeV},&\abs{y}<y_{\gap}=1\ &\l(\to Y_{\gap}=2 y_{gap} = 2\r).
\end{aligned}
\]
Note that the gap is a fixed domain in rapidity in the central region (as it is the case at the Tevatron and the LHC), between $-1$ and $1$ units in rapidity.

The role of a finite energy threshold is twofold: firstly, it reflects the finite resolution of the detector; secondly, prerequisite to the cancellation of the singularities in the theoretical analysis, it leaves the soft emission unconstrained.%
\footnote{The choice of a value larger than the experiment reference $E_{\rm th}\approx 200$ MeV~\cite{Baldenegro21:ColorSingletExchange13TeV} is motivated considering that the hadronization process, here absent, would spread the parent energy among its spawn hence reducing the energy density. Other choices of the \(E_{\rm th}\) value are explored in the following.}
\end{enumerate}

These types of jet-gap-jet events are predominantly generated through color-singlet exchange. Other color representations tend to populate the final state with a lot of central radiation~\cite{MuellerTang92} which destroys the rapidity gap. Due to the large scales provided by the jet transverse momenta, QCD perturbation theory is effective in describing the cross-section of these processes. In addition, the large rapidity separation insures that the center-of-mass energy $\sqrt{\hat{s}}$ of the interacting system is much larger than the momentum transferred $\sqrt{-t}\simeq|\vec{p}_{j_i}|$, so that the process is governed by BFKL dynamics.
\begin{figure}[bh]
  \centering
\includegraphics[width=0.5\linewidth]{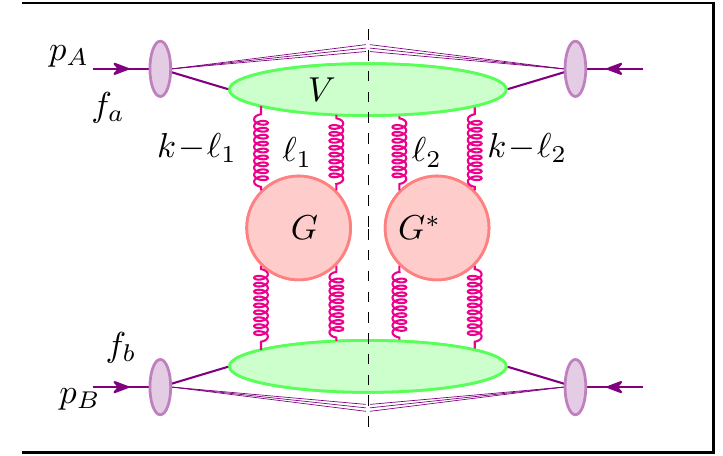}\hspace{10pt}
\vspace{-10pt}
\caption{Schematic of the Mueller-Tang cross-section. The red disks denote the GGFs, the green blobs the impact factors, the small violet blobs the PDFs.}
  \label{fig:MTscheme}
\end{figure}

As implied by the structure shown in Fig.~\ref{fig:MTscheme}, the working hypothesis is that
the jet-gap-jet cross-section is mostly determined by a hard partonic cross
section convoluted with the corresponding parton distribution functions (PDF) $f_{a,b}$
according to standard collinear factorization.  In turn, the partonic cross
section can be factorized into process-dependent impact factors (IFs) $V_{a,b}$ --- which,
in this case, are also called ``jet vertices'', one for each hadron --- and the
universal non-forward colour-singlet gluonic Green-function (GGF) $G$ together with
its complex-conjugate $G^*$.

Each GGF resums to all orders the logarithmically-enhanced energy-dependent terms $\sim(\alpha_s\log(s))^n$ in the perturbative expansion of the amplitude,  which arise in the Regge limit $s\gg(-t)\gg\Lambda_{QCD}$.
In this limit, due to the extreme Lorentz contraction, the dynamics is essentially transverse, and therefore the GGF depends only on the partonic center of mass energy squared $\hat{s}=x_1 x_2 s$ and on the transverse momenta of its legs $\vec\ell$ and $\vec{k}-\vec\ell$,
where $\vec{k}$ is the overall momentum transferred between the subsystems on the opposite sides of the gap, such that
$\vec{k}\simeq \vec{p}_{j_1}\simeq -\vec{p}_{j_2}$ and $t=-\vec{k}^2$.

\begin{figure}[bh]
  \centering
\includegraphics[width=0.35\linewidth]{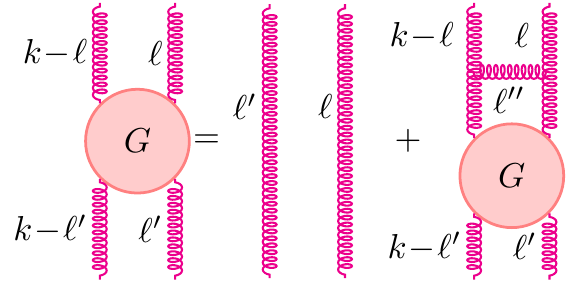}\hspace{10pt}
\vspace{-10pt}
\caption{Schematics of the BFKL recursive integro-differential equation.}
  \label{fig:bfkl-eq}
\end{figure}

According to the BFKL framework, the GGF is the solution of an integral equation (see Fig.~\ref{fig:bfkl-eq}) whose kernel can be computed in perturbation theory:
\begin{equation}\label{eq:bfkl}
    \frac{\partial}{\partial \log s}G(\vec{\ell},\vec{\ell}',\vec{k},s) =
    \delta^2(\vec{\ell}-\vec{\ell}') + \int\dd\vec{\ell}''\;
    K(\vec{\ell},\vec{\ell}'',\vec{k})\,
    G(\vec{\ell}',\vec{\ell}'',\vec{k},s) \;.
  \end{equation}
Currently, the kernel $K=K^{(0)}+\alpha_s K^{(1)}$ is known at leading~\cite{FadinKureaevLipatov75:PomeranchukSingularityAsymptoticFreeTheories,FadinKuraevLipatov76:MultiReggeonYangMillsTheory} and next-to-leading~\cite{FadinFiore05:NonForwardKernelNLO} order. The solution of the BFKL equation with only the leading kernel $K_0$ determines the leading-log GGF, denoted as $G^{\mr{LL}}$ (while $G^{\mr{NLL}}$ denotes the solution with the full, leading and next-to-leading, kernel). The jet vertex couples the external probe with the pomeron, here represented by the GGF in the color singlet representation, and dresses the emission in terms of jets.

In practice, the maximally differential jet-gap-jet cross-section can be written as
\begin{align}
  \label{eq:sigmaNLL}
  \frac{\dd\sigma}{\dd J_1\dd J_{2}\dd^2\vec{k} }
  &= \sum_{a,b}\int_0^1 \dd x_1\; f_a(x_1)\int_0^1\dd x_2\; f_b(x_2)
  \, \frac{\dd\hat{\sigma}_{ab}}{\dd J_1\dd J_{2}\dd^2\vec{k}} \\
  \label{eq:sigmahat}
  \frac{\dd\hat{\sigma}_{ab}}{\dd J_1\dd J_{2}\dd^2\vec{k}}
  &=\frac1{\pi^2}\int \Biggl[\prod_{i=1,2}\!\!\dd^2\vec{\ell}_i \dd^2\vec{\ell}'_i\Biggr]
  \frac{\dd V_a(x_1,\vec{\ell}_1,\vec{\ell}_2,\vec{k})}{\dd J_1}
  \frac{\dd V_b(x_2,\vec{\ell}'_1,\vec{\ell}'_2,\vec{k})}{\dd J_{2}}\\
  &\qquad \times G(\vec{\ell}_1,\vec{\ell}'_1,\vec{k},\hat{s}/s_0)
  \,G(\vec{\ell}_2,\vec{\ell}'_2,\vec{k},\hat{s}/s_0)\,, \nonumber
\end{align}
where $x_{1/2}$ is the momentum fraction and $a/b$ the active parton (\(q,\bar{q}\) with their flavors or \(g\)) in hadron $h_{A/B}$. The jet vertices, unintegrated in the jet variables, are here denoted by \(\frac{\dd V_{a/b}}{\dd J}\).

The various factors depend also on the choice of renormalization scale $\mu_R$, factorization scale $\mu_F$, and energy scale $s_0$. The latter is introduced to define the energy logarithms as $\log(\hat{s}/s_0)$ which stand at the basis of the BFKL factorization. By construction, changing the value of $s_0$ alters both GGFs and jet vertices in such a way that their product is left unchanged up to terms in the current approximation.

The very existence of such a hybrid factorization formula at NLL level
relies on two properties of the partonic cross-sections:
\begin{itemize}
\item[\it (i)] All infrared divergencies cancel in the sum of real and virtual
  contributions, after the initial state collinear singularities are absorbed
  via collinear factorization into the projectile parton densities.
\item[\it (ii)] All \(\log(s)\) enhanced terms are incorporated into the GGFs.
\end{itemize}
Property {\it (i)} has been demonstrated in ref.~\cite{HentschinskiSabioVera14:QuarkImpactFactor,HentschinskiSabioVera14:GluonImpactFactor}, provided the jets are defined through an infrared-safe algorithm. Property {\it (ii)} is valid at LL for any kind of jets but at NLL it crucially depends on the treatment of radiation $X_{1/2}$ outside the jets or, equivalently, on the definition of the gap.  As we will show in Sec.~\ref{sec:break-bfkl-fact}, property {\it (ii)} is violated with the standard gap definition presented above.  Nevertheless, the size of such violation turns out to be smaller than the overall theoretical uncertainties (stemming from the physical scale variations). Therefore, from a practical point of view, we consider phenomenologically adequate the factorization formula presented above.

Having laid out the foundation of the description, we proceed to discuss
the lowest approximation first and successively build our way up to
the complexity needed for the NLL case.

\section{Leading-Log approximation}
\label{sec:lead-log-appr}

\begin{figure}[th]
  \centering
\includegraphics[width=0.6\linewidth]{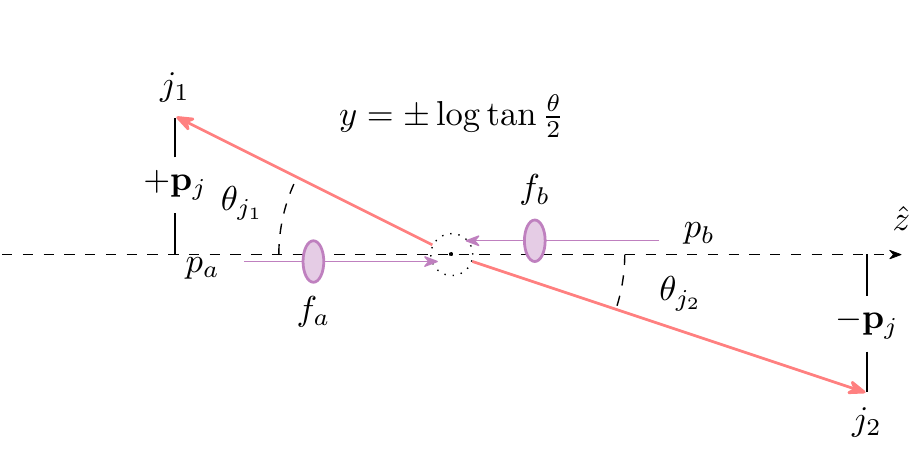}\hspace{10pt}
\vspace{-10pt}
\caption{Geometric picture of a $2\to 2$ scattering process. The two incoming partons move along the $z$ axis before colliding and being deflected, giving rise to two outgoing jets.}
  \label{fig:scat-geom}
\end{figure}

At LL the interaction is purely elastic and involves the scattering of a single constituent parton (quark/gluon)
\[\mathpzc{p}(p_a)+\mathpzc{p}(p_b)\to \mathpzc{p}(p_1)+\mathpzc{p}(p_2)\]
where \(\mathpzc{p}\in\{q,g\}\) are partons, $a/b$ and $1/2$ label the incoming and outgoing particles respectively. As shown in Fig.~\ref{fig:scat-geom}, the gap spans the whole range between the partons.
The rapidity ordering emerging in the high-energy regime assures that each interacting parton forms an independent jet with no overlapping between the two (\(j_i=p_i:i\in\{1,2\}\)) after being slightly deflected in the scattering.

At LO, the IFs corresponding to quark and gluon initiated scattering (\(V_{q/g}\)) differ by a mere multiplicative factor: \(V_q/V_g=C_F/C_A\), where $C_F=(N^2_c-1)/2N_c$ and $C_A=N_c=3$ are the quark and gluon color factors respectively. Their explicit expressions are
\begin{align}
      \frac{\dd V^{\mr{LO}}(\vec{k},x)}{\dd J}&=V^{(0)}_{q/g}\,S^{(2)}_J(\vec{k};x),\\
  V^{(0)}_{q/g}&=\frac{\alpha^2_sC^2_{F/A}}{N_c^2-1}\;,\qquad S^{(2)}_J(\vec{k};x)=x\delta(x-x_j)\delta^{(2)}(\vec{k}-\vec{p}_j).
\end{align}
As for the LL GGF, the solution of the BFKL equation at arbitrary values of the momentum transfer, the so called non-forward BFKL equation (\(k>0\)) was computed in Ref.~\cite{Lipatov85:barePomeron} for the case of colorless particle scattering.
Mueller and Tang~\cite{MuellerTang92} adapted the solution to (colorful) partonic scattering. We adopt the MT prescription in our analysis. More details on the origin of the MT prescription are given in appendix~\ref{sec:ggf}.
The result is a complicated expression for the GGF once projected back to momentum space.

Since the LO IFs carry no dependence on the reggeon momenta \(\vec{\ell}_i\), the integration over such variables in eq.~\eqref{eq:sigmahat} reduces to the integral average of the GGF over external legs:
\begin{equation}\label{calG}
  \mathcal{G}(\vec{k}^2,\hat{s}/s_0)=\int \dd^2\vec{\ell}\dd^2\vec{\ell}'\;G(\vec{\ell},\vec{\ell}',\vec{k},\hat{s}/s_0) \;.
\end{equation}
It turns out that in LLA only the non-analytic contribution --- the Mueller-Tang ``subtraction'' --- survives the average operation, greatly simplifying the form of the GGF:
\begin{equation}   \label{eq:GGFavg}
  \mathcal{G}^{\rm LL}(\vec{k}^2,\hat{s}/s_0)
  =\left(\frac{{k}}{2}\right)^{-2}\sum_{n\in\mathbb{Z}}
  \int_{-\infty}^{\infty}\dd\nu \left(\frac{\hat{s}}{s_0}\right)^{\omega(n,\nu)}R(n,\nu)
\end{equation}
where
$\omega(n,\nu)$ is the eigenvalue function of the LL BFKL kernel:\footnote{$\psi(z)=\frac{\dd}{\dd z} \log{\Gamma(z)}$ is the digamma function.}
\begin{align}
  \label{eq:omegaLL}
  \omega(n,\nu)&=\bar{\alpha}_s\left[2\psi(1)-\psi(1/2+|n|+i\nu)-\psi(1/2+|n|-i\nu)\right] \\
  \label{eq:Rmnu}
  R(n,\nu)&=\frac{\nu^2+n^2}{[\nu^2+(n-1/2)^2][\nu^2+(n+1/2)^2]}
  \;,\qquad \bar{\alpha}_s=\frac{N_c}{\pi}\alpha_s
\end{align}
In conclusion, at LL level, the master formula of eq.~\eqref{eq:sigmahat} becomes
\begin{gather}
  \label{eq:dsigmaLL}
\frac{\dd\hat{\sigma}^{\mr{LL}}_{ab}}{\dd J_1\dd J_{2}\dd^2\vec{k}}=
\frac{\dd V^{\mr{LO}}(\vec{k},x_1)}{\dd J_1}\frac{\dd V^{\mr{LO}}(-\vec{k},x_2)}{\dd J_{2}}
\biggl(\frac{\mathcal{G}^{\rm LL}(\vec{k}^2,\hat{s}/{s_0})}{\pi}\biggr)^2\;.
\end{gather}
The same formula~\eqref{eq:dsigmaLL} holds for the LO impact factors with the NLL GGF~\eqref{calG} with $\mathcal{G}^{\rm LL}\rightarrow \mathcal{G}^{\rm NLL}$.
On the other hand, when coupled with NLO IFs, the GGF must retain its full form as explained in detail in 
appendix~\ref{sec:ggf}.

\section{Beyond LLA}
\label{sec:MTJ-NLO}

In the spirit of the BFKL approach,  all the radiative corrections accompanied by one less $\log(s)$ factor (relative to the LL approximation) must be retained to refine the precision of the prediction by one (logarithmic) approximation order.
This can be achieved by using in eq.~\eqref{eq:sigmahat} the next-to-leading version of the jet vertices~\cite{HentschinskiSabioVera14:QuarkImpactFactor, HentschinskiSabioVera14:GluonImpactFactor, Fadin99:QuarkImpactFactor, Fadin99:GluonImpactFactor} and the GGFs. However, the large number of integrations in the factorization formula and inside the NLO jet vertices $V^{(1)}$ causes the computation of the cross-section to be a formidable numerical task. We can simplify the computation, while retaining the requested NLL accuracy, by perturbatively expanding the jet vertices
$V= V^{(0)}+\alpha_sV^{(1)}$ and thereby splitting the cross-section into two contributions. The first one
combines the NLL gluon Green functions $G^{\rm NLL}$ with the lowest-order jet vertices $V^{(0)}$
and the other one
contains the next-to-leading corrections of the vertices $V^{(1)}$ convoluted with the leading-log Green functions $G^{\rm LL}$. More precisely%
\footnote{In our equations and plots the label NLL
means NLL accuracy, i.e., both leading-log and next-to-leading-log terms are included. [NLO$\,\otimes\,$\dots] means that one of the vertices has only the NLO term $V^{(1)}$, the other vertex being at LO.}
\begin{align}
  \dd\hat{\sigma}^{\rm FULL}_{ab}&\simeq V^{(0)}_a\otimes G^{\rm NLL}(\vec{\ell}_{1,2})\otimes G^{\rm NLL}(\vec{\ell}'_{1,2})\otimes V^{(0)}_b
  &(\mr{LO}\otimes\mr{NLL})\qquad\nonumber\\
   &+\alpha_sV^{(1)}_a(\vec{\ell}_1,\vec{\ell}'_1)\otimes G^{\rm LL}(\vec{\ell}_{1,2})\otimes G^{\rm LL}(\vec{\ell}'_{1,2})\otimes V^{(0)}_b(\vec{\ell}_2,\vec{\ell}'_2)+\{V_a\leftrightarrow V_b\}
   & (\mr{NLO}\otimes\mr{LL})\qquad\nonumber\\
 &+\msout{\alpha^2_sV^{(1)}_a\otimes G^{\rm LL}\otimes G^{\rm LL}\otimes V^{(1)}_b +\alpha_sV^{(1)}_a\otimes G^{\rm NLL}\otimes G^{\rm NLL}\otimes V^{(0)}_b}\dots\;,
\label{eq:NLL-NLO}
\end{align}
where in the last line we show, for the sake of clarity, some contributions that we neglect, being formally NNLL.
Eq.~\eqref{eq:NLL-NLO} also shows on the right the notation that will be used throughout this analysis: LO vertices and NLL GGFs (first line); one NLO vertex $V^{(1)}$ and LL GGFs (second line); ``FULL'' for their sum.

\subsection{Effective NLL GGF}
\label{sec:effetive-nlla}

One of the major tasks in computing the MT jet cross-section in NLLA is the calculation of the NLL GGF. In the LLA, the calculation of the GGF is considerably simplified by exploiting the conformal properties of the LL kernel. This allows the explicit determination of the eigenvalues and eigenfunctions. At NLLA, such properties are no longer true because of running-coupling effects\footnote{It is also not clear what would be the role of the Mueller-Tang prescription at NLLA. See ref.~\cite{BartelsForshaw95} for a discussion in that direction.}.
While waiting for a breakthrough towards the search for a solution of the BFKL equation at NLLA, a partial account of the NLL corrections has already been pursued~\cite{RoyonChevallier09:GapsBetweenJets}.
It is based on the assumption that the \emph{forward} NLL eigenvalue in place of its LL equivalent should capture the bulk of the corrections.

Proceeding along the lines of ref.~\cite{RoyonChevallier09:GapsBetweenJets}, we adopt an effective (averaged) GGF at NLL level by replacing in eq.~\eqref{eq:GGFavg} the LL eigenvalue \(\omega(n,\nu)\) of eq.~\eqref{eq:omegaLL} with the forward NLL eigenvalue $\omega^{\mr{f-NLL}}$ of eq.~\eqref{eq:omegaNLL}, keeping all the rest unchanged:
\[
\mathcal{G}^{\mr{NLL}}=\mathcal{G}^{\mr{LL}}\big|_{\omega\;\to\;\omega^{\mr{f-NLL}}}\;.
\]

It is known that both the zero conformal spin (\(n=0\)) component of the NLL eigenfunction and, to a lesser extent,  the higher conformal spin components are sensitive to collinear corrections. These corrections, albeit formally subleading, can be large and should be taken into account~\cite{Salam98:CollinearResummation}. The collinear improvement~\cite{CiafaloniColferai03:RenormGroupImprovSmall-xGreenFunction} developed to account for these corrections, introduces a scheme dependence in the upgraded NLL kernel. We adopt scheme (4) of ref.~\cite{RoyonMarquet07:MuellerNaveletCollinearScheme,Salam98:CollinearResummation} and indicate this contribution simply as NLL in the following. Other schemes, which were tested, give very similar predictions and will be omitted.

We note that a fully correct account of the NLL corrections can only come from the solution of the NLL GGF equation. However, the non-forward NLL eigenfunction are not known at present. Therefore, following previous analyses, we approximate its computation by keeping the eigenfunction at LL while upgrading only the eigenvalue to NLL.

\subsection{NLO Impact Factors}
\label{sec:nlo-impact-factors}

The other fundamental ingredient of the BFKL factorization formula for MT jets is the so-called jet vertex, i.e., the impact factor for the production of a jet stemming from the interaction of a parton with a pomeron.  The quark-induced and gluon-induced impact factors at NLO were computed in~\cite{HentschinskiSabioVera14:QuarkImpactFactor,HentschinskiSabioVera14:GluonImpactFactor}.
They are considerably more complicated then their LO siblings, both in the analytical structure and in the kinematical configuration of the final state particles.
In fact, while at LO a vertex emits only one parton (necessarily to be identified as one of the jets), two partons can be emitted from each NLO vertex, thus causing the QCD matrix elements and the jet configurations to acquire a nontrivial structure.
\begin{figure}[bh]
  \centering
\includegraphics[width=0.2\linewidth]{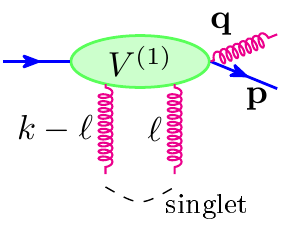}\hspace{10pt}
\vspace{-10pt}
\caption{Structure of the quark-induced NLO impact factor.}
  \label{fig:nloVertex}
\end{figure}
In order to explain these points and their main consequences, let us refer, for definiteness, to the quark-induced vertex where an incoming quark of momentum $n^\mu= x p_q^\mu$ ($p_q^\mu$ being the parent proton momentum) interacts with a pomeron (two gluons with momenta $\ell^\mu$ and $k^\mu-\ell^\mu$ in the colour singlet channel at lowest order) emitting a quark of momentum $p^\mu$ and a gluon of momentum $q^\mu$, as depicted in Fig.~\ref{fig:nloVertex} at amplitude level.

In the high-energy limit, the $t$-channel gluons are essentially transverse, while the outgoing quark and gluon carry away all the longitudinal momentum of the incoming parton\footnote{Notice that \(\vec{p}\) is not an independent variable, as \(\vec{p}+\vec{q}=\vec{k}\)}:
\begin{align}
  \ell & \simeq \vec{\ell} \;, & k &\simeq \vec{k} \nonumber \\
  q &= z n + \bar{z}_q p_b + \vec{q} \;, &
  p &= (1-z)n + \bar{z}_p p_b + \vec{p} \;,
\end{align}
$\bar{z}_{p,q}$ being very small components along the backward hadron's momentum $p_b^\mu$ and determined by the mass-shell conditions.
The structure of the NLO jet vertex is then given by
\begin{align}
  \frac{\dd V^{\mr{NLO}}}{\dd J}(\vec{\ell}_1,\vec{\ell}_2,\vec{k},x)
  &=\frac{\alpha_s^3}{2\pi(N_c^2-1)}\int_0^1\dd z\int\dd^2\vec{q} \;
  S_J(\vec{k},\vec{q},z)\, C_F\frac{1+(1-z)^2}{z} \nonumber \\
  & \times \left\{C_F^2\frac{z^2\vec{q}^2}{\vec{k}^2(\vec{k}-z\vec{q})^2}
  +C_F C_A\,z J_1(\vec{\ell}_{1,2},\vec{k},\vec{q},z) + C_A^2\, J_2(\vec{\ell}_{1,2},\vec{k},\vec{q})
  \right\} \;, \label{eq:nlostruct}
\end{align}
where \(J_1,J_2\) are given in sec.~\ref{sec:NLO-IF-final}.
The NLO jet vertex contains both real and virtual contributions, the latter being included as delta-function contributions at $z=0$ and $\vec{q}=0$. The $P_{qg}(z) = C_F (1+(1-z)^2)/z$ splitting function (last factor in the first line of eq.~\eqref{eq:nlostruct}) stands as an overall factor of the integrand, which presents three different colour structures. The distribution $S_J$ selects the final states contributing to the observables, i.e., it embodies the jet algorithm and the gap restriction.

According to the $k_t$ jet algorithm applied to two outgoing particles, the jet clustering may end up into three, mutually exclusive, configurations: {\it (i)} the two partons are clustered into a composite jet; {\it (ii)} the two partons are not clustered, and the quark is the jet; {\it(iii)} the two partons are not clustered and the gluon is the jet. In the composite jet configuration {\it(i)}, both partons are inside the ``jet cone'' and cannot be found in the gap region. On the contrary, in the single jet configurations {\it(ii) and \it(iii)}, the parton outside the jet could be emitted in the forbidden gap region, if no additional constraints are imposed. Therefore we must supplement this additional constraint to the single-jet configurations. In conclusion, the selection function $S_J^{(3)}$ reads
\be
\begin{aligned}
  \label{eq:jetDistribution}
  S^{(3)}_j(\vec{p},\vec{q},zx;x)& =
  S^{(2)}_j\big(\vec{p},(1-z)x\big) \; \Theta(\abs{\vec{p}}-\abs{\vec{q}})\;
  \Theta_{J}^{k_\perp} \;  \Theta^{{g}}_{\gap}   &\ ({\rm quark-jet})\\
  &+S^{(2)}_j(\vec{q},zx) \; \Theta(\abs{\vec{q}}-\abs{\vec{p}})\;
   \Theta_{J}^{k_\perp} \; \Theta^{{q}}_{\gap} & ({\rm gluon-jet})\\
  &+S^{(2)}_j(\vec{p}+\vec{q},x)\;\bigl(1-\Theta_{J}^{k_\perp}\bigr),&\ ({\rm composite})
\end{aligned}
\ee
where
\begin{align}\label{eq:singleJet}
  & \Theta_{J}^{k_\perp}(\vec{p},\vec{q},z;R_{\rm jet}) = \Theta(\Delta y^2+\Delta\phi^2-R_{\rm jet}^2) \\ & \Delta y = \log \left(\frac{1-z}{z}\frac{|\vec{q}|}{|\vec{p}|}\right)\;, \quad \Delta\phi = \arccos\left(\frac{\vec{p}\cdot\vec{q}}{|\vec{p}|\,|\vec{q}|} \right). \nonumber
\end{align}
\(\Theta_{J}^{k_\perp}\) is 1 in case of single (quark or gluon) jet and 0 for composite (quark and gluon) jet. It clearly depends on the jet algorithm, and eq.~\eqref{eq:singleJet} refers to the (anti-)$k_t$ one.
In the first line of eq.~\eqref{eq:jetDistribution} $S_J^{(2)}$ enforces the quark to be the jet, $\Theta(\abs{\vec{p}}-\abs{\vec{q}})$ makes sure that this is the leading jet in the hemisphere, while
\begin{align}
  \Theta^{q/g}_{\gap}\bigl(y_{q/g},\abs{\vec{k}_{q/g}}\bigr)=&
1-\Theta\l(y_{\gap}-y_{q/g}\r)\Theta\l(\abs{\vec{k}_{q/g}}-E_{\rm th}\r)
\end{align}
rejects the event if the gluon outside the jet ends up in the gap. In the second line of eq.~\eqref{eq:jetDistribution} the gluon and quark roles are interchanged. The third line describes the condition of
a composite jet.

One would argue that, because of the colourless (pomeron) exchange, particle emission in the central region is expected to be dynamically suppressed. Each vertex is expected to emit its two partons in the fragmentation region of the incoming parton, thus constraining partons not to be inside the gap (central region) is unnecessary or at most it should amount to a small effect. However, this is not completely true if the parton outside the jet is the gluon. We will discuss in detail this aspect in the next section.

The original calculation of the NLO jet vertex was done by using Lipatov's effective action for QCD at high-energy~\cite{Lipatov95:EffectiveAction}. We repeated the calculation using standard QCD Feynman rules, and confirmed the correctness of their results~\cite{HentschinskiSabioVera14:QuarkImpactFactor,HentschinskiSabioVera14:GluonImpactFactor} (except for a few typos that we fixed; see Appendix~\ref{sec:NLO-IF-final}).
  However, we preferred to employ an alternative procedure to extract \(\epsilon\)-poles and finite reminders. Although formally equivalent, it yields expressions that are better suited for the numerical analysis. Our procedure is illustrated in Appendix~\ref{sec:singularities}, taking the term with color factor \(C_f^2\) in the quark-induced IF as an example. The complete expressions for the NLO IF after the singularity extraction is given in appendix~\ref{sec:nlo-ifs}.

A final remark concerns the numerical integrations involving such impact factors. The computation of the cross-section involves convoluting GGFs, IFs, PDFs and then integrating the maximally differential cross-section \label{eq:sigmaNLO} according to the specific observable being computed. Going back to the expression~\eqref{eq:nlostruct} of the NLO jet vertex, we note that it involves three internal integrations $\dd z \dd^2 \vec{q}$, which are however constrained by at least three delta functions contained in the jet function $S_J$. Therefore, when convoluting the NLO vertex with the GGFs ($\dd^2\vec{\ell}_1\,\dd^2\vec{\ell}_2$), the opposite LO vertex ($\dd^2\vec{k}$) and the two PDFs ($\dd x_1\,\dd x_2$), one is faced with an eight-dimensional integral. Finally, integrating over the jet variables ($\dd^3 J_1\,\dd^3 J_{2}$) and taking into account four-dimensional momentum conservation, we have to deal with ten-dimensional integrations in order to compute cross-sections at NLO accuracy.

\subsubsection{Breaking of BFKL factorization}
\label{sec:break-bfkl-fact}

Any impact factor which enters a BFKL formula for a physical observable must be
{\it (i)} IR-finite and {\it (ii)} independent of $s$ in the $s\to\infty$ limit. The first requirement is obvious. The cancellation of the $\epsilon$-poles when combining real and virtual correction and subtracting the universal collinear singularities, was proven in~\cite{HentschinskiSabioVera14:QuarkImpactFactor,HentschinskiSabioVera14:GluonImpactFactor},
provided that an IR-safe jet definition is adopted in $S_J$ for the final-state phase space integration. Nevertheless, if only the jet clustering is implemented, without further restrictions, we face a problem: the $z$-integration in eq.~\eqref{eq:nlostruct} is divergent for $z\to 0$ due to the $1/z$ behaviour of the
$P_{gq}$ splitting function. For the quark-induced case,%
\footnote{The gluon-induced case is completely equivalent, due to the
singular behaviour of the corresponding $P_{gg}$ splitting function.}
this happens in the $C_A^2$ term when only the quark generates the jet while the gluon, whose longitudinal momentum is $z$, can be emitted at arbitrary rapidities ($y_g \lesssim y_J$) outside the jet cone. In fact, since $\dd z/z = \dd y$, the gluon emission density appears to be flat in rapidity.
This spurious divergence should not be taken seriously, as it stems from the kinematic region $y_g < 0 \iff z<|\vec{q}|/\sqrt{s}$ which is outside the domain of validity of the calculation: an exact calculation would provide a suppression of the emission probability in that region.
Therefore, we can believe the prediction of a flat rapidity distribution to hold only up to central rapidities: $|\vec{q}|/\sqrt{s} \lesssim z < 1$. In that case, the $\dd z/z$-integration is finite, but gives rise to a term proportional to $\log s$, in conflict with the requirement {\it(ii)} mentioned above. In fact, BFKL factorization assumes that all the energy dependence of the cross-section is taken into account by the GGFs.

The authors of \cite{HentschinskiSabioVera14:QuarkImpactFactor} tackled this problem by imposing an upper limit on the invariant mass \(M^2_{X,\rm max}\) of the forward (and backward) system.  This prescription avoids the occurrence of a $\log s/\vec{p}^2_j$ term at the price of introducing new logarithms of the invariant mass \(\log\bigl(M^2_{X,\rm max}/\vec{p}^2_j\bigr)\).  For phenomenological purposes, such constraint is not viable, since it requires the measurement of all proton remnants, most of which escape
the detector. Furthermore, that prescription differs considerably from the experimental event selection for MT jets.

In the case of MT jets that we are considering, one actually constrains the rapidity of the gluon (and of any parton stemming from the incoming forward quark) to stay above $y_{\gap}$, the gap upper bound, unless its energy is so small that it remains undetected. For such gluons below threshold, the occurrence of a $\log s$ term in the cross-section is unavoidable, therefore we have to admit that BFKL factorization in the NLLA is violated for MT jet processes.
\begin{figure}[th]
  \centering
\includegraphics[width=0.6\linewidth]{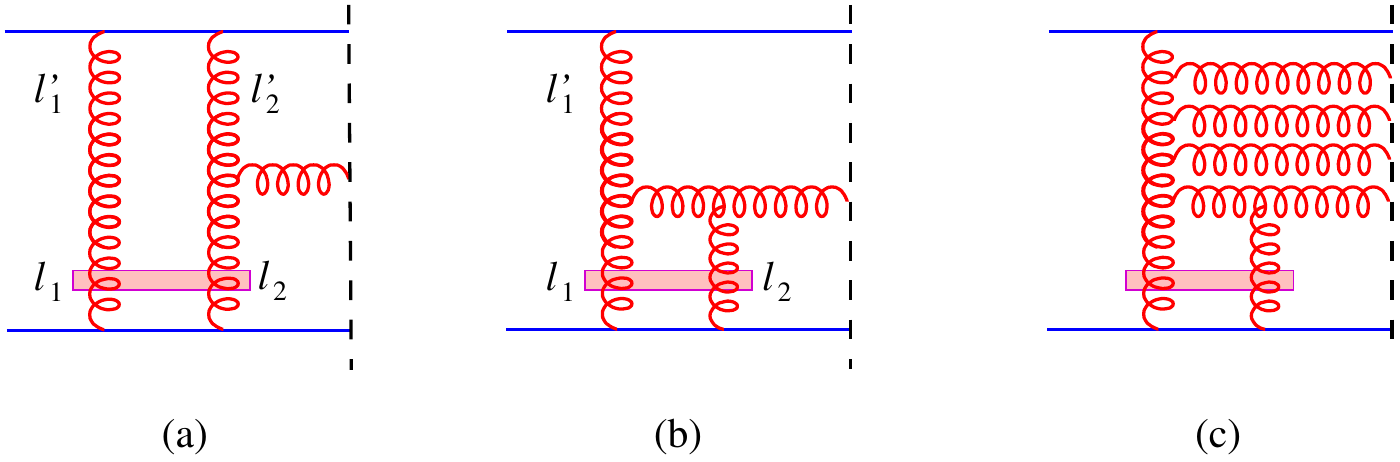}\hspace{10pt}
\vspace{-10pt}
\caption{Diagrams contributing to the $\log s$ term in the jet vertex.
The pink rectangle ranging from $l_1$ to $l_2$ denotes that the two gluons
are in a colour-singlet state. In (a) the two gluons $l'_1=l_1$ and $l'_2$ emerging from the upper quark cannot be in a color-singlet state. In (b) the only gluon $l'_1$ emerging from the upper quark is in a colour-octet state.
The resummation of the logarithmic terms in the impact factors involves diagram with multiple emissions (c).}
  \label{f:logDiagram}
\end{figure}
A question then arises: why a singlet exchange does not dynamically suppress gluon emission in the central region? Intuitively one can explain this fact by looking at Fig.~\ref{f:logDiagram}-a: if the lower $t$-channel gluons $\ell_{1,2}$ are in a colour-singlet state, the two upper gluons $\ell'_{1,2}$ are in a colour-octet state, causing gluon radiation at all rapidities between the two jets. Even more explicit is the diagram in Fig.~\ref{f:logDiagram}-b.

The rescue for our computation comes from the fact that such violation is quantitatively small for the proposed CMS setup. In fact, the function $J_2(\vec{\ell}_{1,2},\vec{k},\vec{q})$ in eq.~\eqref{eq:nlostruct} is uniformly bounded for $E_{\rm th} > |\vec{q}|\to 0$, so that the logarithmic terms vanishes for $E_{\rm th}\to 0$ for lack of phase space. In practice, for $E_{\rm th}\ll |\vec{p}_J|$,
\be
 \left.\frac{\dd V^{\mr{NLO}}}{\dd J}\right|_{\log s} \sim C^2_a \frac{E^2_{\rm th}}{\vec{p}^2_j}\log\frac{s}{\vec{q}^2} \; \delta^{(2)}(\vec{p}-\vec{p}_j) \;.
\ee
Thus, for the sake of our phenomenological analysis, the actual size of the violating term is expected to be small, of order $\sim E^2_{\rm th}/\vec{p}^2_j$. In sec.~\ref{sec:results} we confirm this assumption.

Concerning gluon emission in the rapidity range $y_{\gap}<y\lesssim y_J$, there is no $E_{\rm th}^2/\vec{p}_j^2$ suppression, so that we expect a contribution to the jet vertex of order
\begin{equation}
   \left.\frac{\dd V^{\mr{NLO}}}{\dd J}\right|_{\Delta y} \sim
   C^2_a \Delta y\;\delta^{(2)}(\vec{p}-\vec{p}_j) \;, \qquad
   \Delta y \equiv y_J-y_{\gap} \;.
\end{equation}
However, in the limit $s\to\infty$ which implies $y_J\simeq \log(s/\vec{p}_j^2)\to\infty$, one should consistently require also $y_{\gap}\to\infty$ in such a way that $\Delta y$ remains finite, so that no $\log s$ enhancement comes from this region of phase space.

In practice, the formal divergence, occurring when \(z\to 0\), is removed by imposing a lower rapidity bound on the gluon emission from the violating term \(y_g>0\) (\(y_g<0\) for the bottom IF). Other choices \(y_g>1\; (y_g<-1)\) were also explored and produced no appreciable difference.

From the theoretical point of view, in order to conceptually solve the problem of the violation of factorization, one could think of a modified factorization formula, where the expected higher-order logarithms, stemming from multi-gluon emission diagrams like that in Fig.~\ref{f:logDiagram}-c, are resummed to all orders. However, the condition of energy below detector threshold suppresses such emission with the same $E_{\rm th}^2/\vec{p}_j^2$ relative factor with respect to unconstrained emission. Therefore there is no need to perform such resummation for phenomenological purposes.
\begin{figure}[th]
  \centering
\includegraphics[height=0.25\linewidth,width=0.25\linewidth]{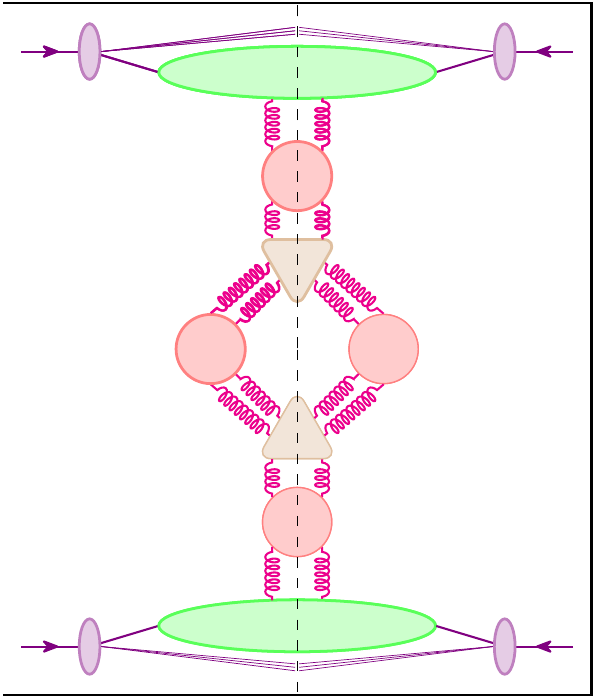}\hspace{10pt}
\vspace{-10pt}
\caption{Mueller-Navelet and Mueller-Tang hybrid structure: a pomeron loop formed connecting two forward-GGFs and two non-forward GGFs (red blobs) by means of two 3-pomeron vertices (yellow triangles).}
  \label{fig:pomeronloop}
\end{figure}
Another option to describe the MT process within the framework of reggeon field theory is to consider the factorization formula depicted in Fig.~\ref{fig:pomeronloop}. The vertical axis along the unitarity cut corresponds to rapidity of emitted particles. The region of the gap is described by two BFKL GGFs in the singlet channel, while the region outside the gap is described by the imaginary part of BFKL GGFs, one for each hemisphere, attached to the corresponding jet vertices for Mueller-Navelet jets~\cite{BartelsColferai02:MuellerNaveletQuarkImpactFactor}. The GGFs are connected by two triple pomeron vertices, so as to build a pomeron loop diagram. Besides the difficulties for a practical implementation of this formula, the major obstacle is that the triple pomeron vertex is not known in NLLA yet.

In conclusion, we adopt the pragmatic attitude to exploit the conventional BFKL factorization formula for MT jets with the impact factors described in this section, expecting a small systematic theoretical error at LHC energies.

\section{Mueller-Navelet jets with rapidity gap}
\label{sec:background}

The main aim of studying MT jets is to find a clear signal of BFKL-type dynamics, which in this case governs the behavior of color-singlet GGFs. Dijet events with the same signature -- no radiation in the gap, but  not involving a singlet exchange -- provide an additional contribution to the cross-section that we want to describe with the present calculation, thus affecting the reliability of our predictions. In this section we want to give an estimate of such contribution.  The simplest diagram for dijet production at high-energy involves a single gluon exchange (a color octet) in the $t$-channel. Being an $\ord{\alpha_s^2}$ term, it might seem that it should dominate over the signal and that its radiative corrections could be important as well.  However, colour octet exchanges are very likely to radiate uniformly in rapidity, as predicted by BFKL dynamics in such channel. Such dijet-inclusive emission --- the so called Mueller-Navelet jets --- is described by another factorization formula which involves the imaginary part of the forward GGF. We can estimate it by computing the cross-section for Mueller-Navelet jets with the constraint of no emission above energy threshold in the rapidity gap. A fully-fledged calculation is not easy; however, following Mueller and
Tang~\cite{MuellerTang92}, a reliable estimate in the LLA is given by
\begin{equation}\label{eq:elasticMN}
  \frac{\dd\sigma^{MN}_{\gap}}{\dd J_1 \,\dd J_{2}}
  \simeq \frac{\dd\sigma^{(0)}}{\dd J_1 \,\dd J_{2}} \,
  \mathrm{e}^{\omega(t) \,Y_{\gap}}
  \simeq \frac{\dd\sigma^{(0)}}{\dd J_1 \,\dd J_{2}} \,
  \left(\frac{E_{\rm th}^2}{\vec{p}_j^2}\right)^{\bar{\alpha}_s Y_{\gap}} \;,
\end{equation}
where
\begin{equation}
  \frac{\dd\sigma^{(0)}}{\dd J_1 \,\dd J_{2}}
  = \left(\frac{2 C_F \alpha_s}{\sqrt{N^2-1}}\right)^2
  \frac{\delta^{(2)}(\vec{p}_{j_1}-\vec{p}_{j_2})}{\vec{p}_{j_1}^2 \,\vec{p}_{j_2}^2}
  \delta(1-x_1/x_{j_1}) \delta(1-x_2/x_{j_2})\label{bornMN}
\end{equation}
is the one-gluon-exchange elastic cross-section and
\begin{align}\nonumber
  \omega(-\vec{p}^2)
  &= -\frac{\bar{\alpha}_s}{4\pi}
  \int\frac{\dd^2\vec{k}}{\vec{k}^2+m_g^2}\,
  \frac{\vec{p}^2}{(\vec{p}-\vec{k})^2+m_g^2}\ &\\
    &\simeq -\bar{\alpha}_s\log \l( \vec{p}^2 \middle/ m_g^2 \r) ,
  &\mr{if}\ m_g\ll |\vec{p}| \label{eq:trajectory}
\end{align}
is the correction to the gluon Regge trajectory for a gluon of mass $m_g$.

The idea is that the jet cross-section with gluon emission only below some energy scale $E_{\rm th}\ll|\vec{p}_j|$ is almost equal to the elastic cross-section with gluons of mass $m_g\simeq E_{\rm th}$, because in the realistic massless case real and virtual corrections at scales $\lambda<E_{\rm th}$ cancel almost exactly. Gluon reggeization~\cite{Fadin98:GeneralNonForwarBfkl} means that the color octet projection of the elastic amplitude can be written as
\begin{equation}\label{eq:reggeAmp}
   \mathcal{A}_{\underline{8}}=\mathcal{A}_{\underline{8}}^{(0)} \left(\frac{s}{-t}\right)^{\omega(t)} = \mathcal{A}_{\underline{8}}^{(0)} \mathrm{e}^{\omega(t) Y} \;.
 \end{equation}
By squaring the amplitude~\eqref{eq:reggeAmp} and exploiting eq.~\eqref{eq:trajectory} one obtains eq.~\eqref{eq:elasticMN}, which is consistent with the fact that in the limit $E_{\rm th}\to0$ virtual corrections completely suppress the purely elastic cross-section.
Actually, since the energy threshold does not extend all the way in between the jets but is required only within the rapidity gap,
we expect the approximation of eq.~\eqref{eq:elasticMN} to provide a lower bound for the estimate of the non-singlet contribution to the jet-gap-jet process cross-section.
Nonetheless, since it rarely exceed $15\%$ of the singlet contribution (see sec.~\ref{sec:results}), we conclude that, notwithstanding large corrections,%
\footnote{
A better estimate could include the NL corrections to the Mueller-Navelet predictions with finite energy threshold. On the other hand, since there is a fixed gap size in the rapidity interval $[-1,1]$, some events show a much larger rapidity difference between the jets, leaving enough phase space for emission. Therefore our observable may be represented by a Mueller-Navelet, then Mueller-Tang and finally Mueller-Navelet process in different regions of rapidity, as depicted in fig~\ref{fig:pomeronloop}.}
it can be safely neglected in first approximation.

\section{Results}
\label{sec:results}

In this section, we present the results of the phenomenological analysis of the MT jet processes at LHC energies.
The goal is to investigate the impact of the NLO IF over the NL corrections (LL\(\otimes\) NLO vs NLL\(\otimes\) LO) and consequently the comparison of the complete FULL NL estimate with respect to the LL predictions.

Let us first define the observables that we study in the analysis.
We investigate the jet distributions with respect to the jet rapidity difference \(Y_{j_{12}}=y_{j_1}-y_{j_2}\), the azimuthal difference between the jets \(\Delta\phi_{j_{12}}=\phi_{j_1}-\phi_{j_2}\) and the transverse energy of the ``second leading jet'' \(p_{j_<}=\min(p_{j_1},p_{j_2})\) defined by
\begin{align}
  \label{eq:Observables}
    \frac{\dd\sigma}{\dd Y_{j_{12}}}=
&\int\!\!\dd^2\vec{p}_{j_1}\dd^2\vec{p}_{j_2}\int\!\!\dd y_{j_1}\dd y_{j_2}\delta{\l(Y_{j_{12}}-\l(y_{j_1}-y_{j_2}\r)\r)}\!\int\!\!\dd^2\vec{k} \frac{\dd\sigma}{\dd J_1\dd J_{2}\dd^2\vec{k}},\\
    \frac{\dd\sigma}{\dd\Delta\phi_{j_{12}}}=&\int\!\! \dd^2\vec{p}_{j_1}\dd^2\vec{p}_{j_2}\delta{\l(\Delta\phi_{j_{12}}-\l(\phi_{j_1}-\phi_{j_2}\r)\r)}
\!\int\!\!\dd y_{j_1}\dd y_{j_2}\!\int\!\!\dd^2\vec{k}\frac{\dd\sigma}{\dd J_1\dd J_{2}\dd^2\vec{k}},\\
    \frac{\dd\sigma}{\dd p_{j_<}}=&\int\!\!\dd^2\vec{p}_{j_1}\dd^2\vec{p}_{j_2}
\Bigl[\delta{\l(p_{j_<}-p_{j_2}\r)}\theta{\l(p_{j_1}-p_{j_2}\r)}+\delta{\l(p_{j_<}-p_{j_1}\r)}\theta{\l(p_{j_2}-p_{j_1}\r)}\Bigr]\\
 &\qquad\times\int\!\!\dd y_{j_1}\dd y_{j_2}\!\int\!\!\dd^2\vec{k}\frac{\dd\sigma}{\dd J_1\dd J_{2}\dd^2\vec{k}}.\n
\end{align}
where the integration extends over the bin sizes. Note that, the azimuthal angular distribution is trivial (\(\Delta\phi_{j_{12}}=\pi\)) if the NLO IFs are not included.

\subsection{Generalities about the phenomenological calculation}

In this section, we report the details of the calculation methods employed for the analysis and we fix the kinematic domain considered in all following phenomenological studies.

The kinematic setup is related to the recent CMS analysis~\cite{Baldenegro21:ColorSingletExchange13TeV}.
At a center-of-mass energy of $\sqrt{s}=13$ TeV the number of active flavors is fixed to $N_f=5$.
The value of the strong coupling constant at the $Z$ boson mass was taken to be $\alpha_S\l(M^2_Z\r)=0.1176$, corresponding to   $\Lambda_{QCD}=221.2$ MeV as the QCD-scale. Following the typical acceptance of a detector at the LHC (CMS or ATLAS), we assume the jets to be reconstructed in the forward region \(1.4<\abs{y_j}<4.7\)  (\(2.8<\abs{Y_{j_{12}}}<9.4\)) with a gap devoid of any energy spanning the central region between $-1$ and $1$ units in rapidity. Note that the reliability of the BFKL approach, which formally neglects momentum conservation constraints, begins to deteriorate approaching the kinematic boundary (\(Y_{j_{12}}\gtrsim 9\)).
In addition, at the other end of the rapidity spectrum \(Y_{j_{12}}\lesssim 4\), the approximation becomes less reliable since we move away from the high-energy regime.

The clustering of particle emission into jets is performed using the (anti-)\(k_t\) jet algorithm\footnote{Actually, since the jet clustering is stopped at the first iteration and there are never more then two partons that can be grouped together, the $k_t$ or anti-\(k_t\) jet algorithms behave identically.} with a jet radius set to \(R_{\rm jet}=0.4\) in the rapidity-azimuth plane (see eq.~\eqref{eq:singleJet}). According to collinear factorization, the matrix element of the interacting partons must be convoluted with
parton distribution functions (PDFs). These, as well as the running \(\alpha_S\) coupling value, were introduced via the CTEQ18 routine~\cite{Hou19:CteqGlobalAnalysis}.

The experimental observables measured in jet-gap-jet events often rely on the ratio between the MT dijet prediction with a color-singlet exchange and the standard inclusive QCD dijet events in order to reduce the systematic uncertainties on the observables~\cite{abe1998dijet, collaborations2021hard}. However, in this paper, we choose to focus on completing the BFKL predictions at the NLL approximation. For a direct comparison with experimental data, an implementation of this process in a general purpose Monte-Carlo generator would be necessary (this will be performed in an upcoming study).
The Monte-Carlo implementation will also allow to simulate the additional soft radiation mechanisms that are not part of the BFKL description but tend to contaminate the gap~\cite{BaldenegroKlasenRoyon22:JetsSeparatedLargeRapidityGap}. In the simplest scenario, the effect of these soft rescatterings is to rescale the overall normalization by the so called gap survival probability~\cite{RoyonKepka10:GapsBetweenJets}. More sophisticated models, where these effects emerge dynamically as consequence of soft radiation mechanisms, have also been proposed (e.g., \cite{MartinRyskin08:RapidityGapSurvivalProb}). In the following, we present the results without any survival probability factor.

Let us now discuss briefly the accuracy of the calculation.
The difference between the Vegas and Suave algorithms from the Cuba multidimensional integration routine~\cite{Hahn05:Cuba} has been used to estimate the numerical uncertainty on the integrals.
Despite the large number of integrals, the numerical uncertainties can generally be pushed down to the order of 1\% by increasing the number of iterations.

Another source of uncertainty affecting our numerical analysis originates from the fact that we approximate the GGF as a multilinear interpolation over a grid. Each node stores the result of the numerical integration of the GGF found in eq.~\eqref{eq:GGF2} over the Mellin variable \(\nu\).
An estimate of the quality of the interpolation can be obtained \textit{a posteriori} by comparing the average of the GGF over its external legs as given by the analytic expression \eqref{eq:GGFavg} and the explicit integration of its interpolated values over the grid
\[\int\dd^2\vec{\ell}\;G(\vec{\ell},\dots)\stackrel{?}{=}\mathcal{G}(\dots),\]
where the right-hand side is given by eq.~\eqref{eq:GGFavg}.
A precision of the order of 2-5\% has been achieved (the larger uncertainties are in the lower rapidity range). Higher precision could be reached by using a finer grid, but we consider it unnecessary in view of the much larger uncertainties originating from varying the physical scales.

The theoretical uncertainties have been estimated as usual by varying the renormalization scale \(\mu_R\), the factorization scale \(\mu_F\) and the BFKL scale \(s_0\) by a factor of \(\half\) and 2. In the following, the result of each scale variation is depicted individually.
Renormalization and factorization scales were chosen as $\mu_F=\mu_R=\abs{\vec{p}_{j_1}}+\abs{\vec{p}_{j_2}}$ for both IFs but were varied independently to estimate the uncertainty. Additional choices for \(\mu_R\) were also  explored and discussed below. 
The BFKL scale was fixed at the  value  \(s_0(k_{j_i})=\abs{\vec{p}_{j_1}}\abs{\vec{p}_{j_2}}\)\footnote{Note that, with this choice, \(\hat{s}/s_0\), the argument of the GGFs is not necessarily equal to the jet rapidity difference (\(\hat{s}/s_0\neq \exp{Y_{j_{12}}}\)) nor to the gap size (\(\hat{s}/s_0\neq \exp{Y_{\rm gap}}\)).}.

\subsection{Predictions on dependence on the rapidity, transverse momentum and difference in azimuthal angle}

\begin{figure}[ph]
  \centering
  \includegraphics[width=0.8\linewidth]{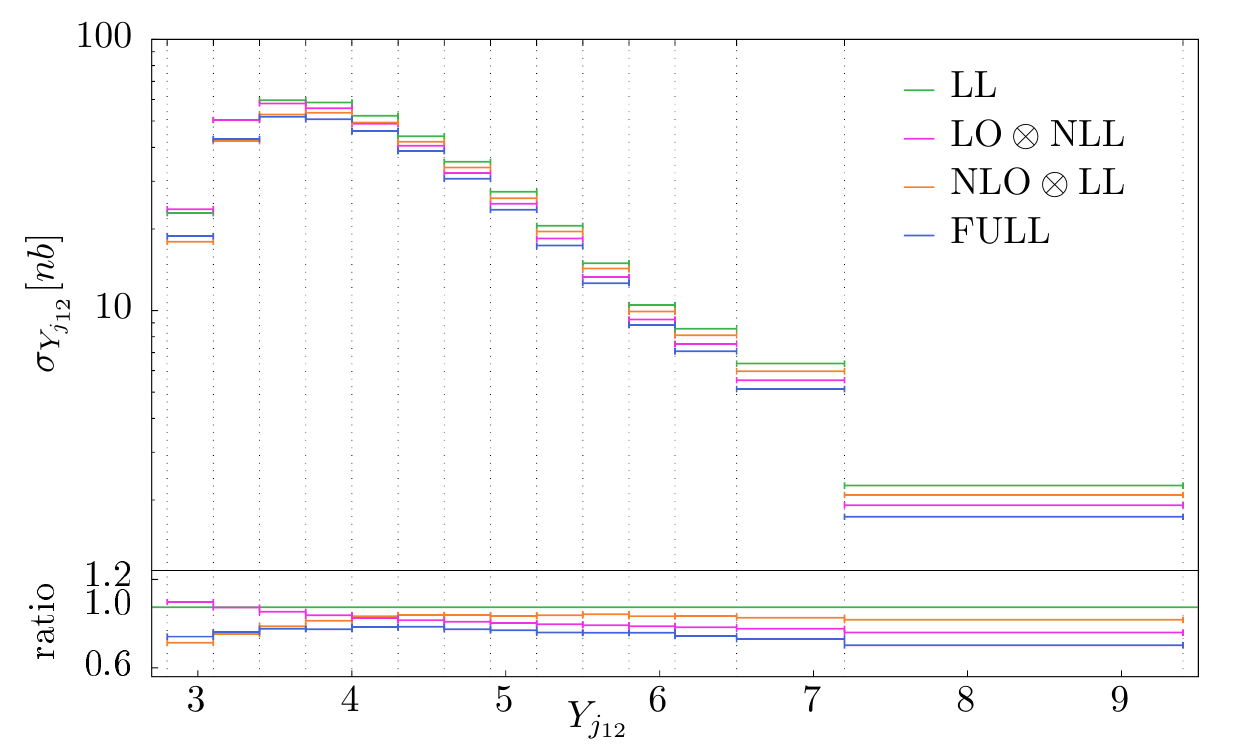}\hspace{10pt}
\caption{Mueller-Tang cross-section $\sigma_{Y_{j_{12}}}$ at LL (green), NLO\(\otimes\)LL (orange), LO\(\otimes\)NLL (pink) and FULL NL (blue).
The  NLO\(\otimes\)LL and LO\(\otimes\)NLL terms include the LL predictions.
We also display the ratios with respect to the LL predictions in the bottom plot. 
The NLO IF and NLL GGF corrections have the effect of reducing the cross-section estimate. The IF NLO corrections have a strong effect on the low rapidity range whereas the NLL GFF ones dominate at higher rapidities.
As a consequence of these two negative contributions, the FULL NL predictions remain below the LL estimate for the whole rapidity range by a factor of 15-20\%.}
\label{fig6.1}
\end{figure}

\begin{figure}[ph]
  \centering
  \includegraphics[width=0.8\linewidth]{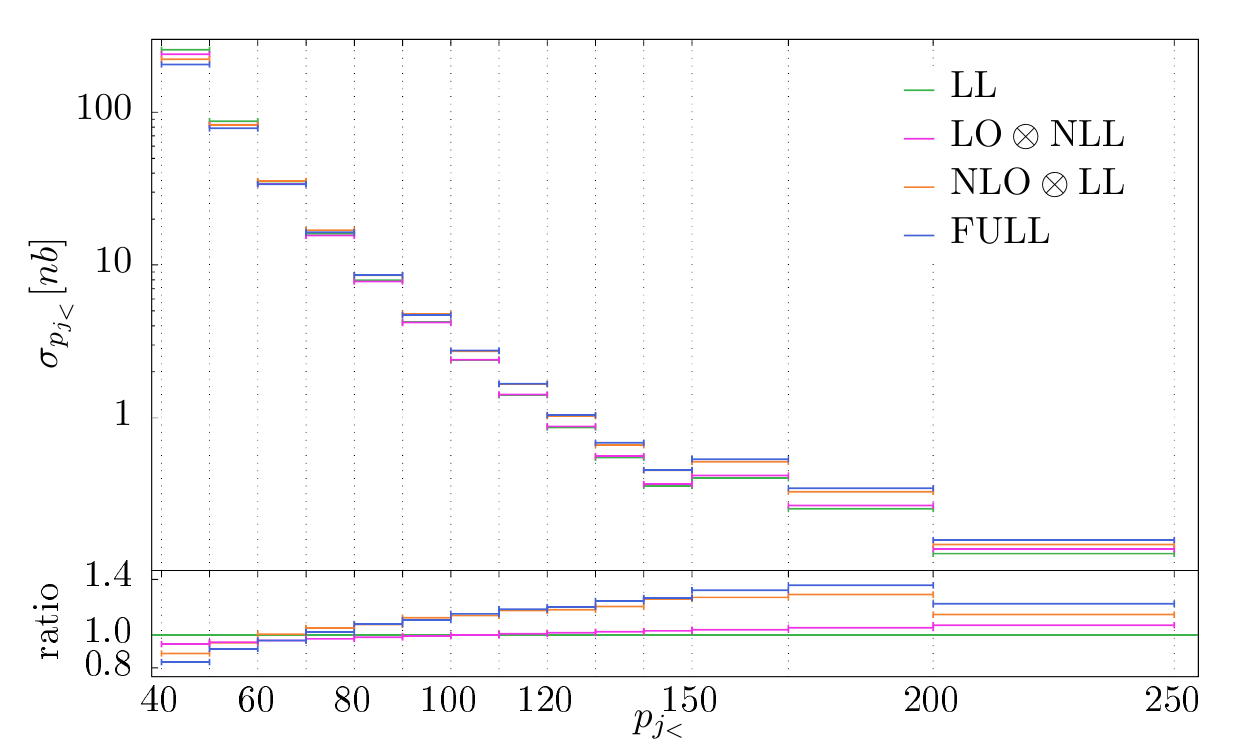}\hspace{10pt}
\caption{Mueller-Tang cross-section \(\sigma_{p_{j_{<}}}\)  at LL (green), NLO\(\otimes\)LL (orange), LO\(\otimes\)NLL (pink) and FULL NL (blue).
The  NLO\(\otimes\)LL and LO\(\otimes\)NLL terms include the LL predictions.
We also display the ratios with respect to the LL prediction in the bottom plot.
The NLO IF corrections change sign around \(p_{j_<}\simeq 70\) GeV. Overall, the effect is to predict a larger cross section at higher momentum relatively to LL.}
  \label{fig6.2}
\end{figure}

In this section, we present and discuss the results of the BFKL calculations on the jet-gap-jet  \(\sigma_{Y_{j_{12}}}\), \(\sigma_{p_{j<}}\) and \(\sigma_{\Delta\phi_{j_{12}}}\) cross sections.
In all figures the plot canvas is splitted in two parts: the top part gives the absolute values of the cross sections in nanobarn (nb) on  logarithmic scale and the bottom plot displays the ratios with respect to the standard value or the LL calculation depending on the plots.
In Figs. \ref{fig6.1} and \ref{fig6.2}, the MT predictions are shown in green for LL, in orange for LL\(\otimes\)NLO, in pink for NLL\(\otimes\)LO and in blue for the FULL NLO, following the notations given in the last paragraph of sec.~\ref{sec:MTJ-NLO}.
The cross sections in pink and orange include the sum with the pure LL results. The same color scheme will consistently be employed throughout the result section.
The bottom portion of the plots shows in general the ratios of the various approximations with respect to the LL one.

The corrections to the GGFs and to the IFs tend to reduce the cross-section with the latter  dominating over the former at small rapidities and vice-versa at higher rapidity separation.

The azimuthal angular distribution is shown in Fig.~\ref{fig6.5}. At LL, it is a delta distribution at $\pi$ by definition, and at NLO, it is strongly peaked towards \(\Delta\phi_{j_{12}}\sim \pi\), with a strong suppression towards the opposite side of the  angular range  \(\Delta\phi_{j_{12}}\sim \pi/2\), where it is compatible with zero.%
\footnote{The reason the distribution cannot reach lower angular values is related to the observable definition. The single emission originating from the LO IF must be balanced by the two partons on the opposite side.
  Transverse momentum conservation prohibits the hardest of the two parton system to be emitted towards the same semi-plane of the opposite jet.}

\subsection{Studies on the dependence on the energy scales}

In this section, we discuss the uncertainties on the BFKL calculations related to the factorization, renormalization and BFKL scale variations.

The theoretical uncertainty is estimated by varying all physical scales (\(\mu_R, \mu_F, s_0\)) that appear as a consequence of the truncation of the perturbative expansion.
Usually, one considers as scale uncertainty the differences between the results of the calculations when the scales are halved and  doubled.
The observation that the sensitivity to the scale variations decreases while increasing the order of the approximation is considered to be an indication of a good convergence behavior. Calculations based on the BFKL approach have not always fulfilled that criterion and we will discuss this in detail in the following.

\begin{figure}[ph]
  \centering
  \includegraphics[width=0.8\linewidth]{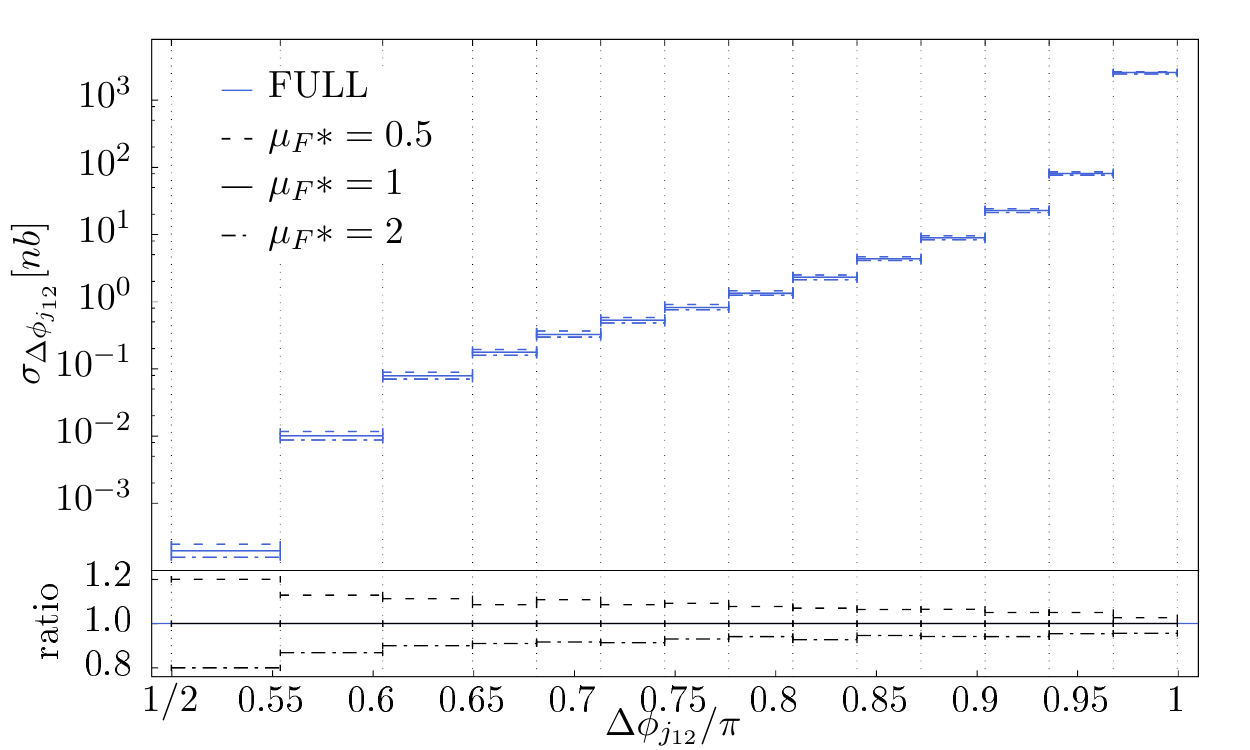}\hspace{10pt}

  \caption{Mueller-Tang cross-section \(\smfrac{\dd\sigma}{d\Delta\phi_{j_{12}}}\) for three different choices of the factorization scale  \(\mu_F=\{2\mu^*_F,\mu^*_F,\mu^*_F/2\}\), where \(\mu^*_F=\abs{\mathbf{p}_{j_1}}+\abs{\mathbf{p}_{j_2}}\). In the lower plot, we display the ratio with respect to $\mu_F=1$. This leads to a systematic uncertainty on the calculation between 5\% and 20\%.}
  \label{fig6.5}
\end{figure}

\begin{figure}[ph]
  \centering
   \includegraphics[width=0.8\linewidth]{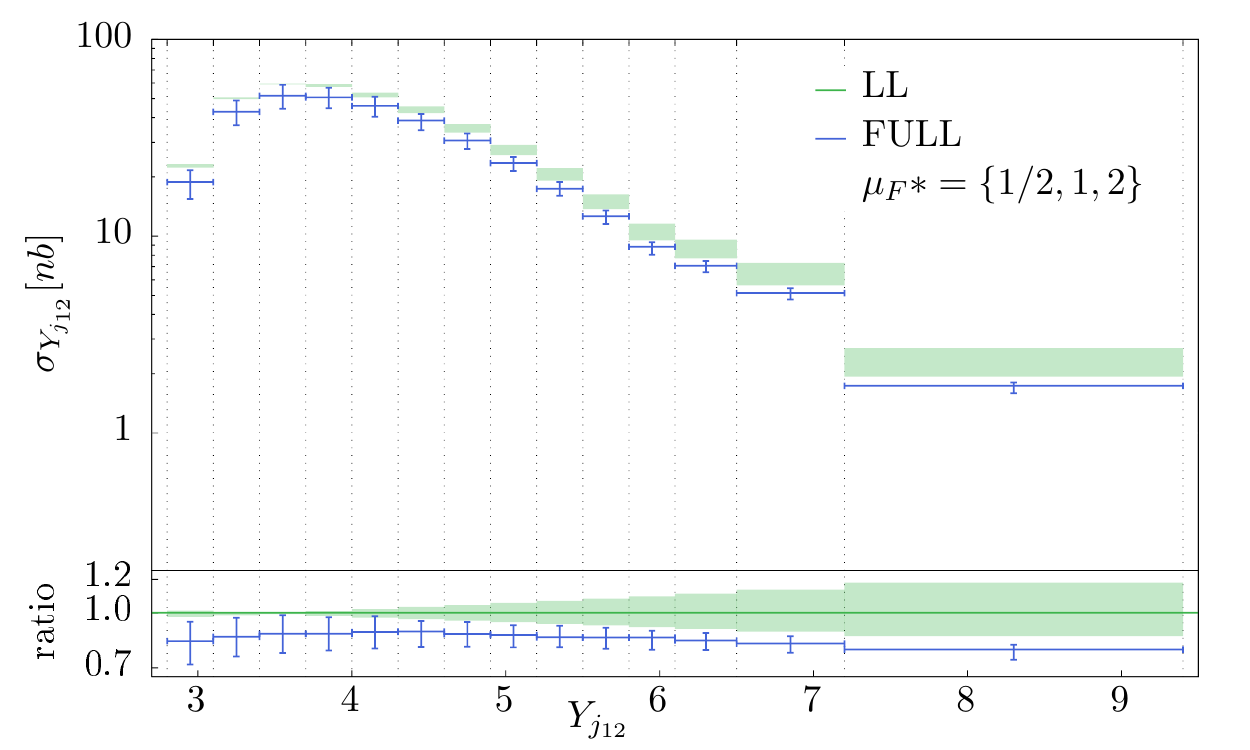}\hspace{10pt}
\caption{Mueller-Tang cross-section \(\sigma_{Y_{j_{12}}}\) at LL (green) and NLO (blue). The vertical bars (respectively the green band) represent the uncertainty due to the factorization scale variation \(\mu_F=\{2\mu^*_F,\mu^*_F/2\}\), where \(\mu^*_F=\abs{\mathbf{p}_{j_1}}+\abs{\mathbf{p}_{j_2}}\) at NLO (respectively LL). The ratio with respect to the LL prediction is shown in the bottom plot.
  The uncertainties due to the $\mu_F$ variations are of the order of 15-20\%.
  The FULL  and LL uncertainty bands do not overlap, albeit by a very small margin, for the whole rapidity range.
The \(\mu_F\) dependence is stronger for the FULL NL predictions compared to the LL estimate except in the high rapidity range. }
\label{fig6.3}
\end{figure}

\begin{figure}[ph]
  \centering
   \includegraphics[width=0.8\linewidth]{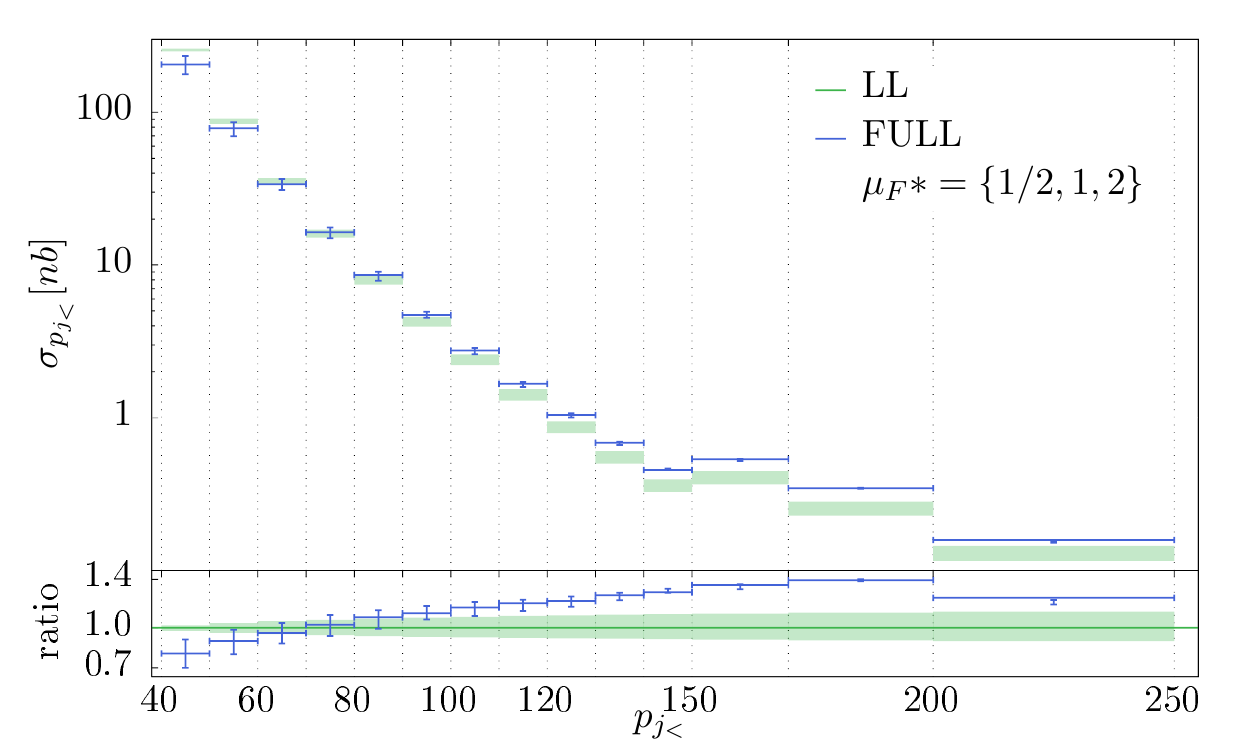}\hspace{10pt}
\caption{Mueller-Tang cross-section \(\sigma_{p_{j_{<}}}\) at LL (green) and NLO (blue). The vertical bars (respectively the green band) represent the uncertainty due to the factorization scale variation \(\mu_F=\{2\mu^*_F,\mu^*_F/2\}\), where \(\mu^*_F=\abs{\mathbf{p}_{j_1}}+\abs{\mathbf{p}_{j_2}}\) at NLO (respectively LL). The ratio with respect to the LL prediction is shown in the bottom plot.
The uncertainties due to the $\mu_F$ variations are of the order of 15-20\%.
The FULL  and LL uncertainty bands do not overlap, albeit by a very small margin, for the whole rapidity range.
The \(\mu_F\) dependence is stronger for the FULL NL predictions compared to the LL estimate except at large ``second leading'' jet momentum. }
\label{fig6.4}
\end{figure}

\begin{figure}[ph]
  \centering
 \includegraphics[width=0.8\linewidth]{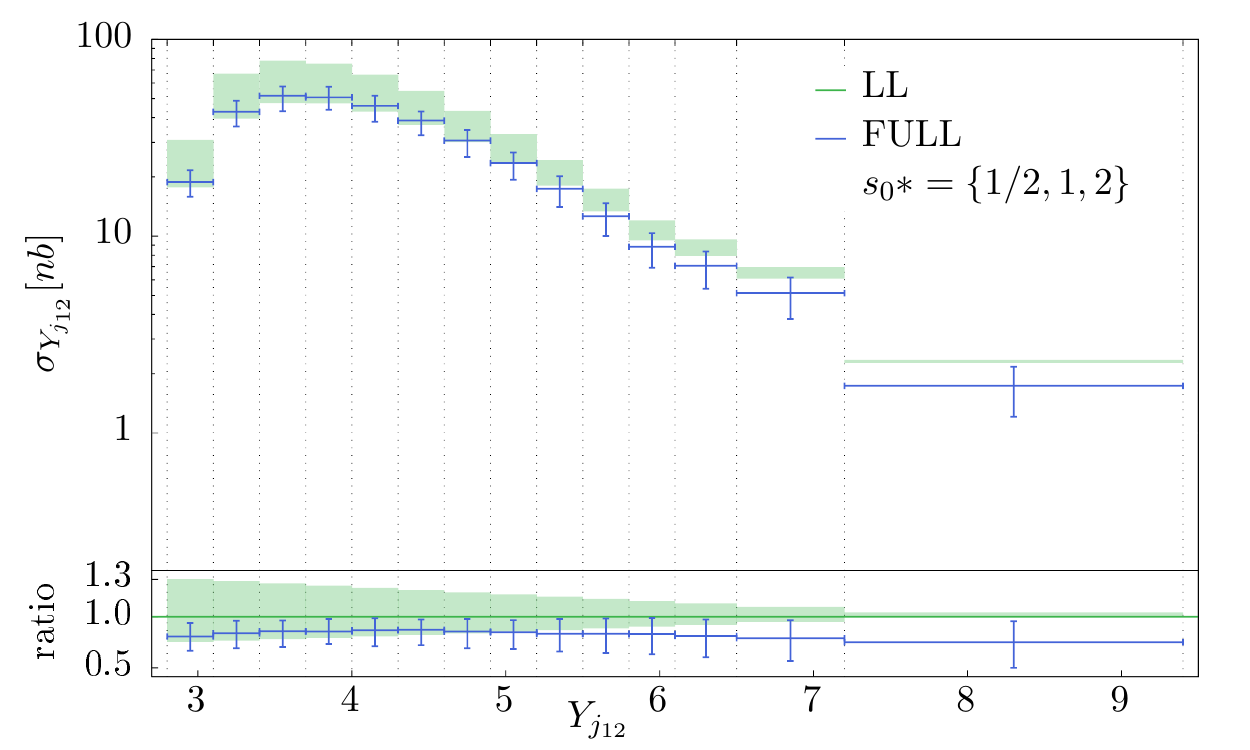}\hspace{10pt}

  \caption{Mueller-Tang cross-section \(\sigma_{Y_{j_{12}}}\) at LL (green) and FULL NLO approximation (blue). The vertical bars (respectively the green band) represent the uncertainty due to the BFKL scale variation \(s_0=\{2s^*_0,s^*_0/2\}\), where $s^*_0=\abs{\mathbf{p}_{j_1}}\abs{\mathbf{p}_{j_2}}$ at NLO (respectively LL). In the bottom plot, we show the ratio FULL NLO to LL. The NLO corrections marginally reduce the uncertainty coming from the choice of the BFKL scale. The systematic uncertainty band for the LL and NLO cross sections are of the order of 15\% at low $Y_{j_{12}}$ and increase up to 30\% at large rapidities.}
  \label{fig6.6}
\end{figure}

\begin{figure}[ph]
  \centering
\includegraphics[width=0.8\linewidth]{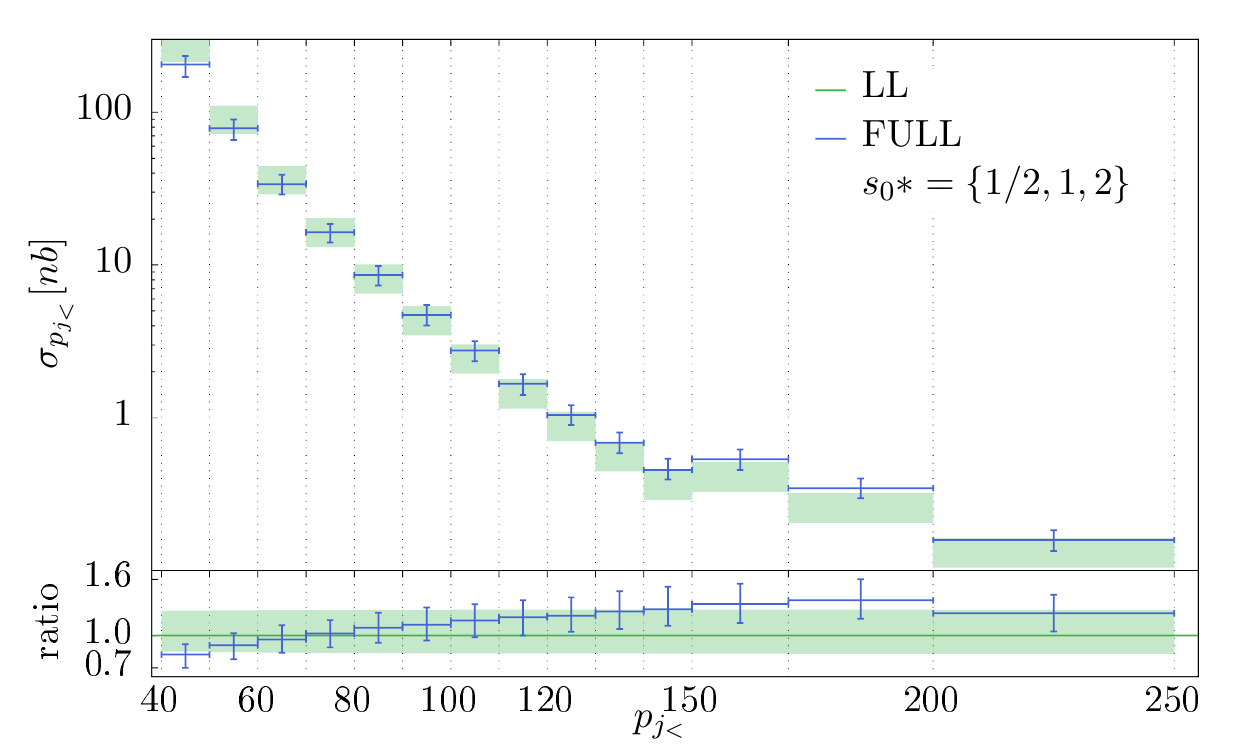}\hspace{10pt}

 \caption{Mueller-Tang cross-section \(\sigma_{p_{j_{<}}}\) at LL (green) and  FULL NLO approximation (blue). The vertical bars (respectively the green band) represent the uncertainty due to the BFKL scale variation \(s_0=\{2s^*_0,s^*_0/2\}\), where $s^*_0=\abs{\mathbf{p}_{j_1}}\abs{\mathbf{p}_{j_2}}$ at NLO (respectively LL). In the bottom plot, we show the ratio FULL NLO to LL.
The NLO corrections marginally reduce the uncertainty coming from the choice of the BFKL scale. The systematic uncertainty band for the LL and NLO cross sections are of the order of 15\% at low $p_{j_<}$ and increase up to 30\% at large momentum.}
  \label{fig6.7}
\end{figure}

\begin{figure}[ph]
  \centering
  \includegraphics[width=0.8\linewidth]{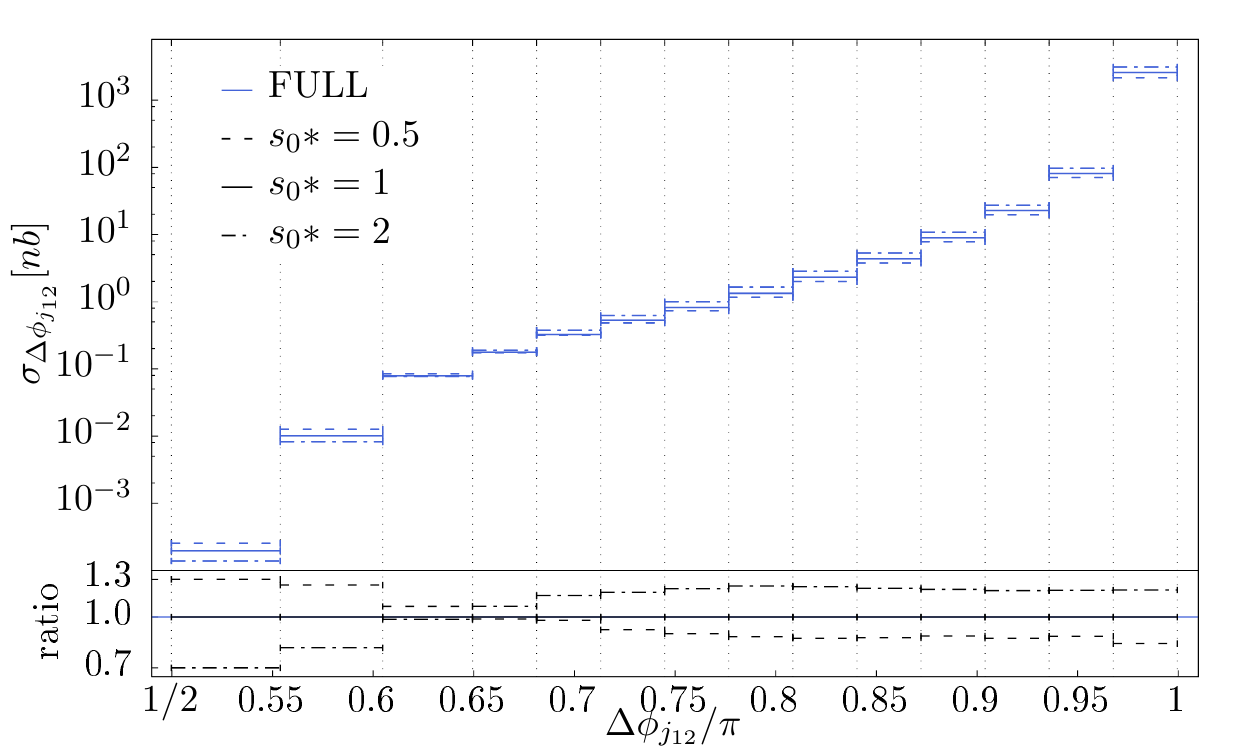}\hspace{10pt}
    \caption{NLO Mueller-Tang cross-section \(\sigma_{\Delta\phi_{j_{12}}}\) for three different choices of the BFKL scale  \(s_0=\{2s^*_0,s^*_0,s^*_0/2\}\), where \(s^*_0=\abs{\mathbf{p}_{j_1}}\abs{\mathbf{p}_{j_2}}\). The effect of varying \(s_0\) is stronger at smaller angles, ranging from 20\% to 30\%. The bottom plot shows the ratios with respect to $s_0=s^*_0$.}
  \label{fig6.8}
\end{figure}

\begin{figure}[ph]
  \centering

\includegraphics[width=0.8\linewidth]{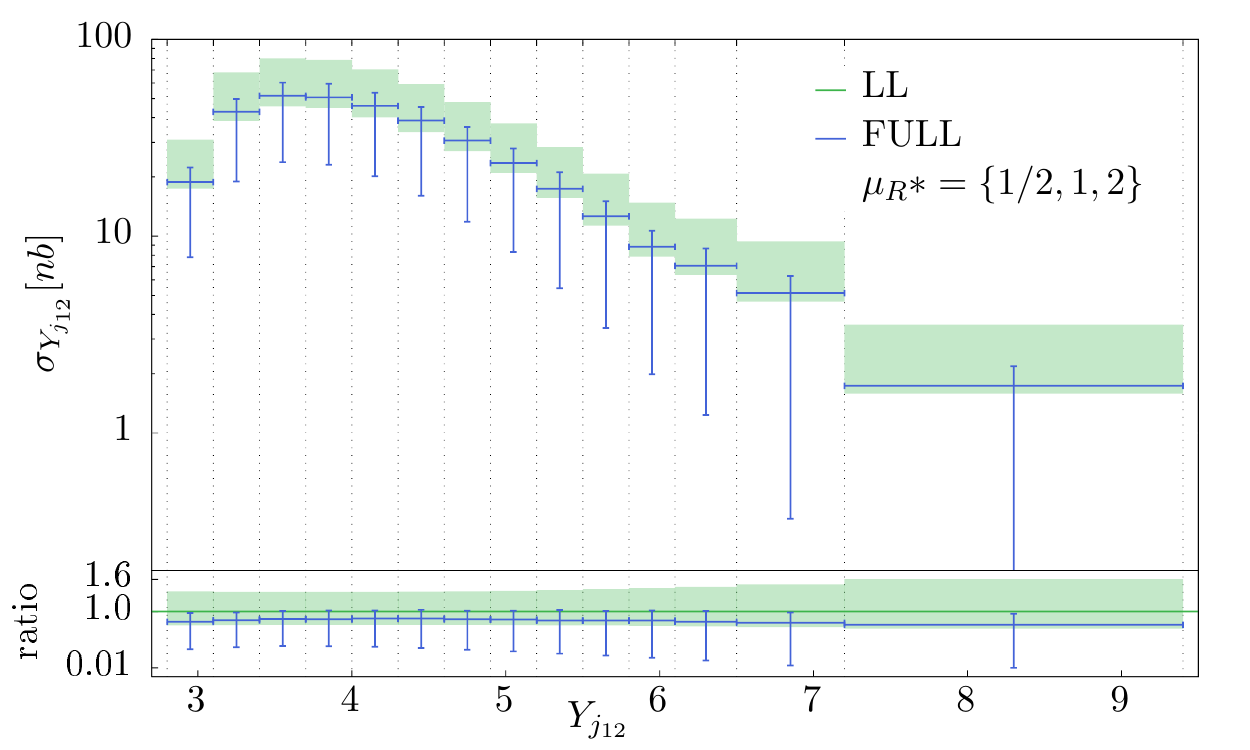}\hspace{10pt}
\caption{Mueller-Tang cross-section \(\sigma_{Y_{j_{12}}}\) at LL (green) and FULL NLO approximation (blue). The  vertical bars (respectively the green band) represent the uncertainty due to the renormalization scale variation \(\mu_R=\{2\mu^*_R,\mu^*_R/2\}\), where \(\mu^*_R=\abs{\mathbf{p}_{j_1}}+\abs{\mathbf{p}_{j_2}}\) at NLO (respectively at LO). The ratio NLO/LL (shown in the bottom plot) is compatible with 1 throughout the whole rapidity range.
The renormalization scale dependence is non-linear. A larger change is observed  when \(\mu_R\) is halved compared to when it is doubled, going from a 20\% to 90\% systematic uncertainty.}
  \label{fig6.9}
\end{figure}

\begin{figure}[ph]
  \centering

\includegraphics[width=0.8\linewidth]{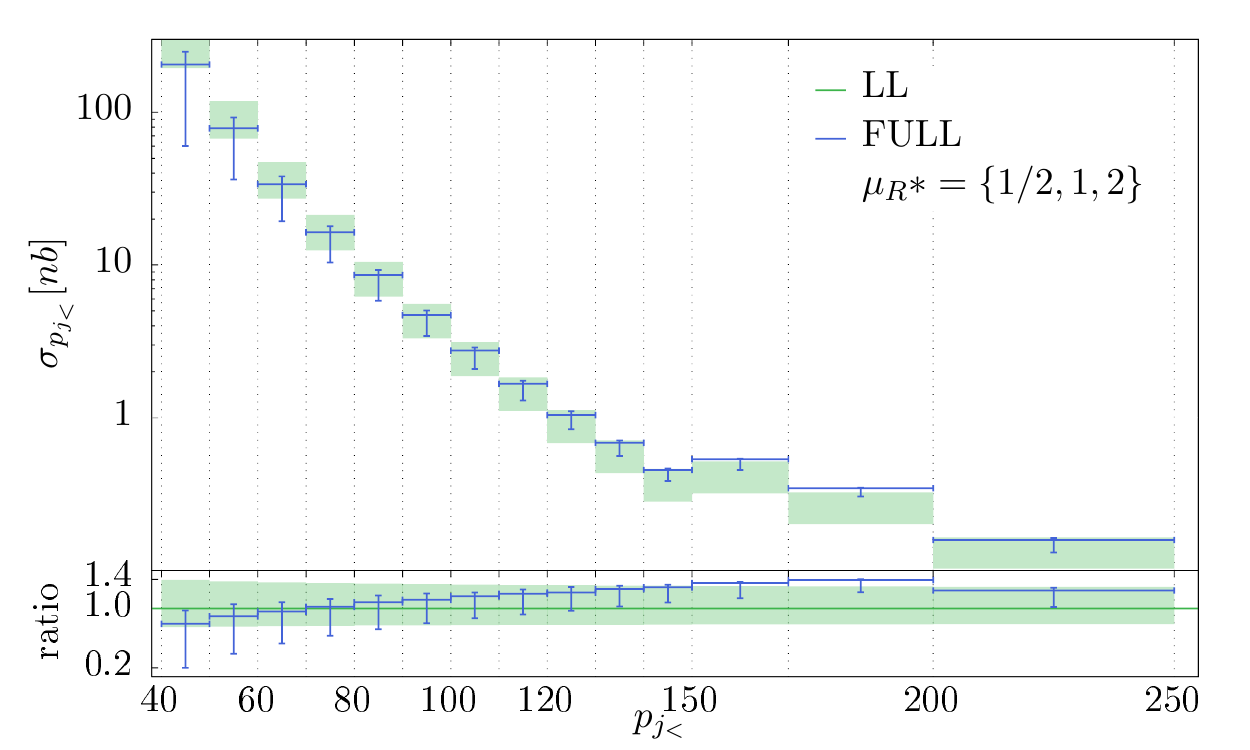}\hspace{10pt}
\caption{Mueller-Tang cross-section \(\sigma_{p_{j_{<}}}\) at LL (green) and FULL NLO approximation (blue). The  vertical bars (respectively the green band) represent the uncertainty due to the renormalization scale variation \(\mu_R=\{2\mu^*_R,\mu^*_R/2\}\), where \(\mu^*_R=\abs{\mathbf{p}_{j_1}}+\abs{\mathbf{p}_{j_2}}\) at NLO (respectively LL). The ratio NLO/LL (shown in the bottom plot) is compatible with 1 throughout the whole rapidity range.
The renormalization scale dependence is non-linear. A larger change is observed  when \(\mu_R\) is halved compared to when it is doubled, going from a 20\% to 90\% systematics. }
  \label{fig6.10}
\end{figure}

\begin{figure}[ph]
  \centering

\includegraphics[width=0.8\linewidth]{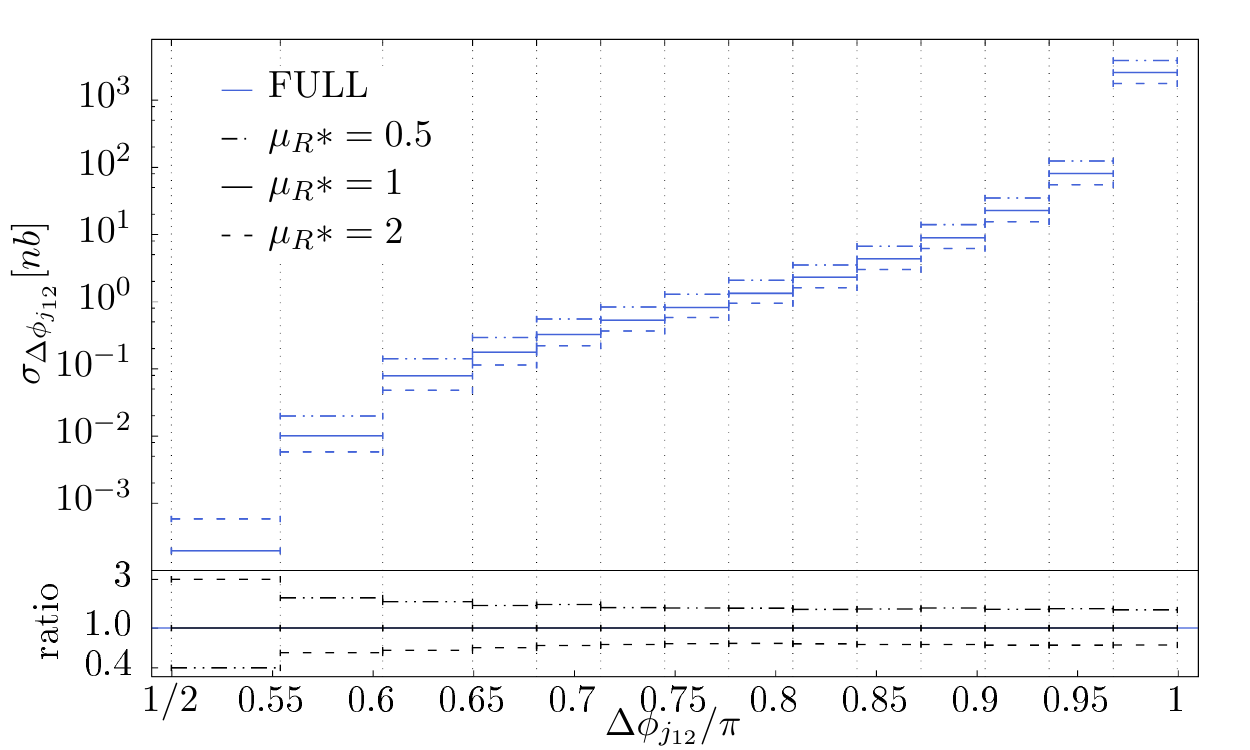}\hspace{10pt}
\caption{NLO Mueller-Tang cross-section \(\sigma_{\Delta\phi_{j_{12}}}\) for three different choices of the renormalization scale  \(\mu_R=\{2\mu^*_R,\mu^*_R,\mu^*_R/2\}\). In the bottom plot, we display the ratio with respect to the default choice $\mu_R=\mu^*_R=1$. The systematic uncertainties related to the $\mu_R$ variations are about 30-50\% and get slightly larger at very low angles.}
  \label{fig6.11}
\end{figure}

We first display the effects of scale variation in two sets of figures. The first set, Figs.~\ref{fig6.3}, \ref{fig6.4}, \ref{fig6.6}, \ref{fig6.7}, \ref{fig6.9} and \ref{fig6.10} display the effects of varying the scales (renormalization, factorization, etc.) on the $Y_{j_{12}}$ and \(p_{j_{<}}\) distributions as a blue vertical error bars for the NLO predictions and as a green band for the LL calculation. At the bottom of each figure, the ratio of the NLL results with respect to the LL calculation is shown.
Figs.~\ref{fig6.5}, \ref{fig6.8}, and \ref{fig6.11}  display the scale variations for the $\Delta\phi_{j_{12}}$ variable for the FULL NLO calculation only, as well as the ratios with respect to the default scale value. The various scale choices are indicated by different line types. The solid line always represents the default value. \(\sigma_{\Delta\phi_{j_{12}}}\) is indeed a Dirac-delta distribution in LLA, which prevents a meaningful comparison with the FULL NLO result.

In Fig.~\ref{fig6.3}, we observe that
the FULL NLO contribution is less sensitive to the choice of the factorization scale only in the higher rapidity region compared to the LL case. On the other hand, Fig.~\ref{fig6.4} shows that the NLO corrections tend to reduce the overall sensitivity to the choice of the scale only at moderate jet transverse momentum.
Figs.~\ref{fig6.9} and \ref{fig6.10} show that the NLO BFKL corrections do not reduce the renormalization scale uncertainty band if the renormalization scale is fixed at the ``natural'' scale \(\mu_R^N\equiv\abs{\vec{p}_{j_1}}+\abs{\vec{p}_{j_2}}\). In particular, when the scale is halved, the prediction is reduced by about 50-60\%, a rather large factor.

All scales \(\mu_R,\mu_F,s_0\) influence the results implicitly through the strong coupling, the PDFs, and the GGF, but also, explicitly, as part of the NLO corrections.
A slightly more monotonous dependence on all scale variations is observed for the $\phi$ angular distribution in Figs.~\ref{fig6.5}, \ref{fig6.8} and \ref{fig6.11}  since the scale variations only appear implicitly on the cross-section at angles \(\Delta\phi_{j_{12}}<\pi\).

Summarizing, typical uncertainties related to the variations of the $\mu_F$, $\mu_R$ and $s_0$ scales are of the order of 10-20\%, 50-60\% and 20\%, respectively. In particular, the variations of $\mu_R$ lead to a larger systematic uncertainty. This sensitivity can be regarded as a symptom of instability of the expansion due to large subleading corrections that demand a tailored treatment.
It is often suggested that a better choice for the renormalization scale (that takes better accounts of the higher and uncontrolled terms of the perturbative expansion) can be found by using dedicated methods. These methods often tend to use a scale further away (much larger) from the ``natural" ones and their interval allowed by the uncertainty definition.
In the following, we will use the principle of minimal sensitivity (PMS) which prescribes to fix the coupling scale at the value where the variation induced on the cross section has a stationary point.

\begin{figure}[ph]
  \centering

\includegraphics[width=0.8\linewidth]{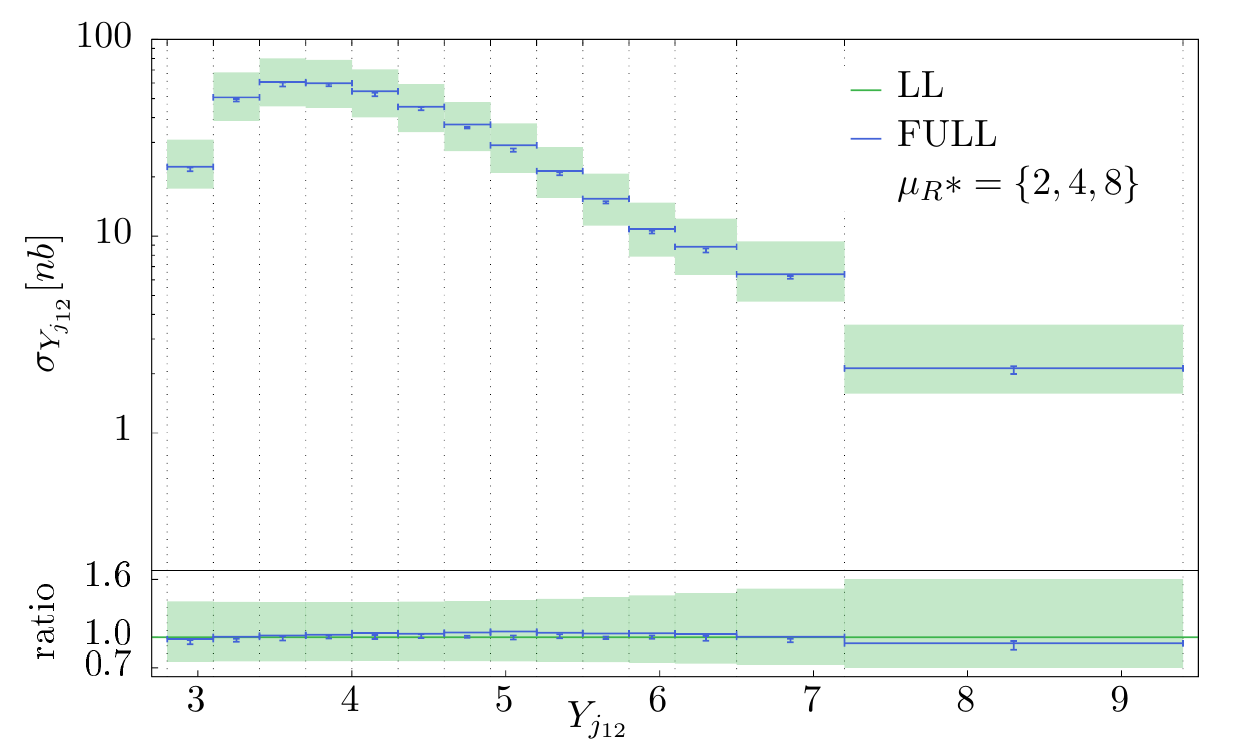}\hspace{10pt}

\caption{Mueller-Tang cross-section \(\sigma_{Y_{j_{12}}}\) at LL (green) and FULL NLO approximation (blue). The vertical bars (respectively the green band) represent the uncertainty due to the renormalization scale variation \(\mu_R=\{\mu^*_R,\mu^*_R/2\}\), where \(\mu^*_R=4\l(\abs{\mathbf{p}_{j_1}}+\abs{\mathbf{p}_{j_2}}\r)\) at NLO (respectively LL).
The ratio FULL/LL (shown in the bottom plot) is compatible with 1 throughout the whole rapidity range. }
  \label{fig6.12}
\end{figure}

\begin{figure}[ph]
  \centering

\includegraphics[width=0.8\linewidth]{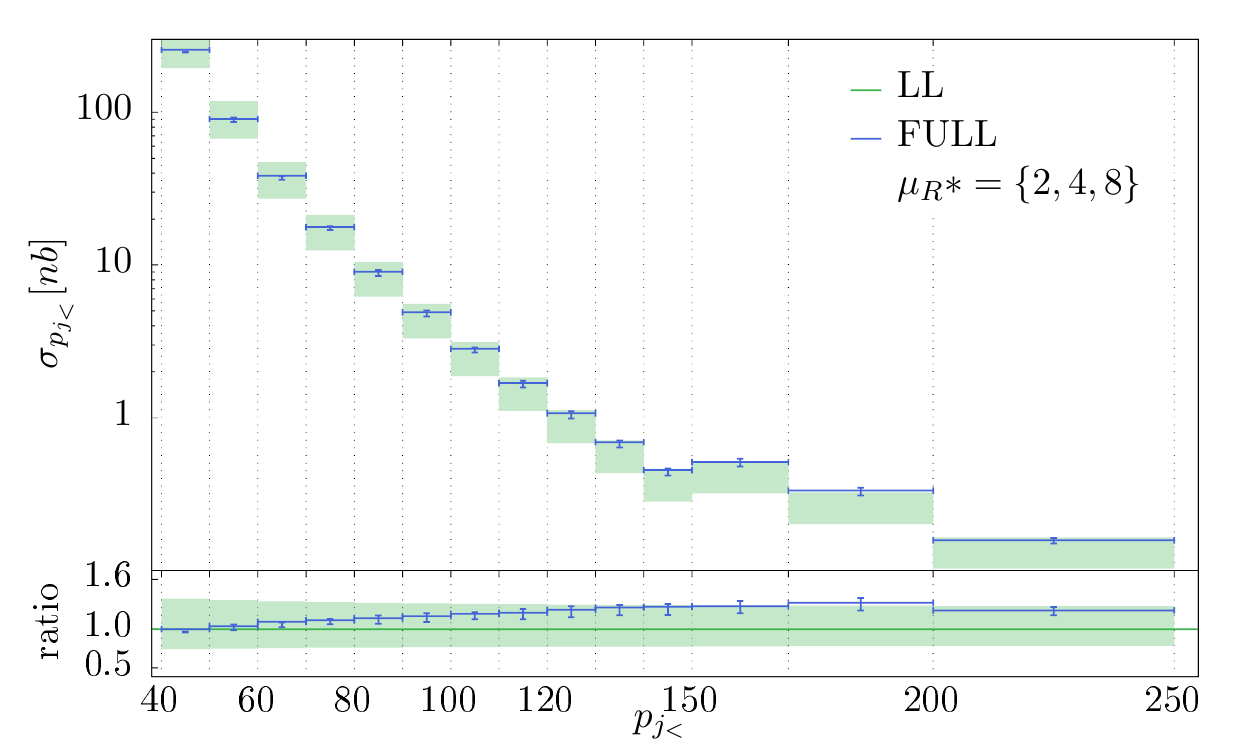}\hspace{10pt}

\caption{Mueller-Tang cross-section \(\sigma_{p_{j_{<}}}\) at LL (green) and FULL NLO approximation (blue). The vertical bars (respectively the green band) represent the uncertainty due to the renormalization scale variation \(\mu_R=\{\mu^*_R,\mu^*_R/2\}\), where \(\mu^*_R=4\l(\abs{\mathbf{p}_{j_1}}+\abs{\mathbf{p}_{j_2}}\r)\), at NLO (respectively LL).
On the bottom plot, the vertical error bars never cross the horizontal line corresponding to the equality of FULL and LL predictions, except for the first few bins.}
  \label{fig6.13}
\end{figure}

\begin{figure}[ph]
  \centering
\includegraphics[width=0.8\linewidth]{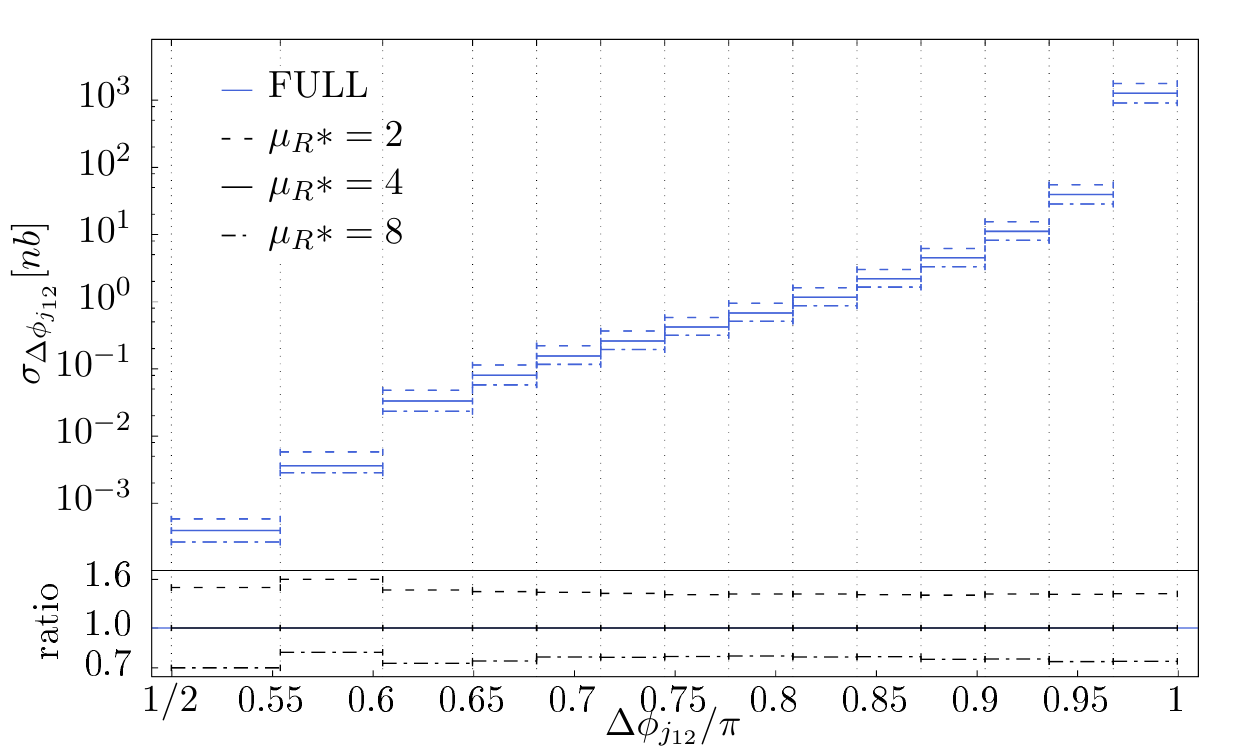}\hspace{10pt}
\caption{Mueller-Tang cross-section \(\sigma_{\Delta\phi_{j_{12}}}\) for three different choices of the renormalization scale  \(\mu_R=\{2\mu^*_R,\mu^*_R,\mu^*_R/2\}\), where \(\mu^*_R=4\l(\abs{\mathbf{p}_{j_1}}+\abs{\mathbf{p}_{j_2}}\r)\).
In the bottom plot we display the ratio with respect to the choice fixed by the PMS tuning $\mu_R=\mu^*_R$.}
\label{fig6.14}
\end{figure}

\begin{figure}[ph]
  \centering
  \includegraphics[width=0.8\linewidth]{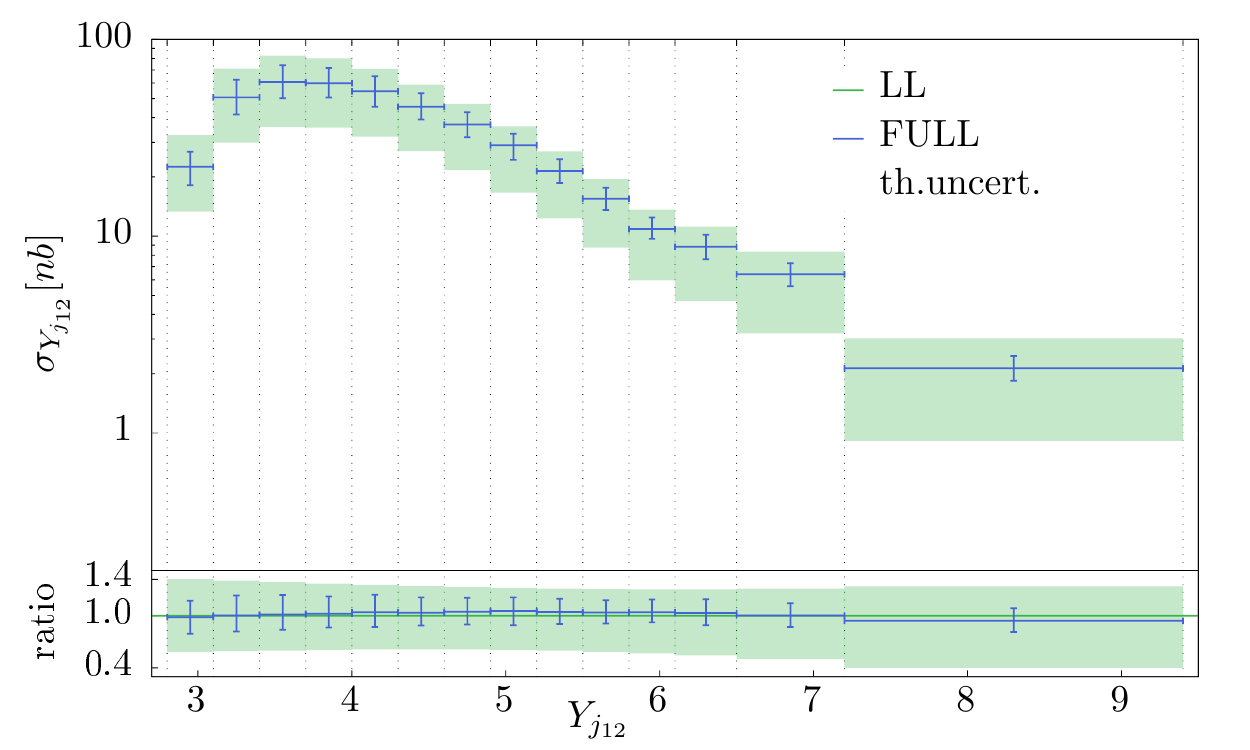}\hspace{10pt}
 \caption{Mueller-Tang cross-section \(\sigma_{Y_{j_{12}}}\) at LL (green) and  NLO (blue). The vertical bars represent the total uncertainty, summing in quadrature the uncertainties coming from the variation of \(\mu_R,\mu_F,s_0\).
\[
  \Delta\sigma^{\rm tot.}=
  \sqrt{(\Delta\sigma_{\mu_R})^2+(\Delta\sigma_{\mu_F})^2+(\Delta\sigma_{s_0})^2}.
\]
The default value \(\mu_R(=\mu^{\rm PMS}_R)\) originates from the PMS method for the FULL NLO contribution whereas it was set to the natural scale \(\mu_R=\mu^{\rm N}_R\) for the LL calculation. With this choice of the renormalization scale, the theoretical uncertainties are about 20\% (compared to about 50\% without this choice). The ratio NLO to LL is shown in the bottom plot. The LL and FULL NLO estimates remain consistent with 1 within the reduced uncertainty band for the whole rapidity range.}
  \label{fig6.15}
\end{figure}

\begin{figure}[ph]
  \centering
 \includegraphics[width=0.8\linewidth]{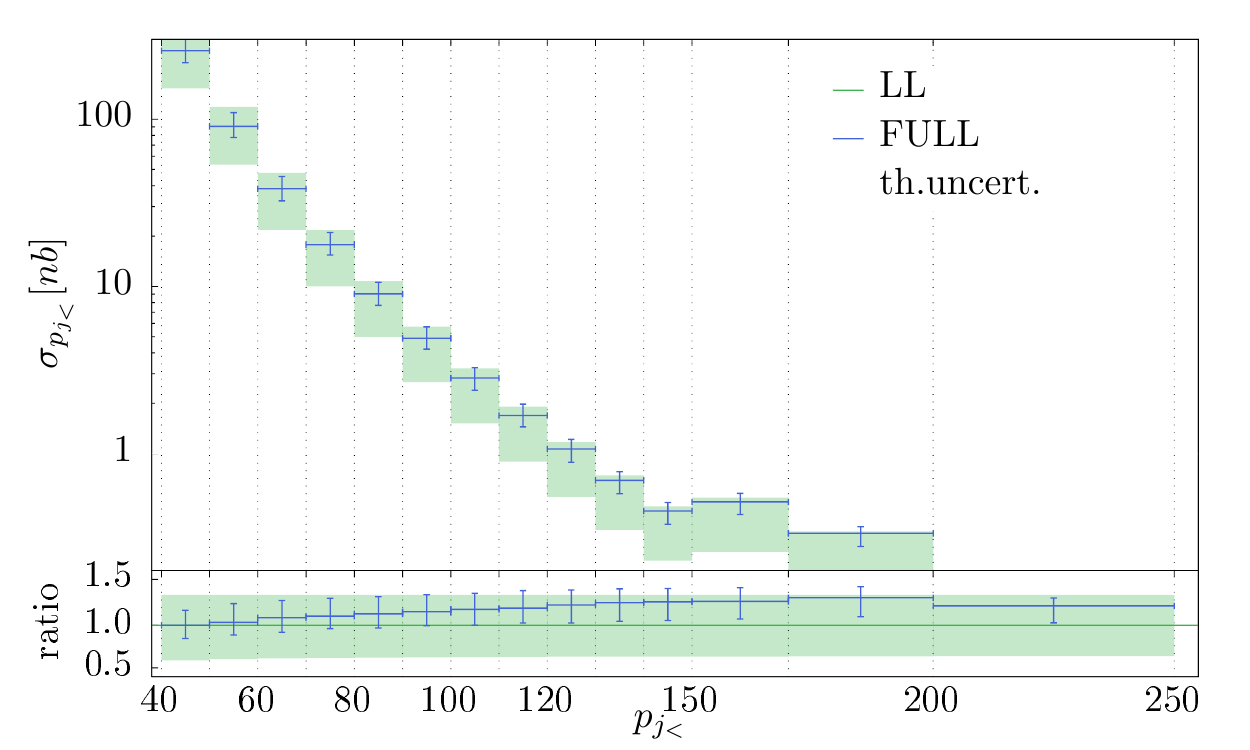}\hspace{10pt}
 \caption{Mueller-Tang cross-section \(\sigma_{p_{j_{<}}}\) at LL (green) and  NLO (blue). The vertical bars represent the total uncertainty, summing in quadrature the uncertainties coming from the variation of \(\mu_R,\mu_F,s_0\).
   \[
     \Delta\sigma^{\rm tot.}=
     \sqrt{(\Delta\sigma_{\mu_R})^2+(\Delta\sigma_{\mu_F})^2+(\Delta\sigma_{s0})^2}.
   \]
The default value of \(\mu_R\) originates from the PMS method \(\mu^{\rm PMS}_R\) for the FULL NLO contribution whereas it was set to the natural scale (\(mu^{\rm N}_R\)) for the LL contribution.
With this choice of the renormalization scale, the theoretical uncertainties are reduced after applying the NLO corrections.}
  \label{fig6.16}
\end{figure}

We found \(\mu_R^{\rm PMS}\simeq 4\mu^{\rm N}_R\) as the stationary point since we observe an inversion in the direction of the cross section variations in  Figs.~\ref{fig6.12} and \ref{fig6.13}. These figures show indeed that the uncertainty band narrows greatly at NLO using this choice of scale compared to the one at LL. Nonetheless, LL and FULL NLO predictions appear to fall within the uncertainties for \(\sigma_{Y_{j_{12}}}\). In \(\sigma_{p_{j_<}}\) the FULL NLO cross-section is also consistent with the LL estimate for the first few bins but tends to become increasingly larger at higher momenta.
The total uncertainty bands, at the PMS scale, accounting for the compound theoretical uncertainties from all the sources combined in quadrature are shown in Fig.~\ref{fig6.15} and \ref{fig6.16}.
The uncertainty band gets reduced to about 15-20\% when the NLO corrections are taken into account.
The cross section as a function of the azimuthal angles \(\Delta\phi_{j_{12}}<\pi\) cannot show (see Fig.~\ref{fig6.14}) a similar reduction since it depends on the renormalization scale only implicitly through the strong coupling constant dependence as an overall factor.

\subsection{Dependence on the gap size and the energy in the gap and influence of the factorization breaking term}

It is interesting to investigate the sensitivity to
the details of the gap definition. It is directly related to the NLO IF corrections and, consequently, was not performed in previous studies.
The parton emission is supressed in the gap region, hence it introduces a dependence on the gap definition and its parameters \(E_{\rm th}\) and \(Y_{\rm gap}\). Their values are not arbitrary but should be fixed to reproduce the experimental prescriptions.

\begin{figure}[ph]
  \centering
  \includegraphics[width=0.8\linewidth]{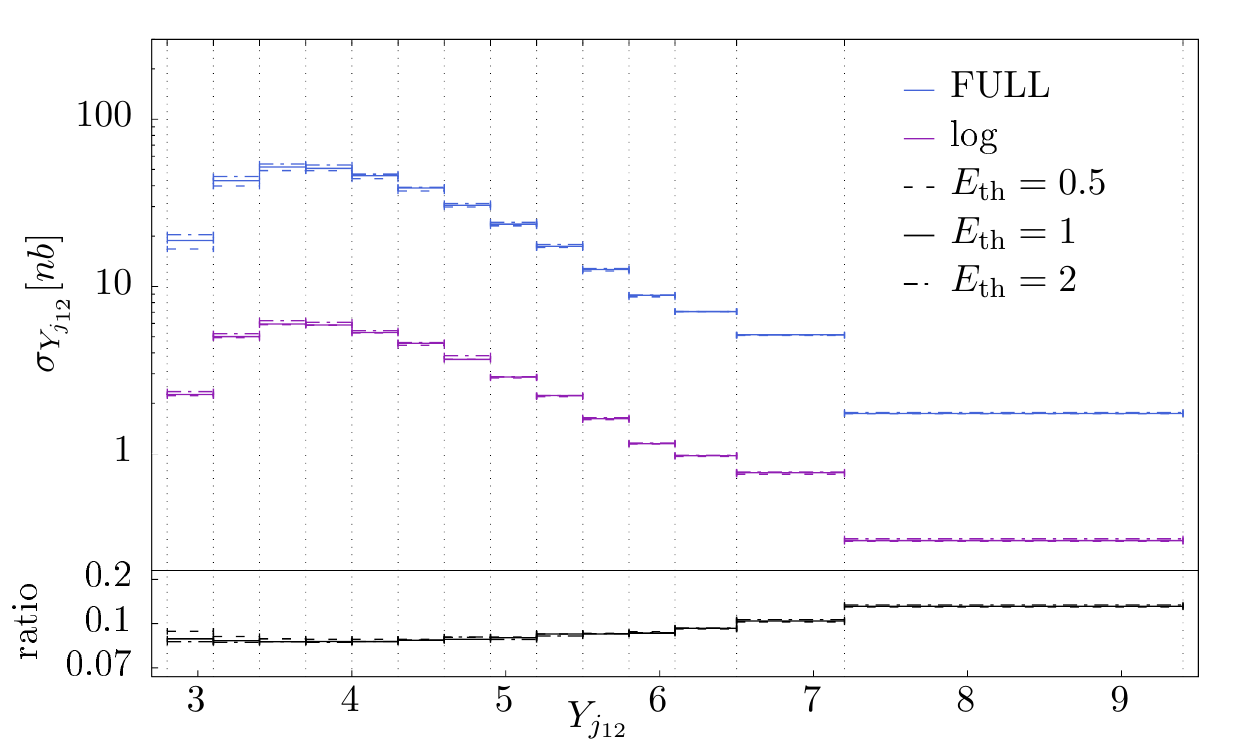}\hspace{10pt}
 \caption{Mueller-Tang cross-section \(\sigma_{Y_{j_{12}}}\) at FULL NLO (blue) and the contribution of the factorization breaking  logs term (violet), as a function of rapidity. In the bottom plot, we display the ratio of the factorization breaking term with respect to the NLO cross section. The effect of this term is small (less than 10\%) except at highest rapidity where it reaches up to 15\%. We also display three different choices for the transverse energy threshold (\(E_{{\rm th}}=\{1/2,1,2\}\) GeV) in the gap region using three line styles (full, dashed and dooted dashed). The effect of modifying the threshold value is minimal over the full rapidity range (except for the very first bin). The weak sensitivity to the energy threshold  is due to the fact that the gap region does not extend over the whole rapidity separation between the jets.
}
  \label{fig6.17}
\end{figure}

\begin{figure}[ph]
  \centering
   \includegraphics[width=0.8\linewidth]{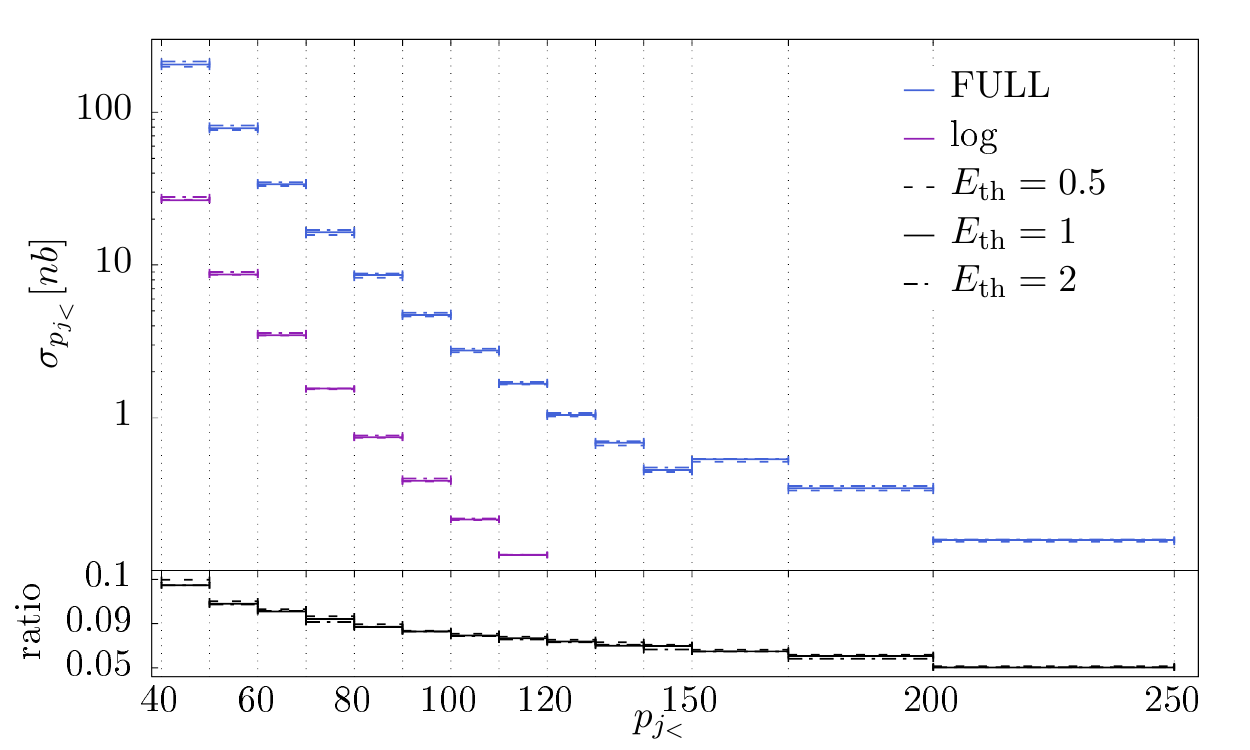}\hspace{10pt}
 \caption{Mueller-Tang cross-section \(\sigma_{p_{j_{<}}}\) at FULL NLO (blue) and the contribution of the factorization breaking  logs term (violet), as a function of jet $p_T$. In the bottom plot, we display the ratio of the factorization breaking term with respect to the NLO cross section.
 The effect of this term is small (less than 5\%) except at highest rapidity where it reaches up to 15\%. We also display three different choices for the transverse energy threshold (\(E_{{\rm th}}=\{1/2,1,2\}\) GeV) in the gap region  using three line styles (full, dashed and dotted dashed).
 The effect of modifying the threshold value is minimal over the full rapidity range (except for the very first bin).
The very weak sensitivity to the varying energy threshold  is due to the fact that the gap region does not extend over the whole rapidity separation between the jets.
}
  \label{fig6.18}
\end{figure}

\begin{figure}[ph]
  \centering
  \includegraphics[width=0.8\linewidth]{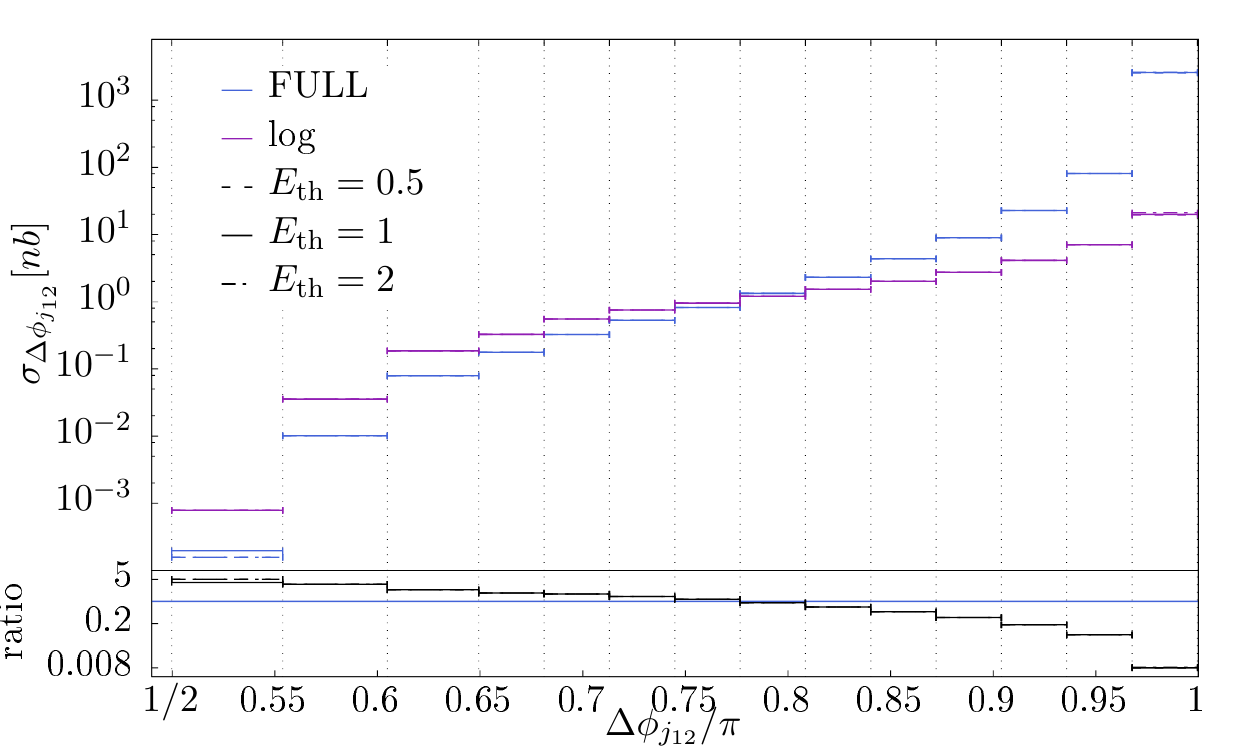}\hspace{10pt}
 \caption{NLO Mueller-Tang cross-section \(\sigma_{\Delta\phi_{j_{12}}}\) at FULL NLO (blue) and the contribution of the violating logs term (violet). Different dash types refer to three different choices for the transverse energy threshold (\(E_{{\rm th}}=\{1/2,1,2\}\) GeV) in the gap region. The sensitivity to the energy threshold is minimal.
}
  \label{fig6.19}
\end{figure}

In addition, it is quite interesting to study the sensitivity of the term responsible for the violation of the BFKL factorization to the gap definition.
Figs.~\ref{fig6.17}, \ref{fig6.18} and \ref{fig6.19} show the effects of this ``\(\log s\)'' term compared to the FULL NLO calculation and they are displayed in purple and blue respectively. On the same plots we show (using the same colors) the effects of modifying the energy thresholds in the gap and the gap size as dashed and dotted-dashed lines.

For instance, for the MT measurements as performed in the CMS experiment, the threshold on the electromagnetic calorimeter energy is set up to \(E_{\rm th}=200\) MeV.
Since our study does not include hadronization effects, the spread of the primary parton energy over the detector area is not included in our study. We chose to fix the threshold level at the higher value of \(E_{\rm th}=1\) GeV. As it can be observed in Figs.~\ref{fig6.17}, \ref{fig6.18} and \ref{fig6.19}, the dependence of our results on the exact value of the threshold parameter is weak. The results do not vary much when one changes the value of the energy in the gap. This is a feature of the MT dynamics. When the interaction is mediated by the exchange of a pomeron, a color-singlet object, radiation is rarely emitted at large scattering angles.
The energy threshold variation has a small but visible effect only at very low rapidity differences, where the distance between the jet and the edge of the gap is minimal.

Let us now discuss the impact on our results of the ``\(\log s\)'' term, responsible for the BFKL factorization breaking. In fact, it amounts to a small fraction of the total BFKL NLO cross section. It grows from a ratio of 8\% at low rapidities to 20\% at the rapidity boundary, motivating our conclusions that the impact of the BFKL factorization can be neglected at LHC energies. On the other hand, from a pure theoretical point of view, the fact that the ratio ``\(\log s\)''/FULL raises with rapidity implies the non-reliability of BFKL factorization at larger center-of-mass energies, and will be an issue to compute the MT cross section accurately at very high center-of-mass energies.

Strictly speaking, already in the current kinematics, the violation cannot be overlooked in certain configurations. Fig.~\ref{fig6.19} shows that the ``\(\log s\)'' term can effectively be considered as a negligible contribution to the cross section
only approaching the ``back-to-back'' configuration. At smaller angles \(\l(\Delta\phi_{j_{12}}\lesssim \smfrac{3}{4}\pi\r)\), the BFKL factorization is ``maximally'' violated as ``\(\log s\)''\(>{\rm FULL}\). The estimate of the NLO MT cross section giving is not reliable in this kinematical region. Fortunately, the cross section over the full kinematical range is dominated by the kinematical region where violation is small.

\begin{figure}[ph]
  \centering
   \includegraphics[width=0.8\linewidth]{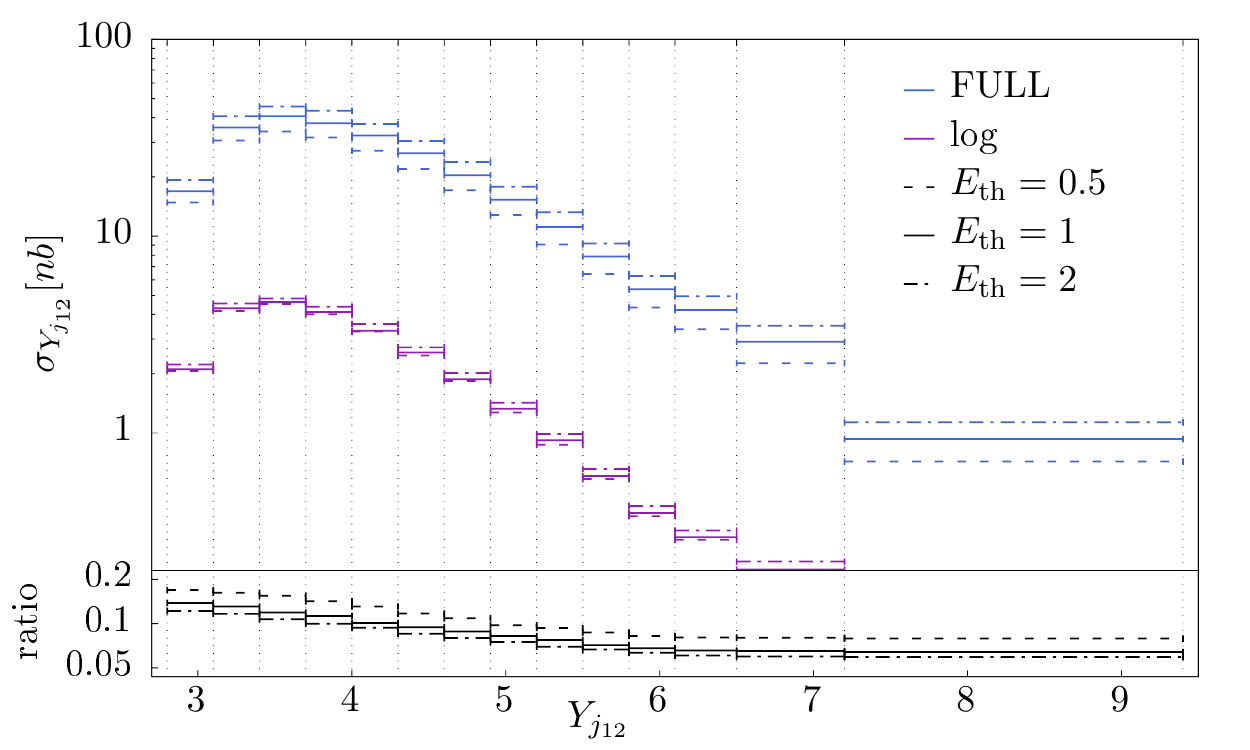}\hspace{10pt}
  \caption{NLO Mueller-Tang cross-section \(\sigma_{Y_{j_{12}}}\) (blue) and contribution of the violating logs term (violet) in case of a dynamic gap definition. Different line types refer to three different choices for the transverse energy threshold (\(E_{{\rm th}}=\{1/2,1,2\}\) GeV). The ratio relative to the default choice \(E^*_{\rm th}=1\) GeV is shown in the bottom plot.
With this gap choice, the factorization breaking effect never exceeds 10\% of the total and steadily decreases towards higher rapidity. The cross-section is less sensitive to the choice of the threshold energy while using a dynamic gap definition.
}
  \label{fig6.20}
\end{figure}

\begin{figure}[ph]
   \centering
    \includegraphics[width=0.8\linewidth]{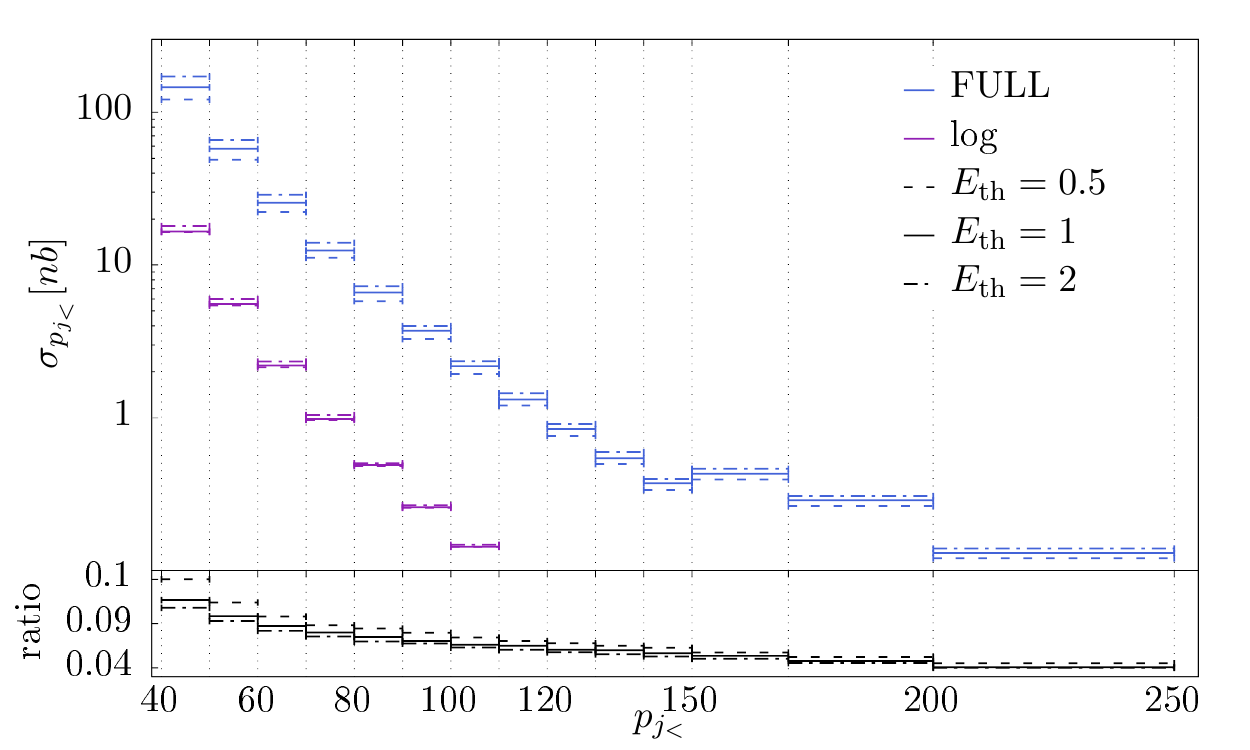}\hspace{10pt}
   \caption{NLO Mueller-Tang cross-section \(\sigma_{p_{j_{<}}}\) (blue) and contribution of the violating logs term (violet) in case of a dynamic gap definition. Different line types refer to three different choices for the transverse energy threshold (\(E_{{\rm th}}=\{1/2,1,2\}\) GeV). The ratio relative to the default choice \(E^*_{\rm th}=1\) GeV is shown in the bottom plot.
 }
\label{fig6.21}
 \end{figure}

\begin{figure}[ph]
  \centering
  \includegraphics[width=0.8\linewidth]{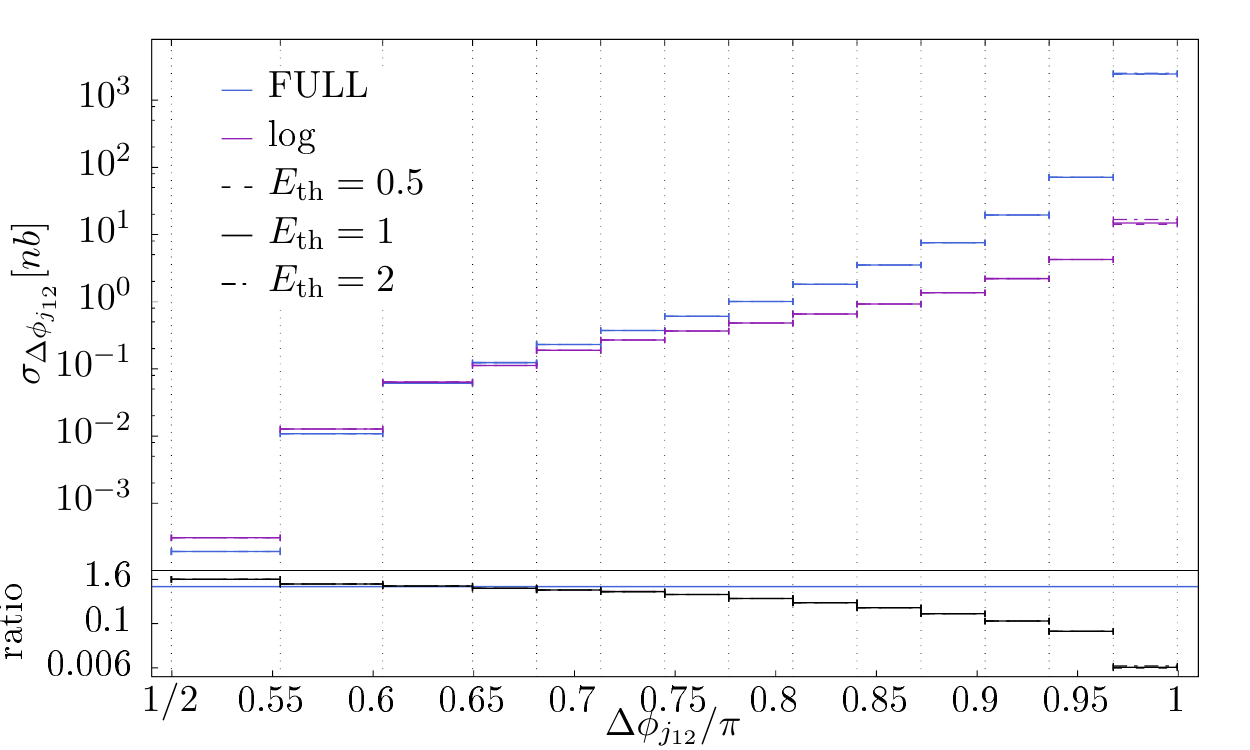}\hspace{10pt}
\caption{NLO Mueller-Tang cross-section \(\sigma_{\Delta\phi_{j_{12}}}\) (blue)  together with the contribute of the factorization breaking logs term (violet) using the dynamic gap definition. Different line types refer to three different choices for the transverse energy threshold (\(E_{{\rm th}}=\{1/2,1,2\}\) GeV) in the gap region. The ratio with respect to the default value \(E^*_{\rm th}=1\) GeV is shown in the bottom plot. With a dynamic gap, the importance of the violating ``\(\log s\)'' term is reduced. The validity of the BFKL hypothesis is extended to lower azimuthal angles  \(\Delta\phi\simeq \frac{3}{4}\pi\).
}
  \label{fig6.22}
\end{figure}

A solution to reduce the relative size of the factorization breaking term, that should also work at higher energies, can be found by modifying the gap definition from a central fixed gap size to a dynamical gap definition, where its size grows with the jet rapidity separation. A dynamical gap extending in the region
\[Y_{\rm gap^*}=\max\l(y_{j_2}-y_0,y_{\rm gap}\r)-\min\l(-y_{j_1}+y_0,-y_{\rm gap}\r),\]
where \(y_0=0.4\) works as a buffer zone between the jet edges and the beginning of the gap to avoid interfering with the jet collinear region. With such a definition, the gap size is wider than with the CMS definition, and the cross sections are reduced by about a factor of 2 as shown in Figs.~\ref{fig6.20}, \ref{fig6.21} and \ref{fig6.22}. The ``\(\log s\)'' contribution never exceeds 10\% of the total MT cross section and it is further reduced at higher rapidity differences. Similarly, throughout the whole azimuth angular range the impact of factorization breaking remains a small fraction of the total BFKL cross section as shown in Fig.~\ref{fig6.22}. The wider extension of the dynamical gap brings a stronger, but still weak, threshold dependence that is not confined to the low rapidity range.

\begin{figure}[ph]
  \centering
    \includegraphics[width=0.8\linewidth]{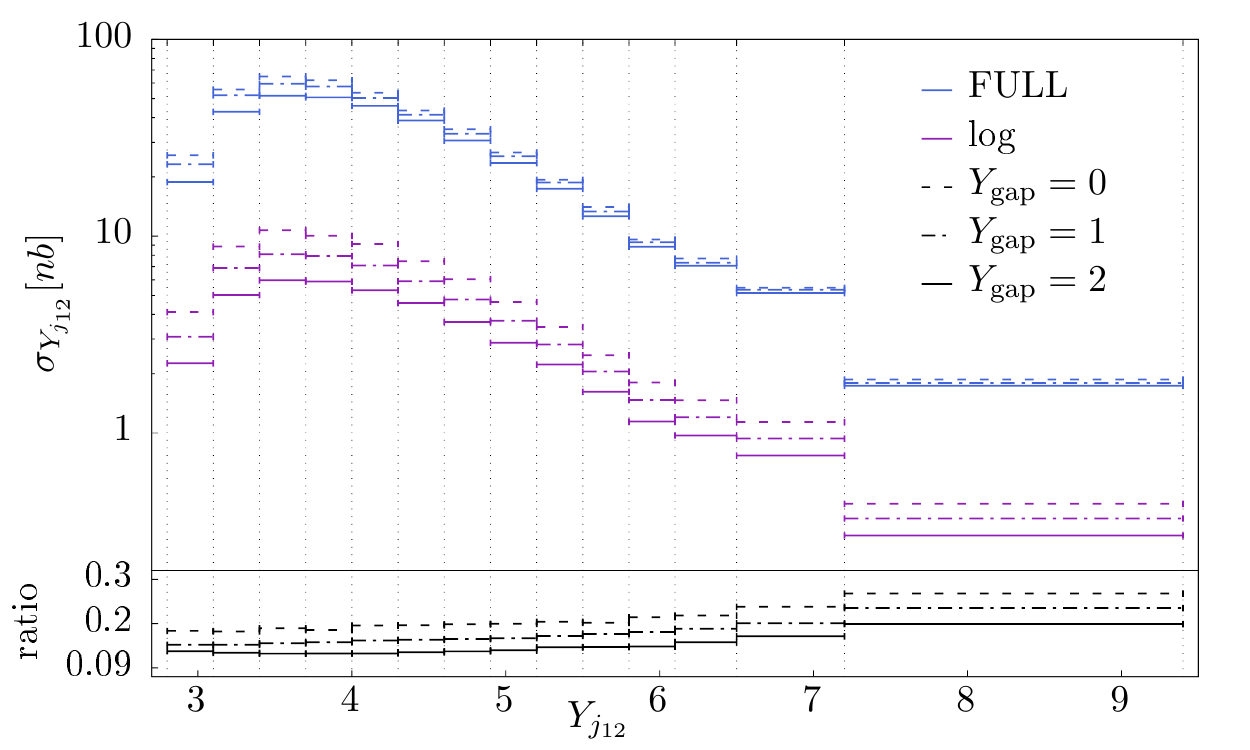}\hspace{10pt}

 \caption{NLO Mueller-Tang cross-section \(\sigma_{Y_{j_{12}}}\)  (blue)  and contribution of the factorization breaking term only in logs (violet).
   Different line types refer to three different choices for the size of the rapidity gap  (\(Y_{{\rm gap}}=\{0,1,2\}\)). The ratio relative to the default choice \(Y^*_{\rm gap}=2\) is shown in the bottom plot. We observe that the effect of the gap imposition is relatively stronger on the ``\(\log s\)'' term compared to the total as one would expect. The gap constraint reduces the size of the violation of the BFKL factorization by a factor of 2 with respect to the case of no gap.}
  \label{fig6.25}
\end{figure}

\begin{figure}[ph]
   \centering
   \includegraphics[width=0.8\linewidth]{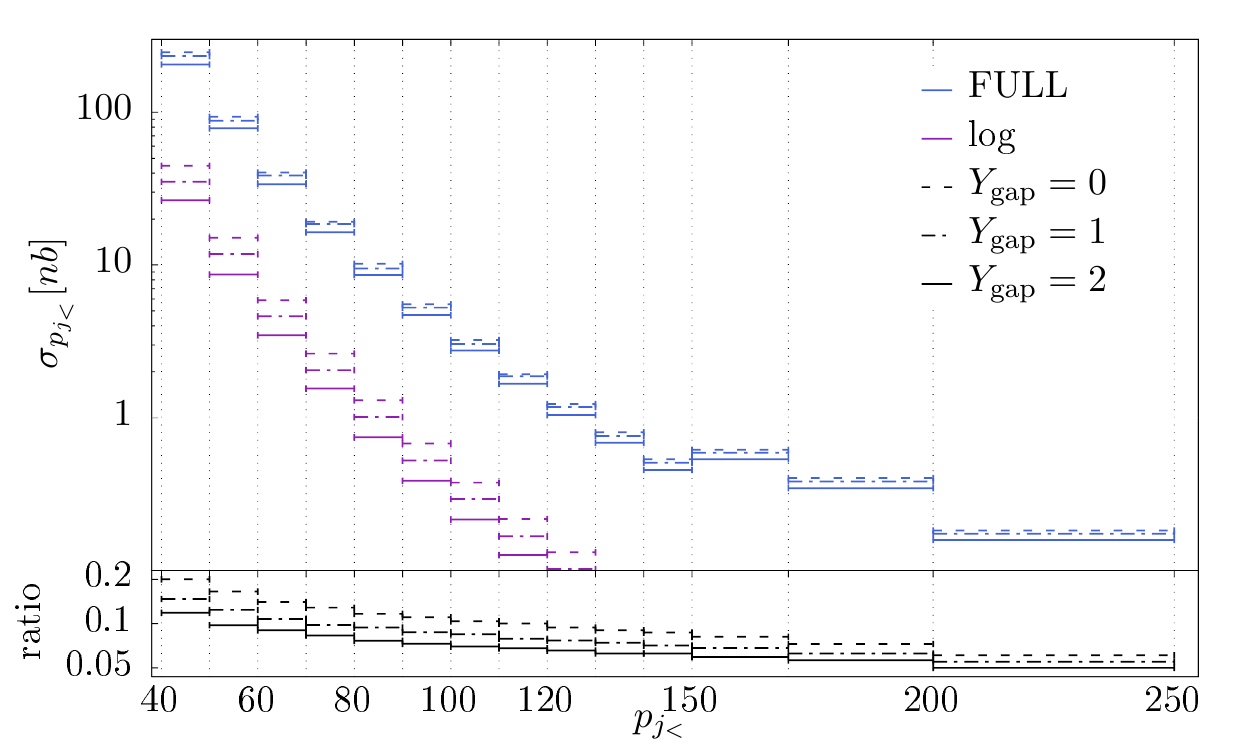}\hspace{10pt}

\caption{NLO Mueller-Tang cross-section \(\sigma_{p_{j_{<}}}\)  (blue)  and contribution of the factorization breaking term only in logs (violet).
 Different line types refer to three different choices for the size of the rapidity gap (\(Y_{{\rm gap}}=\{0,1,2\}\)). The ratio relative to the default choice \(Y^*_{\rm gap}=2\) is shown in the bottom plot.}
  \label{fig:MT-JETS-PTj2-ygap-log}
\end{figure}

\begin{figure}[ph]
  \centering
    \includegraphics[width=0.8\linewidth]{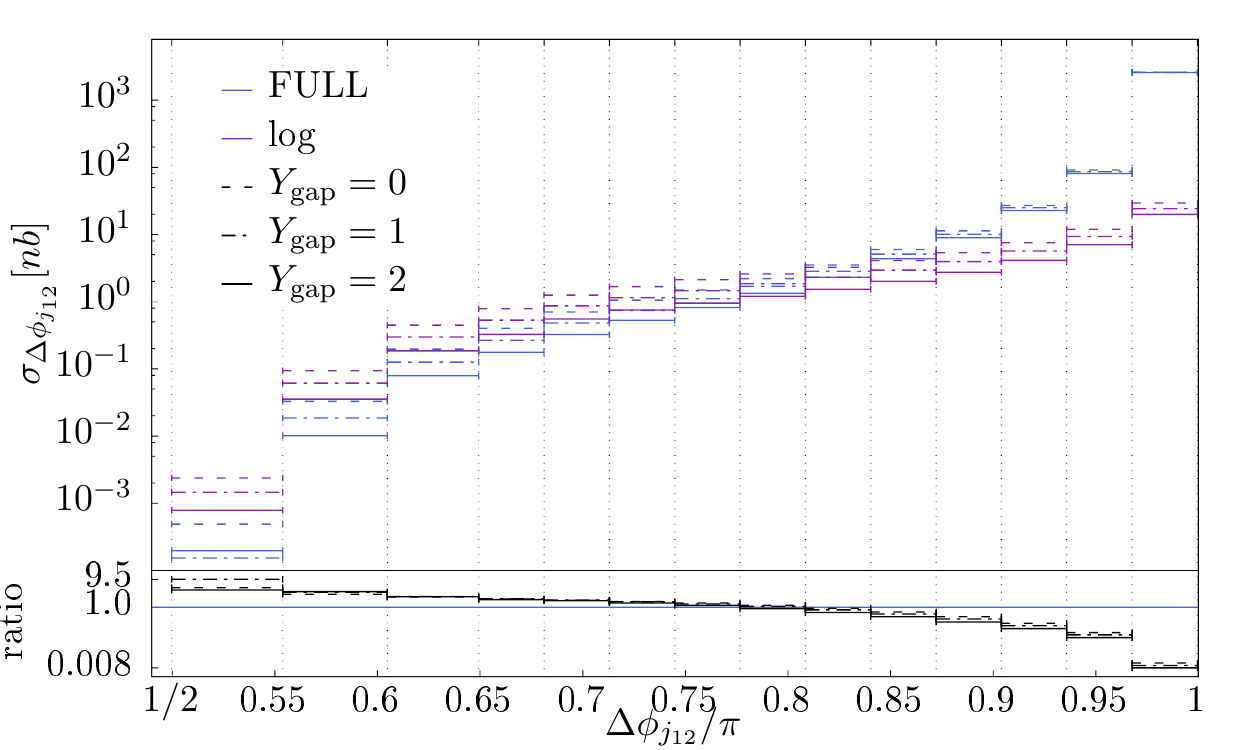}\hspace{10pt}

 \caption{NLO Mueller-Tang cross-section \(\sigma_{\Delta\phi_{j_{12}}}\)  (blue) and contribution of the factorization breaking term in logs (violet).
Different line types refer to three different choices for the gap rapidity size  (\(Y_{\rm gap}=\{0,1,2\}\)). the ratio relative to the default choice \(Y^*_{\rm gap}=2\) is shown in the bottom plot.}
  \label{fig6.27}
\end{figure}

Figs.~\ref{fig6.25}, \ref{fig:MT-JETS-PTj2-ygap-log} and \ref{fig6.27} are further illustrations of the gap size requirement on the ``\(\log s\)'' contribution. Requiring a larger gap size reduces the impact of the factorization breaking term by about a factor 2. The ``\(\log s\)'' contribution is sensitive to the gap definition since its impact is larger in the central rapidity region.

\subsection{Comparison between the MT and Mueller-Navelet cross sections}

\begin{figure}[ph]
  \centering
  \includegraphics[width=0.8\linewidth]{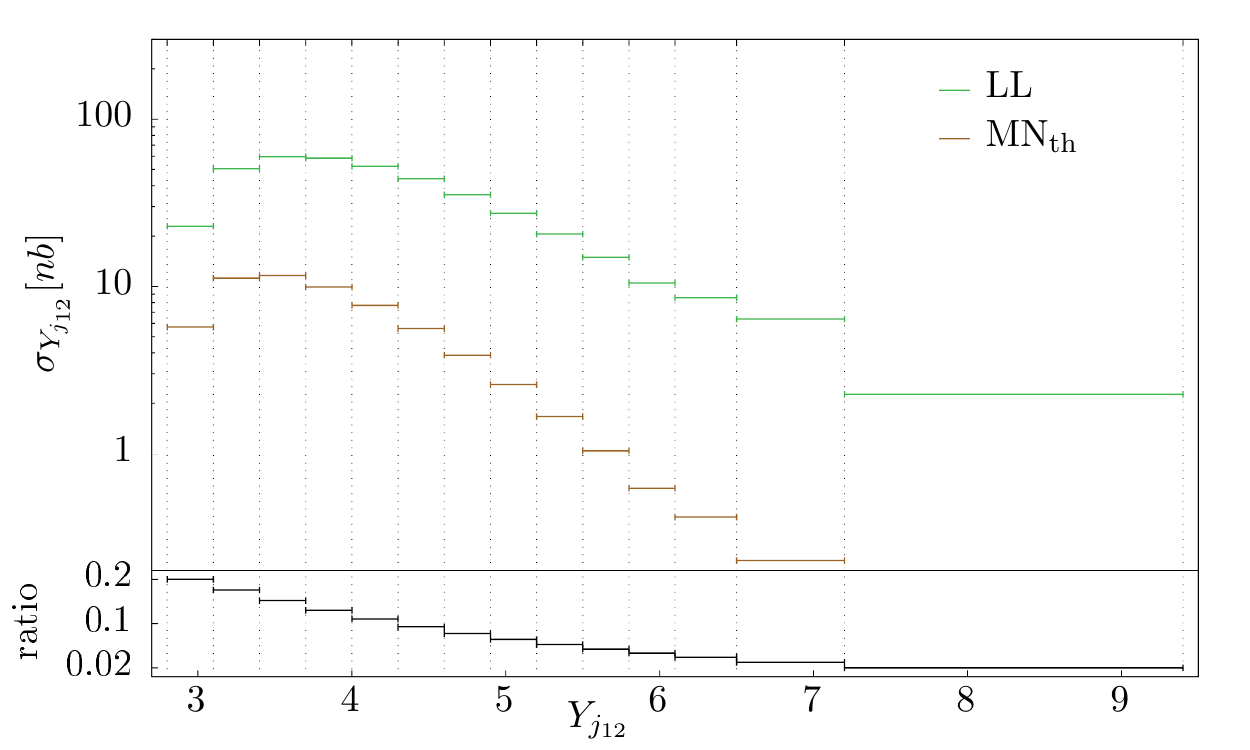}\hspace{10pt}

  \caption{Mueller-Tang cross-section \(\sigma_{dY_{j_{12}}}\)
 and Mueller-Navelet cross-section with energy threshold constraint, all at LL approximation and integrated over rapidity bins.
    On the top canvas are reported the absolute values while the bottom canvas shows the ratio FULL/MN\textsubscript{th}.
    The ratio decrease with rapidity passing from 20\% at \(Y_{j_{12}}\simeq 3-4\) to 2\% at \(Y_{j_{12}}\gtrsim 7\).}
  \label{fig6.28}
\end{figure}

\begin{figure}[ph]
  \centering
  \includegraphics[width=0.8\linewidth]{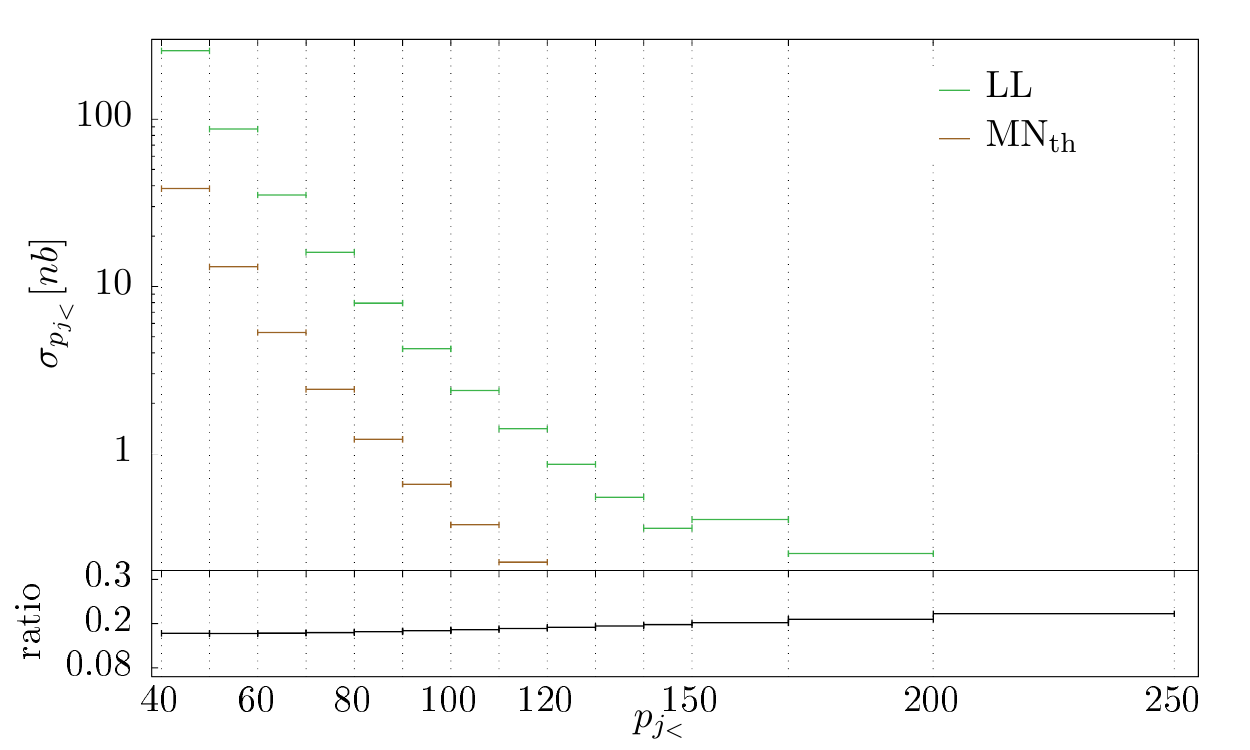}\hspace{10pt}
  \caption{Mueller-Tang cross-section \(\sigma_{p_{j_{<}}}\)
 and Mueller-Navelet cross-section with energy threshold constraint, all at LL approximation and integrated over rapidity bins.
    On the top canvas are reported the absolute values while the bottom canvas shows the ratio FULL/MN\textsubscript{th}.
    The ratio is nearly flat in \(p_{j_2}\) and stays consistently around 20\%.}
  \label{fig6.29}
\end{figure}

Finally, in Figs.~\ref{fig6.28} and \ref{fig6.29}, we compare the Mueller-Tang and Mueller-Navelet predictions at LL.
The same \(2\to 2\) scattering event resulting in two jets separated by a (large) rapidity distance \(Y_{j_{12}}\) can be interpreted as a MT process or as a MN scattering event if the intra-jet radiation can escape detection (for instance if it does not pass the calorimeter threshold).  As expected, the tendency of the MN radiation to increase in multiplicity as the rapidity range widens is reflected in a drop of the MN/MT ratio toward larger rapidities. At small rapidities \(Y_{j_{12}}\lesssim 4\) the MN\(_{\rm th}\) contamination can be sizable, of about 20\%. It is thus important to stress that the BFKL MT calculation is mainly valid at high rapidities for large gaps and large spearation intervals between jets.
We want to stress that the MN\(_{\rm th}\) vs MT comparison should be seen only as an estimate since it is only a LL approximation. Moreover, the central gap configuration differs greatly from the theoretical picture.


\


\section{Conclusions}
\label{sec:conclusions}

We performed a phenomenology analysis of the Mueller-Tang jet process taking into account the NLO corrections to the impact factors (IFs) which represent the new aspects of this work.
In particular, we provide the dijet cross-sections as a function of the rapidity difference (\(Y_{j_{12}}\)), the azimuthal-angular distance (\(\Delta\phi_{j_{12}}\)) and the transverse momentum of the ``second leading'' jet (\(p_{j_<}\)).

The corrections due to the newly included NLO IFs and due to the NLL corrections to the gluon-Green function (GGF) are similar in size but carry a different differential dependence on the observables. The NLO IFs dominate at large \(p_{j_<}\) and moderate \(Y_{j_{12}}\). They also break the elastic symmetry proper to all the other contributions inducing a non-trivial angular jet distribution. At FULL NL approximation, the cross-section  \(\sigma_{\Delta\phi_{j_{12}}}\) is strongly peaked towards the ``back-to-back'' configuration.
Similarly to other NLO BFKL estimate for other processes (see, e.g., \cite{Ducloue:2013bva,Caporale:2014gpa}), the predictions are rather sensitive to the choice of the physical scales. Such sensitivity is greatly reduced if the renormalization scale (\(\mu_R\)) is fixed applying the principle of minimal sensitivity, of which we performed an estimate. The stationary point of the transformation induced by the scale variation on the cross-section was found to be four times larger then the ``natural'' scale choice.

Besides the clear interest that resides in completing the NLO description, which was the prime motivator of the phenomenology analysis,
understanding the role played by the NLO IF corrections merit an interest on its own, in particular from the point of view of the foundation of the BFKL approach.
Unexpectedly, the presumed BFKL factorization which dictates that all BFKL-like \(\log|s/t|\) factors should fit into the well known resummation encapsulated into the GGF fails to hold. We observed that, the BFKL factorization is formally violated when the usual gap encompassing only the central rapidity is imposed. The real (colorful) emission, part of the NLO IF, breaks the color-singlet structure of the interaction, which leads to the unfactorizable \(\log{s}\) factors.

Despite for the enhancing effect of the spurious \(\log{s}\) factor, the violating term is small compared to the totalmat LHC energies. Actually, only for those events (the overwhelming majority) where the jets azimuthal-angle is close to the peak at \(\Delta\phi_{j_{12}}=\pi\) the effect of the violation is negligible. Below intermediate angle values the violation can even exceed the total cross-section.

We showed that a ``dynamical'' gap, widening up as the jet rapidity distance increases, further reduces the size of the violation. The effect it is not limited to current energies and extends the validity to smaller angles between the jets.
A rigorous reformulation of the BFKL structure to accommodate the violating terms is left for a future study.

In general, the sensitivity to the energy threshold (\(E_{\rm th}\)) capping the radiation in the gap region and the rapidity gap extension (\(Y_{\rm gap}\)) is weak.
This feature is expected since, in the BFKL construction, the interaction is
cast in terms of a color-singlet exchange. It follows that, the radiation seldom enters the central region. In most of the events the gap arises naturally from interaction dynamics with no need of being explicitly enforced.

Finally, we confirm that the MT color-singlet exchange process is favored over the potential competitors involving other color structures at LHC kinematics.
In particular, a LL estimate shows that colour-octet exchanges with no emission in the gap region are suppressed with respect to octect exchanges.

In conclusion, our results show that a phenomenological analysis of MT processes at the LHC is feasible at full NLO BFKL accuracy.
Our analysis represents the starting point for more refined predictions that will include soft effects and require the implementation inside a Monte Carlo.

\appendix

\section{Gluon Green-function}
\label{sec:ggf}

As noted by Lipatov~\cite{FadinFiore05:NonForwardKernelNLO}, the non-forward gluon Green function can be inferred from the solution of a modified BFKL equation.
The key to the solution is the realization that, with a slight modification, the equation can be diagonalized when projected over the space of conformally symmetric functions of the coordinate (Fourier conjugated to the momentum space).
The conformal BFKL equation coincides with the original one under the assumption of colorless colliding particles, which is valid for inclusive observables.
It is to be expected that the conformal solution must be mended when the interacting partons can be resolved out from the screening of the hadronic matter at large transferred momentum. In the case of partons, as shown by Mueller and Tang~\cite{MuellerTang92}, a non-analytic (divergent) term appears but, since it is an artifact of the technique used to extract the solution with unclear physical origin, it must be subtracted away\footnote{If instead the non-analytic terms are left there, the results is a rather bizarre null coupling between the pomeron and the quarks. It is to be demonstrated that the Mueller Tang prescription is also solution of the BFKL equation (see ref.~\cite{BartelsForshaw95} for a more extensive discussion).}.

In the complex impact parameter space \(\rho_i=x_i+iy_i\) Lipatov's conformal eigenfunctions read
\begin{equation}
  \label{eq:EigenFunction}
    E^{h\bar{h}}(\rho_i,\rho_i^*)=\left(\frac{\rho_1-\rho_2}{\rho_1\rho_2}\right)^h\left(\frac{\rho_1^*-\rho_2^*}{\rho_1^*\rho_2^*}\right)^{\bar{h}}
\end{equation}
where \(h=1/2+n/2+i\nu\), \(\bar{h}=1/2-n/2+i\nu\)
with
\(n\in\mathbb{Z}\), \(\nu\in\mathbb{R}\).
In the momentum-representation become
\begin{equation}
     \label{eq:EigenFunction2}
  \begin{split}
 \tilde{E}^{h\bar{h}}({\ell}_1,{\ell}_2)=&\int_{}^{}\frac{\dd^2\rho_1}{(2\pi)^2}\frac{\dd^2\rho_2}{(2\pi)^2}E^{h\bar{h}}(\rho_1,\rho_2)e^{i{\ell}_1\cdot\rho_1+i{\ell}_2\cdot\rho_2},\\
    =&\tilde{E}_A^{h\bar{h}}({\ell}_1,{\ell}_2)+\tilde{E}_{\delta}^{h\bar{h}}({\ell}_1,{\ell}_2),\ \ \ \vec{\ell}_1+\vec{\ell}_2=\vec{k}
  \end{split}
\end{equation}
where, the transverse vectors have been expressed in complex notation \(\vec{p}\to p_x+i\,p_y=p e^{i\phi}\).
The analytic \(\tilde{E}^{h\bar{h}}_A\) and the singular part \(\tilde{E}^{h\bar{h}}_{\delta}\) are given by
\begin{equation}
  \begin{split}
    \tilde{E}_A^{h\bar{h}}({\ell}_1,{\ell}_2)=&\frac{h\bar{h}(1\!-\!h)(1\!-\!\bar{h})\Gamma(1\!-\!h)\Gamma(1\!-\!\bar{h})}{i^n(4\pi)^2} \Big[\l(\frac{\ell_1}{2}\r)^{\bar{h}-2}\l(\frac{\ell_2^*}{2}\r)^{h-2}\!\;_2F_1\left(1\!-\!h,2\!-\!h,2;-\frac{\ell_1^*}{\ell_2^*}\right)\!\;\\
    &\times\quad _2F_1\left(1\!-\!\bar{h},2\!-\!\bar{h},2;-\frac{\ell_2}{\ell_1}\right)+(-1)^n\{1\rightarrow 2\}\Big],\\
    E_{\delta}^{h\bar{h}}({\ell}_1,{\ell}_2)=&\left[\delta^{(2)}({\ell}_1)+(-1)^n\delta^{(2)}({\ell}_2)\right]\frac{i^{n}}{4\pi}\left(\frac{k}{2}\right)^{\bar{h}-1}\left(\frac{k^*}{2}\right)^{h-1}\frac{\Gamma(1-\bar{h})}{\Gamma{(h)}}.
  \end{split}
\end{equation}
Following the Mueller-Tang prescription\cite{MuellerTang92}, only the analytic part is kept in the  non-forward gluon Green-function%
\footnote{In \cite{Motyka01:nonForwardBFKLconformalSpin} definition for the momentum space eigenfunction differs by a factor  \((2\pi)^2\) compared to eq. \ref{eq:EigenFunction2} }
\begin{equation}\label{eq:Gintegral}
  {G}\left(\frac{\hat{s}}{s_0},\vec{\ell},\vec{\ell}',\vec{k}\right)=
(2\pi)^2\sum_{n\in\mathbb{Z}}\int \dd\nu \left(\frac{\hat{s}}{s_0}\right)^{\omega(n,\nu)}\!\!R(n,\nu)\; {\tilde{E}_A}^{*\ n,\nu}({\ell},{k})\; \tilde{E}_A^{n,\nu}({\ell}',{k}),
\end{equation}
where $R(n,\nu)$ is a normalization factor given in eq.~\eqref{eq:Rmnu} and $\omega(n,\nu)$ is the LL BFKL eigenvalue of eq.~\eqref{eq:omegaLL}.

The GGF when both IF are at LO is remarkably simple \cite{RoyonChevallier09:GapsBetweenJets}
\begin{equation}   \label{eq:GGFavgBis}
\begin{split}
  {\mathcal{G}}(\vec{k}^2,\hat{s}/s_0)&=\int\dd^2\vec{\ell}\,\dd^2\vec{\ell}'\;G(\vec{\ell},\vec{\ell}',\vec{k},\hat{s}/{s_0})\\
  &=\frac{4}{\vec{k}^2}\sum_{m\in\mathbb{Z}}\int\dd\nu \left(\frac{\hat{s}}{s_0}\right)^{\omega(m,\nu)}R(m,\nu)
  \end{split}
\end{equation}
    whereas if only one IF is left at LO
\begin{equation}
  \label{eq:GGF}
  \begin{split}
    \overline{G}\left(\ell_1,\ell_2,\cfrac{\hat{s}}{s_0}\right)&=\int \dd^2\vec{\ell}'{G}\left(\vec{\ell}_1,\vec{\ell}',\vec{k},\cfrac{\hat{s}}{s_0}\right)\\
    &=\sum_{m\in\mathbb{Z}}\frac{(-1)^{m+1}}{8} \int\dd\nu \left(\frac{\hat{s}}{s_0}\right)^{\omega(m,\nu)}\frac{\nu^2+m^2}{\cosh(\pi\nu)}\left(\frac{k^*}{2}\right)^{-h}\left(\frac{k}{2}\right)^{-\bar{h}}\\
    &\qquad\qquad \times \bigg[{\left(\frac{\ell_1}{2}\right)}^{\bar{h}-2}\left(\frac{\ell_2^*}{2}\right)^{h-2}\!\!\phantom{.}_2F_1\left(1\!-\!h,2\!-\!h,2;-\frac{\ell^*_1}{\ell^*_2}\right)\!\!\phantom{.}\\
    &\qquad\qquad\qquad\times \phantom{.}_2F_1\left(1\!-\!\bar{h},2\!-\!\bar{h},2;-\frac{\ell_2}{\ell_1}\right)+\{1\rightarrow 2\}\bigg],
  \end{split}
\end{equation}
where again $\vec{\ell}_1+\vec{\ell}_2=\vec{k}.$

Thanks to the conformal symmetry, the dependence on the exchanged momentum \(k\) can be factorized out completely, reducing the integral to depend only from the complex radius \(R=r e^{i\delta}\), where \(r=|\ell_1/\ell_2|\) and    \(\delta=\phi_1-\phi_2=\mathrm{Arg}(\ell_1/\ell_2)\):
\begin{gather}
  \label{eq:GGF2}
  \begin{split}
    \abs{k}^{2}\overline{G}\left(\frac{\hat{s}}{s_0},R,k\right)=
    &\left(\frac{1}{2}\right)^{-4}\sum_{m\in\mathbb{Z}}\frac{(-1)^{m+1}}{8}\int_{-\infty}^{+\infty} \!\!\!\dd\nu \frac{\nu^2+m^2}{\cosh(\pi\nu)} \left(\frac{\hat{s}}{s_0}\right)^{\omega(m,\nu)}\\
    &\qquad \qquad \times \big(g^{(m,\nu)}(R)+g^{(m,\nu)}(1/R)\big),\\
  \end{split}
  \shortintertext{where}
  \begin{split}
    g^{(m,\nu)}(R)=&{\left(\frac{1}{1+R^{-1}}\right)}^{\bar{h}-2}{\left(\frac{1}{1+R^*}\right)}^{{h}-2}\!\!\phantom{.}\\
    &\qquad \qquad \times\, _2F_1\left(1\!-\!h,2\!-\!h,2;-R^*\right)\!\!\phantom{.}_2F_1\left(1\!-\!\bar{h},2\!-\!\bar{h},2;-R^{-1}\right).
  \end{split}
\end{gather}
The integrand is real and even in \(\nu\), and \(G(R)=G(R^{-1})=G(R^{*})\)
\footnote{By exploiting symmetry properties of the Gauss hypergeometric function (see eq. 15.3.3 of \cite{AbramovitzStegun65}), it is easy to show that
  \[
    g^{(m,\nu)}(R)=g^{(-m,-\nu)}(1/R);\ \ \ g^{(m,\nu)}(R)=g^{(-m,-\nu)}(R)
  \]
  The number of (costly) function evaluation of the  hypergeometric functions can be reduced to only two for each $R$ and \(\nu\) making use of the contiguous relations\cite{SeguraTemme:HypergeomContiguous, AbramovitzStegun65} which allow to recover all the conformal spin (\(m,0,\pm 1\dots \pm M\)) contributions from the direct computation of only two contiguous conformal spin (\(n=0,1\) for example).
  Nonetheless, extending the sum over large conformal spins has been quite problematic due to numerical rounding problems.
  Although, the sum over conformal spins converges monotonically for \(Re\left[g^{(m,\nu)}(R)\right]\) the sum grows dramatically with \(m\) for \(Im\left[g^{(m,\nu)}(R)\right]\) so that at each step a growing portion of the available precision bits is eaten up by the calculation of the uninteresting imaginary part (double precision already fails at \(m=4\)).
 A representation for the GGF where the real part is computed directly bypassing altogether the imaginary part would solve the problem for any \(m\).
However, as such representation in not available, we solved the problem by employing  \(Arb\)\cite{Johansson2017:Arb-arbitraryPrecisionArithmetic}, a ``C library for arbitrary-precision ball arithmetic'' and we increased the number of bits up to 2000 to reach \(\max(m)=150\).}
\begin{equation}
  \label{eq:GGFeq}
  \begin{split}
    \abs{k}^{2}\mathcal{G}\left(\frac{\hat{s}}{s_0},R,k\right)=&8\sum_{m\in\mathbb{Z}}(-1)^{m+1}\!\!\int_0^\infty \dd\nu \frac{\nu^2+m^2}{\cosh(\pi\nu)} \left(\frac{\hat{s}}{s_0}\right)^{\omega(m,\nu)}\!\!\!Re\left[g^{(m,\nu)}(R)\right]\\
  \end{split}
\end{equation}

\subsection{NLL eigenvalue function}

The forward BFKL eigenvalue at NLL reads\cite{FadinFiore05:NonForwardKernelNLO}%
\footnote{\[ \omega^{\mr{f-NLL}}(n,\nu):=\bar{\alpha}_s\chi(n,\nu)\]}
\begin{equation}\label{eq:NLLeigenvalue}
  \begin{split}
    \chi_1(n,\nu)&=
  \gamma_K^{(2)} \chi_0(n,\nu) + \frac{3}{2}
  \zeta(3)-\frac{\beta_2}{2}\chi_0^2(n,\nu)\\
 &+\frac{1}{4}\chi''_0(n,\nu) -\frac{1}{2}\left( \varPhi(\abs{n},\nu)+\varPhi(\abs{n},-\nu) \right)\\
&+\frac{\pi^2 \sinh{\pi \nu}}{8 \nu \cosh^2{\pi \nu}} \biggl\{
    -\delta_{n 0} \left[ 3+\left(1+\frac{N_F}{N_C^3} \right)
      \frac{11+12\nu^2}{16\left(1+\nu^2 \right)} \right]\\
    &\qquad\qquad\quad\qquad+\delta_{\abs{n} 2} \left(1+\frac{N_F}{N_C^3} \right)
    \frac{1+4\nu^2}{32\left(1+\nu^2 \right)} \biggr\}
  \end{split}
\end{equation}
where $\chi''_0(\nu)=\psi''\left(\frac{1}{2}+i\nu\right)+\psi''\left(\frac{1}{2}-i\nu\right).$ and
\begin{multline}\label{eq:Phidef}
  \varPhi(n,\nu)=\sum_{k=0}^\infty
  \frac{{(-1)}^{k+1}}{k+i\nu+\frac{1+n}{2}} \bigg\lbrace
  \psi'(k+n+1)-\psi'(k+1)+
  \\
  \left.  +{(-1)}^{k+1}\left( \beta'(k+n+1)+\beta'(k+1) \right) +
    \frac{\psi(k+1)-\psi(k+n+1)}{k+i\nu+\frac{1+n}{2}} \right\rbrace
  \, ,
\end{multline}
where
\begin{equation}\label{eq:defbetaf}
  \beta(z)=\frac{1}{2}\left(\psi\left(\frac{1+z}{2}\right)- \psi\left(\frac{z}{2}\right)\right)\, ,
\end{equation}
Finally, the two-loop QCD cusp anomalous dimension in the dimensional reduction scheme is
\begin{equation}\label{eq:cuspdef}
  \gamma_K^{(2)}=\frac{1}{3}\left(5 b+1\right)-\frac{\zeta(2)}{2}=\frac{1}{4}\left(\frac{67}{9}-\frac{10 N_F}{9 N_C} -2 \zeta(2) \right)\, .
\end{equation}

The collinearly improved NLL eigenvalue for arbitrary conformal spin can be found in appedix of ref.~\cite{RoyonMarquet07:MuellerNaveletCollinearScheme}. For convenience, the result for scheme (4) is reported verbatim
\begin{equation}
\label{eq:omegaNLL}
\begin{split}
  \chi^{\underline{4}}(n,\gamma,\omega) &=
\chi_{LL}(n,\g)-f(n,\g)
+[1-\bar\alpha A(n)]f(n+\omega+2\bar\alpha_s B(n),\g)\\
&+\bar\alpha_s\biggl\{\chi_1(n,\g)+A(n)f(n,\g)
+\lr{B(n)+\f{\chi_{LL}(n,\g)}2}\times\\
&\qquad\qquad\qquad\qquad\qquad\quad\left[\lr{\g+\f{n}2}^{-2}+\lr{1-\g+\f{n}2}^{-2}\right]\biggr\}
\end{split}
\end{equation}
with
\be
f(n,\g)=\f1{\g+\f{n}2}+\f1{1-\g+\f{n}2}\, , \quad \gamma=1/2-\nu\, .
\ee
In this scheme, $A(n)$ and $B(n)$ are given by:
\begin{gather}
A(n)=-d_1(n)-\f12\left[\psi'(n+1)-\psi'(1)+\f1{(n+1)^2}\right];
\quad B(n)=-d_2(n)+\f12[\psi(n+1)-\psi(1)]\, ,\\
\shortintertext{with}
\begin{split}
d_1(n)=&\f{1+5b}3-\f{\pi^2}8+b[\psi(n+1)-\psi(1)]+
\f18\left[\psi'\lr{\f{n+1}2}-\psi'\lr{\f{n+2}2}+4\psi'\lr{n+1}\right] \\
&\qquad\qquad-\lr{67+13\f{N_f}{N_c^3}}\f{\delta_{0n}}{36}
-\lr{1+\f{N_f}{N_c^3}}\f{47\delta_{2n}}{1800}
\end{split}
\shortintertext{and}
d_2(n)=-\f{b}2-\f12[\psi(n+1)-\psi(1)]
-\lr{11+2\f{N_f}{N_c^3}}\f{\delta_{0n}}{12}-\lr{1+\f{N_f}{N_c^3}}\f{\delta_{2n}}{60}\ .
\end{gather}

\section{NLO IFs}

\subsection{Singularity cancellation}
\label{sec:singularities}

In ref.~\cite{HentschinskiSabioVera14:QuarkImpactFactor} a phase-space splitting parameter \(\lambda\) was introduced to separate the collinear divergent configurations into \(\epsilon\)-poles in dimensional regularization and the correspondent finite reminders. The procedure is independent of the new parameter only in the limit \(\lambda\to 0\).
The \(\lambda\) value must be chosen carefully as a trade-off between accuracy of the solution and swift numerical convergence which is increasingly hampered approaching that limit.

On a separate note, the suppression of the \(1/\Delta\)(\({\Delta}={k}-z{q}\)) divergence relies on vanishing  of the integral of the quark-quark splitting function \(\int\!\!\dd z P^{(+)}_{qq}(z)=0\). When dealing with highly dimensional numerical integrations it is preferable to work with expressions where singular configurations are explicitly removed at the level of the integrand as opposed to let the integrator figure out that the singularity is integrable.

These two drawbacks can be improved upon by extracting the collinear singularities as done in ref.~\cite{BartelsColferai02:MuellerNaveletQuarkImpactFactor}.

\subsubsection{Color factor $C^{2}_{f}$}
Let us examine the \(C_f^2\) term and apply the modified subtraction prescription to extract its singularities.
The bold notation for transverse vectors is dropped here. The vectorial nature of \(k,q,p,\Delta\) is understood.
Starting from eq.~(48) of ref.~\cite{HentschinskiSabioVera14:QuarkImpactFactor}:
\begin{gather}
  \label{eq:all}
  \biggl(\frac{\dd V_r^{(1)}}{\dd J}\biggr)_{C^2_f}=C^2_fH^\alpha_q\int\!\! \dd z P^{(\epsilon)}_{gq}(z) \int\!\!\dd_\epsilon q\frac{z^2k^2}{q^2\Delta^2}S^{3}_J(p,q,zx;x),\\
  \shortintertext{where}
    \begin{aligned}
      {H^{\alpha_s}_q=C_f^2H^{\alpha_s},}&\quad H^{\alpha_s}=\frac{\alpha_{s,\epsilon}}{2\pi}h^{(0)},\\ {h^{(0)}=\frac{\alpha^2_{s,\epsilon}2^\epsilon}{\mu^{4\epsilon}\Gamma^2(1-\epsilon)(N^2_c-1)},}&\quad  \alpha_{s,\epsilon}=\frac{g^2\mu^{2\epsilon}\Gamma(1-\epsilon)}{(4\pi)^{1+\epsilon}},\\
      \dd_\epsilon q =\frac{\dd^{2+2\epsilon}q}{\pi_\epsilon},&\quad  \pi_\epsilon=\pi^{1+\epsilon}\Gamma(1-\epsilon)\mu^{2\epsilon}\\
    \end{aligned}
  \intertext{and the pole of the quark-gluon splitting function is extracted as}
  P^{(\epsilon)}_{g\leftarrow q}(z)=C_f\frac{\mathcal{P}_{gq}^{(\epsilon)}(z)}{z}.
\end{gather}
 Clearly, there is a collinear divergence when the gluon is emitted along the beam axis \(\vec{q}\to 0\).

The other denominator \(\vec{\Delta}^2\) mixes transverse and longitudinal coordinates and its contribution is better examined with the variable change \(q\to zq\). The soft limit is reached sending \(z\to 0\).
\begin{equation}
  \begin{split}
  \left(\frac{\dd V_r^{(1)}}{\dd J}\right)_{C^2_f}\!\!\!=&H^\alpha_q C_f^3\!\int\!\! \dd z \frac{\mathcal{P}_{gq}^{(\epsilon)}(z)}{z^{1-2\epsilon}}\!\int\!\!\dd_\epsilon q \frac{k^2}{q^2\!+\!(k\!-\!q)^2}\left(\frac{1}{(q-k)^2}+\frac{1}{q^2}\right) S^{3}_J(k\!-\!zq,zq,zx;x).
  \end{split}
\end{equation}
Due to the intertwining of collinear and soft configurations, it is convenient to apply a method for the singularity cancellation split in two phases:
First, the soft configuration \(z\to 0\) is integrated in \(\epsilon\)-regularization and extracted subtracting it from the reminder. Secondly, starting from the reminder that is now free of soft divergences, a similar procedure is applied once more to remove the initial \(q\to 0\) and final \(p\to 0\) state collinear singularities. The last step introduces a phase space slicing parameter \(\lambda\) for each collinear pole. Finally, what is left is the finite reminder \(\mathcal{I}_f\) where the singular configurations are explicitly subtracted away.
The contributions in 4-\(\epsilon\)-dimensions of soft, initial state and final state collinear poles are indicated respectively as \({\mathcal{I}_s}_{\{z\to 0\}},
{\mathcal{I}_c^{\rm in.}}_{\{q\to 0\}},{\mathcal{I}_c^{\rm fin.}}_{\{p\to 0\}}\).
\begin{gather}
  \left(\frac{\dd V_r^{(1)}}{\dd J}\right)_{C_f^2}=H^\alpha_qC^3_f \left[{\mathcal{I}_s}_{\{z\to 0\}}(\epsilon,k)+{\mathcal{I}_c}_{\{q\to 0,p\to 0\}}(\epsilon,k)+{\mathcal{I}_f}(k)\right],\\
  \begin{split}
    \mathcal{I}^{\rm fin.}_s=& \int \frac{\dd z}{z^{1-2\epsilon}} \int\frac{\dd_\epsilon q}{(k-q)^2}I(0,q),\\
    \mathcal{I}^{\rm fin.}_c=& \int \frac{\dd z}{z^{1-2\epsilon}} \int\frac{\dd_\epsilon q}{p^2}\left[I(z,k)-I(0,k)\right]\theta(\lambda^2-z^2p^2),\\
    \mathcal{I}^{\rm fin.}_f=& \int \frac{\dd z}{z} \int\frac{\dd^2q}{\pi p^2}\left[I(z,q)-I(0,q)-(I(z,k)-I(0,k))\theta(\lambda^2-z^2p^2)\right],\\
    \mathcal{I}^{\rm in.}_s=& \int \frac{\dd z}{z^{1-2\epsilon}} \int\frac{\dd_\epsilon q}{q^2}I(0,q),\\
    \mathcal{I}^{\rm in.}_c=& \int \frac{\dd z}{z^{1-2\epsilon}} \int\frac{\dd_\epsilon q}{q^2}\left[I(z,0)-I(0,0)\right]\theta(\lambda^2-z^2q^2),\\
    \mathcal{I}^{\rm in.}_f=& \int \frac{\dd z}{z} \int\frac{\dd q}{\pi q^2}\left[I(z,q)-I(0,q)-(I(z,0)-I(0,0))\theta(\lambda^2-z^2q^2)\right],\\
  \end{split}
  \shortintertext{where}
I(z,q)=\mathcal{P}_{gq}^{(\epsilon)}(z)\frac{k^2}{q^2+p^2}S^{(3)}_J(k-zq,zq,zx;x).
\end{gather}
The singularities are extracted as poles in \(\epsilon\)%
\footnote{The transverse integrals can be reduced to the form
  \begin{equation}
    \label{GammaInt}
    \begin{split}
      I_1(\epsilon,k/\mu)=&\int_{}^{}\frac{d^{2+2\epsilon}q}{\mu^{2\epsilon}\pi^{1+\epsilon}}\frac{{k}^2}{{q}^2(k-q)^2}= \frac{\Gamma(\epsilon)\Gamma(1-\epsilon)}{\Gamma(2\epsilon)}\left(\frac{{k}^2}{\mu^2}\right)^\epsilon=
      \left(\frac{k^2}{\mu^2}\right)^\epsilon\left(\frac{2}{\epsilon}-\frac{\pi^2}{3}\epsilon+o(\epsilon^2)\right),\\
     I_2(\epsilon,\lambda/\mu)=&\int_{}^{}\frac{d^{2+2\epsilon}q}{q^2\mu^{2\epsilon}\pi^{1+\epsilon}}\theta(\lambda^2-q^2)=
     \frac{1}{\epsilon\Gamma(1+\epsilon)\Gamma(1-\epsilon)}\left(\frac{{\lambda}^2}{\mu^2}\right)^\epsilon=
     \left(\frac{1}{\epsilon}+o(\epsilon^2)\right)\left(\frac{\lambda^2}{\mu^2}\right)^\epsilon.
   \end{split}
 \end{equation}}.

\begin{equation}
  \begin{split}
  \mathcal{I}^{\rm fin.}_s=&\frac{\mathcal{P}_{gq}(0)}{2\epsilon}I_1(\epsilon,\mu)S^{(2)}_J(k,x)=\frac{1}{\epsilon}\left(\frac{k^2}{\mu^2}\right)^\epsilon\left[\frac{1}{\epsilon}-\frac{\pi^2}{6}\epsilon+o(\epsilon^2)\right]S^{(2)}_J(k,x) \\
  \mathcal{I}^{\rm fin.}_c=& \!\int\!\! \frac{\dd z}{z^{1-2\epsilon}}I_2(\epsilon,\lambda/z\mu)\left[\mathcal{P}_{gq}^{(\epsilon)}(z)-\mathcal{P}_{gq}(0)\right]S^{(2)}(k,x)\\
  =&\frac{1}{\epsilon}\left(\frac{\lambda^2}{\mu^2}\right)^\epsilon\left[-\frac{3}{2}+\frac{\epsilon}{2}+ o(\epsilon^2)\right]S^{(2)}_J(k,x) \\
  \mathcal{I}^{\rm in.}_s=& \mathcal{I}^{\rm fin.}_s\\
  \mathcal{I}^{\rm in.}_c=& \!\int\!\! \frac{\dd z}{z^{1-2\epsilon}}I_2(\epsilon,\lambda/z\mu)\left[\mathcal{P}_{gq}^{(\epsilon)}(z)S^{(2)}_J(k,(1-z)x)-\mathcal{P}_{gq}(0)S^{(2)}(k,x)\right]\\
  =& \frac{1}{\epsilon}\left(\frac{\lambda^2}{\mu^2}\right)^\epsilon\bigg[\!\int\!\! \dd z\left(\frac{\mathcal{P}_{gq}(1-z)}{1-z}\right)_+S^{(2)}_J(k,zx)\\
  &\qquad\qquad\qquad +\epsilon \!\int\!\! \dd z(1-z)S^{(2)}_J(k,zx)-\frac{3}{2}S^{(2)}_J(k,x)\bigg],\\
  \mathcal{I}^{\rm fin.}_f=&\!\int\!\! \frac{\dd z}{z}\!\int\!\!\frac{\dd^2q}{\Delta^2}\biggl[\frac{k^2}{q^2+\Delta^2}\left(\mathcal{P}_{gq}(z)S^{(3)}_J(p,q,zx;x)-\mathcal{P}_{gq}(0)S^{(2)}_J(k,x)\right)\\
  &\qquad\qquad\qquad-\left(\mathcal{P}_{gq}(z)-\mathcal{P}_{gq}(0)\right)\theta(\lambda^2-\Delta^2)S^{(2)}_J(k,x)\biggr],\\
  \mathcal{I}^{\rm in.}_f=& \!\int\!\! \frac{\dd z}{z}\!\int\!\!\frac{\dd^2q}{q^2}\biggl[\frac{k^2}{q^2+\Delta^2}\left(\mathcal{P}_{gq}(z)S^{(3)}_J(p,q,zx;x)-\mathcal{P}_{gq}(0)S^{(2)}_J(q,x)\right)
  -\\
  &\qquad\qquad\qquad\theta(\lambda^2-q^2)\left(\mathcal{P}_{gq}(z)S^{(2)}_J(k,(1-z)x)-\mathcal{P}_{gq}(0)S^{(2)}_J(k,x)\right)\biggr].\\
  \end{split}
\end{equation}
An identical procedure is applied to the corresponding analogous terms in the gluon-induced IF.

\subsubsection{Color factor $C^2_a$}
Let us examine the term \(\sim C^2_a\) giving rise to the \(\log\!\!\!\ s\) factor.

The \({p}\to 0\) initial state collinear singularity (\(\lim_{\vec{p}\to 0}J_{2}\sim 1/p^2\)), corresponding to a quark emission collinear to its incoming parent, is canceled by the PDF counter term.
\begin{gather}\label{eq:J2pDiverg}
  \left(\frac{\dd V_r^{(1)}}{\dd J}\right)_{C^2_a}\!\!\!\!=H^\alpha_g \left[{\mathcal{I}_c}_{\{p\to 0\}}(\epsilon,k)+\mathcal{I}_f(q)\right],\\ \shortintertext{where}
  \begin{split}
  \mathcal{I}_c=&\frac{1}{1+\epsilon}\int \frac{\dd z}{z^{2\epsilon}}P_{gq}(z,\epsilon)\int \dd_\epsilon q\frac{\theta(\lambda^2-p^2)}{p^2}S^{(2)}_J(k,zx)\\
  =&\frac{1}{\epsilon}\left(\frac{\lambda^2}{\mu^2}\right)^\epsilon\int \dd z P_{gq}(z)S^{(2)}_J(k,zx)+ 2C_f\int \frac{\dd z}{z} (z-1) S^{(2)}_J(k,zx),\\
  \mathcal{I}_f
=&\int\dd zP_{gq}(z)\int \frac{\dd^2 q}{\pi}\bigg(J_{2}(k,q,\ell_2,\ell_2)S^{(3)}_J(p,q,zx;x)\\
  &\qquad\qquad\quad\qquad\qquad -\frac{\theta(\lambda^2-p^2)}{p^2}S^{(2)}_J(k,zx)\bigg)
  \end{split}
\end{gather}
where \(H^\alpha_g=h^{(0)}_g\frac{\alpha_s}{2\pi}\), \(h^{(0)}_g=C^2_ah^{(0)}(1+\epsilon)\).

\subsection{NLO IF final expressions}
\label{sec:NLO-IF-final}

In this section, we report explicitly all the expressions for the NLO IF employed in the analysis. They are equivalent to those of ref.~\cite{HentschinskiSabioVera14:GluonImpactFactor, HentschinskiSabioVera14:QuarkImpactFactor} besides for a few typos that we corrected there: the rescaling \(\vec{q}\to\vec{q}/(1-z)\) should result in an additional factor \((1-z)^2\) in the numerator of eq.~(78) and, subsequently, on the 7th line of eq.~(89); the angle in eq.~(50) has the wrong sign. It is correct in~\cite{Fadin99:QuarkImpactFactor}; in eq.~(41) the signs of the \(\propto C_a\) are incorrect but they get corrected later in eq.~(43);

\subsubsection{Quark-Induced}
\label{sec:nlo-ifs}

\(\frac{\dd V}{\dd J}_\mr{f.r.}\) refer to all the residual finite reminders once all the poles associated to the divergences are cancelled between real and virtual corrections as well as all the counter terms.

\begin{equation}
  \frac{\dd V^{(1)}_q(P_1,\ell_1,\ell_2;J_1;s_0)}{\dd J_1}=H^{\alpha}_q\frac{\dd V^{(1)}_{\mr{v}}}{\dd J_1}+H^{\alpha}\frac{\dd V^{(1)}_{\mr{f.r.}}}{\dd J_1}+H^{\alpha}\frac{\dd V^{(1)}_{\mr{r}}}{\dd J_1} \, .
\end{equation}

The first term has no integration
\begin{equation}
\begin{split}
  \frac{\dd V^{(1)}_{\mr{v}}}{\dd J_1}&=
C_f\l(\frac{3}{2}\ln\frac{{k}^2}{\mu^2_R}+\frac{\pi^2}{6}-4\r)+\frac{\beta_0}{4}\left[\frac{10}{3}-\ln\l(\frac{{\ell}^2_1}{\mu^2_R}\r)-\ln\l(\frac{({\ell}_1-{k})^2}{\mu^2_R}\r)\r]\\
&+\frac{C_a}{2}\Biggl[\frac{3}{2}\frac{\l({\ell}^2_1-({\ell}_1-{k})^2\r)}{{k}^2}\ln\l(\frac{({\ell}_1-{k})^2}{{\ell}^2_1}\r)-6\frac{\abs{{\ell}_1}\abs{{\ell}_1-{k}}}{{k}^2}\phi_1\sin\phi_1\\
    &\quad\qquad-\frac{3}{2}\l(\ln\l(\frac{{\ell}^2_1}{{k}^2}\r)+\ln\l(\frac{({\ell}_1-{k})^2}{{k}^2}\r)\r)-\ln\l(\frac{{\ell}^2_1}{{k}^2}\r)\ln\l(\frac{({\ell}_1-{k})^2}{s_0}\r)\\
&\quad\qquad-\ln\l(\frac{({\ell}_1-{k})^2}{{k}^2}\r)\ln\l(\frac{{\ell}^2_1}{s_0}\r)
-2\phi^2_1+\pi^2+\frac{7}{3}\Biggr]+\{1\to 2\}
\end{split}
\end{equation}
with \(\beta_0=\smfrac{11}{3}N_c-\smfrac{2}{3}n_f\) and  $\phi_i=\angle(\vec{\ell}_i,\vec{k}-\vec{\ell}_i) \equiv
\arccos\big(\frac{(\vec{k}-\vec{\ell}_i)^2-\vec{k}^2-\vec{\ell}_i}{2\abs{\vec{\ell}_i}\abs{\vec{\ell}_i-\vec{k}}}\big)$ for  $i=1,2$. The second term involves a single one dimensional integral
\begin{equation}
\begin{split}
  \frac{\dd V^{(1)}_{\mr{f.r.}}}{\dd J_1}&=C^3_f\left(\frac{3}{2}\ln(\mu^2_R/\lambda^2)+\frac{1}{2}+\frac{3}{2}\ln(\mu^2_R/\lambda^2)-\frac{\pi^2}{3}\right)S^{(2)}_J(k,x)\\
  &\qquad+\int\!\!\dd z S^{(2)}_J(k,zx)\biggl[C^2_f\l(\ln(\lambda^2/\mu^2_F)P_{qq}(z)+C_f(1-z)\r)\\
  &\qquad \qquad\quad+C^2_a\l(\ln(\lambda^2/\mu^2_F)P_{gq}(z)+2C_f\frac{z-1}{z}\r)\biggr]\, .
\end{split}
\end{equation}
The final term involves a one dimensional integral and a two dimensional momentum integral.
\begin{equation}
\begin{split}
  \frac{\dd V^{(1)}_{\mr{r}}}{\dd J_1}&= C^2_f\!\int\!\frac{\dd z}{z} \!\int\!\frac{\dd^2q}{\pi} \biggl\{
\frac{1}{\Delta^2} \biggl[\frac{k^2}{q^2+\Delta^2}\left(\mathcal{P}_{gq}(z)S^{(3)}_J(p,q,zx;x)-\mathcal{P}_{gq}(0)S^{(2)}_J(k,x)\right)\\
  &\quad\qquad\qquad\qquad\qquad-\theta(\lambda^2-\Delta^2)\left(\mathcal{P}_{gq}(z)-\mathcal{P}_{gq}(0)\right)S^{(2)}_J(k,x)\biggr]\\
   &\qquad\qquad+\frac{1}{q^2}\biggl[\frac{k^2}{q^2+\Delta^2}\left(\mathcal{P}_{gq}(z)S^{(3)}_J(p,q,zx;x)-\mathcal{P}_{gq}(0)S^{(2)}_J(q,x)\right)\\
   &\quad\qquad\qquad\qquad\qquad-\theta(\lambda^2-q^2)\left(\mathcal{P}_{gq}(z)S^{(2)}_J(k,(1-z)x)-\mathcal{P}_{gq}(0)S^{(2)}_J(k,x)\right)\biggr]\\
   &\qquad\qquad+C_aC_f\l(J_1(k,q,\ell_1,z)+J_1(k,q,\ell_2,z)\r){P_{gq}(z)S^{(3)}_J(p,q,zx;x)}\\
&\qquad\qquad+C^2_a\biggl(J_2(k,q,\ell_1,\ell_2)\underline{S^{(3)}_J(p,q,zx;x)}\\
&\qquad\qquad\qquad\qquad\qquad-\frac{\theta(\lambda^2-p^2)}{p^2}S^{(2)}_J(k,zx)\biggr)P_{gq}(z)\biggr\}
\end{split}
\end{equation}
where
  \begin{align}
  \label{eq:Jcf2}
 J_{1} (\vec{q}, \vec{k}, \vec{\ell}_i, z)   & = \frac{1}{4}
\bigg[
 2 \frac{\vec{k}^2}{\vec{p}^2}
\bigg(\frac{(1-z)^2}{\vec{\Delta}^2} - \frac{1}{\vec{q}^2} \bigg)
-
\frac{1}{\vec{\Sigma}_i^2}
\bigg(
\frac{(\vec{\ell}_i - z \vec{k})^2}{\vec{\Delta}^2} -
\frac{\vec{\ell}_i^2}{\vec{q}^2}
\bigg)
\notag \\
& \qquad \qquad \qquad \qquad
-
\frac{1}{\vec{\Upsilon}_i^2}
\bigg(
  \frac{(\vec{\ell}_i - (1-z) \vec{k})^2}{\vec{\Delta}^2}
-
\frac{(\vec{\ell}_i - \vec{k})^2}{\vec{q}^2}
\bigg)
\bigg];
\notag \\
 J_{2} (\vec{q}, \vec{k}, \vec{\ell}_1, \vec{\ell}_2)  &=
\frac{1}{4} \bigg[
\frac{\vec{\ell}_1^2}{ \vec{p}^2 \vec{\Upsilon}^2_1}
+
\frac{( \vec{k} - \vec{\ell}_1)^2}{ \vec{p}^2 \vec{\Sigma}^2_1}
 +
\frac{\vec{\ell}_2^2}{ \vec{p}^2 \vec{\Upsilon}^2_2}
+
\frac{( \vec{k} - \vec{\ell}_2)^2}{ \vec{p}^2 \vec{\Sigma}^2_2}
\notag \\
&
- \frac{1}{2}
\bigg(
\frac{(\vec{\ell}_1 - \vec{\ell}_2)^2}{\vec{\Sigma}_1^2 \vec{\Sigma}_2^2}
+
\frac{(\vec{k} - \vec{\ell}_1 - \vec{\ell}_2)^2}{ \vec{\Upsilon}_1^2 \vec{\Sigma}_2^2   }
+
\frac{(\vec{k} - \vec{\ell}_1 - \vec{\ell}_2)^2}{  \vec{\Sigma}_1^2 \vec{\Upsilon}_2^2  }
+
\frac{(\vec{\ell}_1 - \vec{\ell}_2)^2}{\vec{\Upsilon}_1^2 \vec{\Upsilon}_2^2}
\bigg)
 \bigg].
\end{align}
and
\begin{align}
 \vec{\Delta} & = {\bm q} - z \vec{k}, &
  \vec{\Sigma}_i & = \vec{q} - \vec{\ell}_i,  & \vec{\Upsilon}_i & = \vec{q} - \vec{k} + \vec{\ell}_i & i &= 1,2.
\end{align}

\subsection{Gluon-Induced}
When is a gluon to be scattered in the proton two emission channels contribute \(g\to gg\) and \(g\to q\bar{q}\). The finite reminder includes all the pieces resulting from the cancellation of divergences between real and virtual corrections as well as all the counter terms.

\begin{equation}
  \frac{\dd V^{(1)}_g(P_{1,2},\ell_1,\ell_2;s_0)}{\dd J_1}=H^{\alpha_s}_g\frac{\dd V^{(1)}_{\mr{v}}}{\dd J_1}+H^{\alpha_s}\frac{\dd V^{(1)}_{\mr{f.r.}}}{\dd J_1}+H^{\alpha_s}\frac{\dd V_{\mr{r}}^{(1)}{}_{g\to q\bar{q}}}{\dd J_1}+H^{\alpha_s}_g\frac{\dd V_{\mr{r}}^{(1)}{}_{g\to g{g}}}{\dd J_1}
\end{equation}
The first term is again only a function of momenta
\begin{equation}
    \begin{split}
      \frac{\dd V^{(1)}_{\mr{virt.}}}{\dd J_1}&=\frac{\beta_0}{4}+C_a\l(\frac{2}{3}\pi^2-\frac{3}{4}\r)
  -\l[\frac{\beta_0}{2}+\frac{n_f}{6}\l(1+\frac{1}{C^2_a}\r)\r]\l(\ln\frac{{\ell}^2_1}{{k}^2}+\ln\frac{({k}-{\ell}_1)^2}{{k}^2}\r)\\
  &+\frac{C_a}{2}\l[\ln\frac{{k}^2}{{\ell}^2_1}\ln\frac{{\ell}^2_1}{s_0}+\ln\frac{{k}^2}{({k}-{\ell}_1)^2}\ln\frac{({k}-{\ell}_1)^2}{s_0}
  +\ln^2\frac{{\ell}_1}{({k}-{\ell}_1)^2}\r]\\
  &-\l[\frac{n_f}{6}\l(1+\frac{1}{C^2_a}\r)+\frac{\beta_0}{4}\r]\frac{({\ell}^2_1-({k}-{\ell}_1)^2)}{{k}^2}\ln\frac{{\ell}^2_1}{({k}-{\ell}_1)^2}\\
  &-\l(\frac{n_f}{C^2_a}+4C_a\r)\frac{({\ell}^2_1({k}-{\ell}_1)^2)^{1/2}}{{k}^2}\phi_1\sin\phi_1\\
  &+\frac{1}{3}\l(C_a+\frac{n_f}{C^2_a}\r)\bigg[8\frac{({\ell}^2_1({k}-{\ell}_1)^2)^{3/2}}{(k^2)^3}\phi_1\sin^3\phi_1\\
  &\qquad-2\frac{{\ell}^2_1({k}-{\ell}_1)^2}{({k}^2)^2}\l(2-\frac{({\ell}^2_1-({k}-{\ell}_1)^2)}{{k}^2}\ln\frac{{\ell}^2_1}{({k}-{\ell}_1)^2}\r)\sin^2\phi_1\\
&\qquad+\frac{({\ell}^2_1({k}-{\ell}_1)^2)^{1/2}}{{k}^2}\bigg(2-6\frac{({\ell}^2_1({k}-{\ell}_1)^2)^{1/2}}{{k}^2}\phi_1\sin\phi_1\\
&\qquad-\frac{1}{2}\frac{({\ell}^2_1-({k}-{\ell}_1)^2)}{{k}^2}\ln\frac{{\ell}^2_1}{({k}-{\ell}_1)^2}\bigg)\cos\phi_1\bigg]-C_a\phi^2_1 +\{1\leftrightarrow 2\}
\end{split}
\end{equation}

The second term involves a single one-dimensional integral
\begin{equation}
\begin{split}
    \frac{\dd V_{f.r.}^{(1)}}{\dd J}&=
    C_a^2\left[\frac{n_f}{6}-C_a\frac{\pi^2}{3}
      -C_a\frac{11}{6}\ln\left(\frac{\lambda^2}{\mu_R^2}\right)+\frac{N_f}{3}\ln\left(\frac{\lambda^2}{\mu_R^2}\right)
      -\ln{\left(\frac{\mu_F^2}{\mu_R^2}\right)}\frac{\beta_0}{2}\right] S^{(2)}_J(k,x)\\
    &+\!\int\!\!\dd z \bigg[C^2_a\ln\left(\frac{\lambda^2}{\mu_F^2}\right){P^{(2)}_{gg}}_+(z)+C^2_a\ln\left(\frac{\lambda^2}{\mu_F^2}\right)P^{(1)}_{gg}(z)\\
    &\qquad\qquad+2n_fC^2_f\l(\ln\left(\frac{\lambda^2}{\mu_F^2}\right)P_{qg}(z)+\frac{1}{2}\r)\bigg]S^{(2)}_J(k,zx).
    \end{split}
  \end{equation}
And the final two terms involve a one-dimensional integral and a two-dimensional momentum integral
\begin{equation}
\begin{split}
    \frac{\dd V_r^{(1)}{}_{g\to q\bar{q}}}{\dd J_1}=&n_f\!\int\!\!\dd z P_{qg}(z)\!\int\! \frac{\dd^2q}{\pi} \biggl[
    2C^2_f\frac{1}{q^2}\bigg(\frac{k^2}{q^2+p^2}S^{(3)}_J(p,q,zx;x)\\
    &\qquad\qquad\qquad\qquad-\theta(\lambda^2-q^2)S^{(2)}_J(k,(1-z)x)\bigg)\\
    &\qquad\qquad+C^2_a\frac{1}{q^2}\left(\frac{z^2k^2}{q^2+\Delta^2}-\frac{k^2}{q^2+p^2}\right)S^{(3)}_J(p,q,zx;x)\\
    &\qquad\qquad+C^2_a\frac{1}{\Delta^2}\left(\frac{z^2k^2}{q^2+\Delta^2}S^{(3)}_J(p,q,zx;x)-\theta(\lambda^2-\Delta^2)S^{(2)}_J(k,x)\right)\\
    &\qquad\qquad-\left(J_1(z,k,q,\ell_1)+J_1(z,k,q,\ell_2)\right)S^{(3)}_J(p,q,zx;x)\\
    &\qquad\qquad+\frac{1}{C^2_a}\left(J_2(k,q,\ell_1,\ell_2)-\frac{1}{p^2}\frac{k^2}{q^2+p^2}\right)S^{(3)}_J(p,q,zx;x)\biggr],\\
\end{split}
\end{equation}
and
\begin{equation}
\begin{split}
  \frac{\dd V_r^{(1)}{}_{g\to gg}}{\dd J_1}&=\!\int\!\frac{\dd z}{z}\!\int\!\frac{\dd^2q}{\pi}\biggl\{ \frac{1}{q^2}\biggl[\frac{k^2}{q^2+\Delta^2}\left(\mathcal{P}^{(1)}_{gg}(z)S^{(3)}_J(p,q,zx;x)-\mathcal{P}^{(1)}_{gg}(0)S^{(2)}_J(q,x)\right),\\
  &\qquad\qquad\qquad-\theta(\lambda^2-q^2)\left(\mathcal{P}^{(1)}_{gg}(z)S^{(2)}_J(k,(1-z)x)-\mathcal{P}^{(1)}_{gg}(0)S^{(2)}_J(k,x)\right)\biggr]\\
    &\qquad+\frac{1}{\Delta^2} \biggl[\frac{k^2}{q^2+\Delta^2}\left(\mathcal{P}^{(1)}_{gg}(z)S^{(3)}_J(p,q,zx;x)-\mathcal{P}^{(1)}_{gg}(0)S^{(2)}_J(k,x)\right)\\
    &\qquad\qquad\qquad-\left(\mathcal{P}^{(1)}_{gg}(z)-\mathcal{P}^{(1)}_{gg}(0)\right)\theta(\lambda^2-\Delta^2)S^{(2)}_J(k,x)\biggr]\\
    &\qquad+P^{(1)}_{gg}(z)\biggl[\left(J_1({k},{q},{\ell}_1,z)+J_1({k},{q},{\ell}_2,z)\right)S^{(3)}_J({p},{q},zx;x)\\
    &\qquad\qquad\qquad+J_2({k},{q},{\ell}_1,{\ell}_2)\underline{S^{(3)}_J({p},{q},zx;x)}-\frac{\theta(\lambda^2-p^2)}{p^2}S^{(2)}_J(k,zx)\biggr]\biggr\}.
\end{split}
\end{equation}

\bibliographystyle{JHEP}
\bibliography{mt-jet}

\end{document}